\documentclass[aps,prd,showpacs,amsmath,twocolumn,superscriptaddress,floatfix,nofootinbib,notitlepage]{revtex4-2}
 
\usepackage{bm,color,xcolor,amsfonts,amsmath,amssymb,epsfig,esint,orcidlink,slashed,braket}

\usepackage[utf8]{inputenc}

\usepackage[T2A]{fontenc}
\usepackage{bbm}
\usepackage{xspace}
\usepackage{enumitem}
\usepackage{tikz}
\usepackage{lipsum}
\usetikzlibrary{calc}
\usetikzlibrary{positioning}
\usepackage[german]{varioref}
\RequirePackage[compat=1.1.0]{tikz-feynman}
\RequirePackage[compat=1.1.0]{tikz-feynhand}
\usepackage{float}
\usepackage{commath}
\usepackage{booktabs}
\usepackage[normalem]{ulem}

\allowdisplaybreaks

{\large \textsc{Todo:} \normalsize \begin{itemize} \color{blue}}%
{\end{itemize}}

\newcommand{\bbnP}{[\bar{b}b]_1(nP)}
\newcommand{\bbarb}{[\bar{b}b]_1}
\newcommand{\tauEB}{\hat{\cal J}}

\newcommand{\dtauinvEB}{\Bigl[\partial_E\hat\tau^{-1}\Bigr]_{E=-E_B}}
\newcommand{\psibb}{\Psi^{(\bar{b}b)}}
\newcommand{\Mbb}{M^{(\bar{b}b)}}

\newcommand{\GeV}{\mathrm{GeV}}

\newcommand{\vep}{{\bm p}}

\newcommand{\veq}{{\bm q}}
\newcommand{\ver}{{\bm r}}

\newcommand{\veA}{{\bm A}}
\newcommand{\vetau}{{\bm \tau}}
\definecolor{green}{RGB}{44,160,44}
\definecolor{omeRevision}{RGB}{0,105,105}
\newcommand{\be}{\begin{equation}}
\newcommand{\ee}{\end{equation}}
\newcommand{\bea}{\begin{eqnarray}}
\newcommand{\eea}{\end{eqnarray}}
\newcommand{\beas}{\begin{eqnarray*}}
\newcommand{\eeas}{\end{eqnarray*}}

\newcommand{\veL}{{\bm L}}
\newcommand{\veS}{{\bm S}}
\newcommand{\ven}{{\bm n}}

\newcommand{\dqr}{\frac{q^2dq}{2\pi^2}}
\newcommand{\dqrpr}{\frac{q^{\prime2}dq'}{2\pi^2}}

\newcommand{\intk}{\int\frac{q^2dq}{(2\pi)^3}}

\newcommand{\QQ}{\bar{b}b}

\renewcommand{\vec}{\bm}

\newcommand{\Hoo}{H_{\rm MM}}

\newcommand{\mG}{m_\phi}
\newcommand{\EG}{E_\phi}

\newcommand{\En}{E_n}

\newcommand{\X}{X(3872)}
\newcommand{\Xb}{X_b}
\newcommand{\gev}{GeV}

\newcommand{\veep}{{\bm \epsilon}}
\newcommand{\veepst}{{\bm \epsilon}^{\prime*}}

\newcommand{\unit}[1]{~(#1)}
\newcommand{\BOchiEFT}{BO$\chi$EFT\xspace}

\newcommand{\VQQ}{V^{(\bar{b}b)}}
\newcommand{\VOME}{V^{\rm (OME)}}
\newcommand{\calH}{{\cal H}}
\newcommand{\hatH}{H}
\newcommand{\tilH}{\tilde{{\cal H}}}

\newcommand{\fzj}{\affiliation{Institute for Advanced Simulation
and Cluster of Excellence ``Color meets Flavor'',\\ Forschungszentrum J\"ulich, D-52425 J\"ulich, Germany}}

\newcommand{\itp}{\affiliation{Institute of Theoretical Physics, Chinese Academy of Sciences, Beijing 100190, China}}

\newcommand{\ucas}{\affiliation{School of Physical Sciences, University of Chinese Academy of Sciences, Beijing 100049, China}}

\newcommand{\tum}{\affiliation{
Technical University of Munich, TUM School of Natural Sciences, Physics Department, 85748 Garching, Germany}}

\newcommand{\scnt}{\affiliation{Southern Center for Nuclear-Science Theory (SCNT), Institute of Modern Physics,\\ Chinese Academy of Sciences, Huizhou 516000, China}}

\newcommand{\hiskp}{\affiliation{Helmholtz-Institut f\"ur Strahlen- und Kernphysik, Universit\"at Bonn, 53115 Bonn, Germany}}

\graphicspath{{Figs.dir/}}

\begin{document} 

\title{Born-Oppenheimer chiral effective field theory and its application to $\Xb$}

\author{Yu Zhang\orcidlink{0009-0007-2951-6536}}\email{zhangyu@itp.ac.cn}
\itp\ucas

\author{Feng-Kun~Guo\orcidlink{0000-0002-2919-2064}}
\email{fkguo@itp.ac.cn}
\itp\ucas\scnt

\author{Christoph~Hanhart\orcidlink{0000-0002-3509-2473}}\email{c.hanhart@fz-juelich.de}
\fzj

\author{Alexey Nefediev\orcidlink{0000-0002-9988-9430}}\email{nefediev@hiskp.uni-bonn.de}
\hiskp

\author{Paul P\"utz\orcidlink{0009-0007-0783-6520}}\email{paul.puetz@tum.de}
\tum

\begin{abstract}
Following the discovery of the exotic state $\X$ in the charmonium spectrum, the possible existence of an analogous state, $\Xb$, in the bottomonium sector has attracted considerable attention of the hadronic physics community. In this work, we lay the ground for the development of Born-Oppenheimer chiral effective field theory (\BOchiEFT), that allows us to include both quark and hadron degrees of freedom as well as the full complexity of the meson-exchange potential in a systematic way. For illustration, the formalism to leading order (LO) is applied to the system of $\chi_{b1}$ states and in particular the $\Xb$. The two parameters that are present at this order are constrained using experimental information on the bottomonium spectrum together with lattice input on static quark--antiquark potentials, including string breaking effects. The framework identifies the observed $1^{++}$ state at 10.513~GeV with a predominantly molecular $\Xb$, the bottom counterpart of the $\X$, and predicts an additional predominantly compact bottomonium state at 10.544~GeV, about 60~MeV below the $B\bar{B}^*$ threshold.
\end{abstract}

\maketitle

%\tableofcontents

\section{Introduction}

The discovery of the enigmatic state $\X$ by the Belle Collaboration in 2003~\cite{Belle:2003nnu} marked the beginning of a new era in hadron spectroscopy, ushering in the systematic study of exotic hadrons, namely, states that do not fit into the conventional quark-model classification of $q\bar q$ mesons and $qqq$ baryons. Recent reviews of the experimental and theoretical status of exotic hadrons can be found in Refs.~\cite{Ali:2017jda,Guo:2017jvc,Olsen:2017bmm,Albuquerque:2018jkn,Liu:2019zoy,Guo:2019twa,Brambilla:2019esw,Chen:2022asf,Liu:2024uxn,Chen:2024eaq,Wang:2025sic,Dai:2026fkg,Bai:2026atm}. The $\X$ is a charmonium-like state containing a $c\bar{c}$ pair but potentially exhibiting a more intricate internal structure than a conventional quark--antiquark meson. Given its well-established quantum numbers, $J^{PC}=1^{++}$, and that no charged partners have been found, the Particle Data Group (PDG) lists it as $\chi_{c1}(3872)$.\footnote{According to the PDG convention, hadron names are assigned based solely on their quantum numbers and do not imply any assumption about the underlying structure of a state.}

Since the gluon coupling to quarks is only weakly flavour-dependent through the running of the strong coupling constant, one expects the existence of the $\X$ in the charmonium spectrum to imply the presence of an analogous state in the bottomonium sector, conventionally denoted as $\Xb$. However, the implementation of heavy-quark symmetry in theoretical approaches to doubly-heavy exotic hadrons is far from straightforward. In particular, the heavy-quark-mass dependence of key ingredients, such as short-range hadron--hadron interactions and nonperturbative effects associated with the unitary cut, can be highly nontrivial. 
Nevertheless, a variety of model calculations suggest the existence of the $\Xb$ state in the energy region between 10 and 11~\gev. Examples include calculations using compact tetraquarks~\cite{Ebert:2005nc,Ebert:2008se,Ali:2009pi,Matheus:2006xi,Wang:2019mxn}, the unquenched quark model, in which the $\Xb$ emerges predominantly as a $B\bar{B}^*$ state~\cite{Ortega:2021zgk},\footnote{We use $B\bar{B}^*$ as a shorthand notation for a properly normalised $C$-even combination of $B\bar{B}^*$ and $\bar{B}B^*$; see footnote~\ref{footnoteC}.} and a dynamically generated resonance pole~\cite{Zhou:2018hlv}. 
It is stressed in Ref.~\cite{Karliner:2014lta} that the $\chi_{b1}(3P)$ listed in the Review of Particle Physics (RPP)~\cite{ParticleDataGroup:2026mpi} might actually be the $\Xb$ or a mixture of the latter with the generic $\chi_{b1}(3P)$---a scenario close to what we find in our effective field theory (EFT) study below.

A promising systematic approach to doubly heavy systems is based on the Born--Oppenheimer (BO) separation \cite{Born:1927} between the slow dynamics of the heavy quarks and the fast dynamics of the light quarks and gluons. The BO approximation has a long history in the study of heavy-quark hadrons. One of its earliest applications to a multiquark system dates back to around the turn of the 1970s and 1980s~\cite{Chao:1979tg} and then the BO description was provided for heavy hybrid quarkonium~\cite{Griffiths:1983ah,Juge:1997nc,Juge:1997nd,Juge:1999ie,Guo:2008yz}, also including spin-dependent operators up to the order $1/m_Q$~\cite{Soto:2017one} and $1/m^2_Q$~\cite{Brambilla:2018pyn,Brambilla:2019jfi}. Later, the BO approach was extended to heavy exotic $XYZ$ mesons~\cite{Braaten:2014qka}, and a series of lattice quantum chromodynamics (QCD) studies applied the BO approach to heavy tetraquarks and, later, to bottomonium resonances~\cite{Bicudo:2015vta,Bicudo:2015kna,Bicudo:2016ooe,Bicudo:2017szl,Bicudo:2019ymo,Bicudo:2020qhp,Bicudo:2022ihz}. A systematic formulation of the BO approach in the language of EFT was subsequently developed in Ref.~\cite{Brambilla:2017uyf} and dubbed BOEFT.
Heavy meson--antimeson channels and their leading couplings to
quarkonium were incorporated into BOEFT in
Ref.~\cite{TarrusCastella:2022rxb}, which also derived the
matching to a threshold EFT.
The application of BOEFT to the $1^{++}$ channel was discussed in Ref.~\cite{Berwein:2024ztx}, while the first prediction for $\Xb$ within BOEFT, but neglecting the mixing with generic bottomonium, was presented in Ref.~\cite{Braaten:2024tbm}. Further predictions for the $\chi_{b1}$ spectrum including the $\Xb$ within BOEFT can be found in 
Refs.~\cite{Brambilla:2024imu,Alasiri:2026vfz,Brambilla:2026zfw,Brambilla:2026jqz}, while its inclusive hadroproduction was studied in Ref.~\cite{Brambilla:2026ujo}. 

Predictions based on hadronic EFTs, where the $\Xb$ is treated as a molecular state, depend strongly on the assumed heavy-quark-mass scaling of the short-range potential in the $B\bar{B}^*$ channel. In Refs.~\cite{Tornqvist:1993ng,Nieves:2011zz,Guo:2013sya,Karliner:2013dqa,Ozpineci:2013zas,Karliner:2015ina,Yamaguchi:2019vea}, the short-range potential is assumed to be independent of the mass of the heavy quark, $m_Q$, leading to an $\Xb$ molecule that is much more deeply bound than the $\X$. The
underlying assumption comes from
the observation that the potential from the exchange of light mesons in the $t$ and $u$ channels is independent of the heavy quark mass. However, contributions from $s$-channel exchanges of heavy quarkonia, which introduce an explicit $m_Q$-dependence into the potential, are typically not considered in such potentials. 
In Ref.~\cite{Cincioglu:2016fkm} the hadronic framework with a contact term is enlarged by the coupling to such an $s$-channel bottomonium, with the predicted $\Xb$ mass depending on the coupling strength of $B\bar{B}^*$ to the bottomonium.
By contrast, 
inspired by renormalisation group arguments,
Ref.~\cite{AlFiky:2005jd} adopts a $1/m_Q$ scaling of the short-range potential in the $D\bar{D}^*$ and $B\bar{B}^*$ systems, which implies a $1/m_Q^3$ scaling of the binding energy and, therefore, a much more weakly bound $\Xb$ than the $ \X$. 

A systematic heavy-hadron chiral EFT ($\chi$EFT) approach to doubly heavy systems 
was formulated in Ref.~\cite{Baru:2011rs} and subsequently applied in Refs.~\cite{Du:2021zzh,Du:2019pij} to the $T_{cc}(3875)$ and the pentaquarks $P_{c\bar{c}}$, respectively (see also the review~\cite{Meng:2022ozq}). In the meantime, it was demonstrated in Ref.~\cite{Baru:2018qkb} that hadronic EFTs do not allow different heavy-quark sectors to be related in a systematically controllable manner. Thus, while being a fully developed EFT built along the lines of the chiral nuclear force (for reviews see Ref.~\cite{Epelbaum:2008ga,Epelbaum:2019kcf}), the heavy-hadron $\chi$EFT does not allow one to connect the $c\bar{c}$ sector to the $\bar{b}b$ sector, because of the arguments provided in Ref.~\cite{Baru:2018qkb}. Moreover, in its present formulation it does not allow one to study the interplay of the two-hadron states with the more compact $\bar{Q}Q$ structures.
In this work, we further develop the above-mentioned framework to overcome its shortcomings. Building on Refs.~\cite{Baru:2010ww,Hanhart:2011jz,Guo:2016bjq,Hammer:2016prh,Hanhart:2022qxq}, we present what can be viewed as the leading order (LO) calculation in an effective field theory that combines the BO concept with the concepts underlying $\chi$EFT developed for nuclear physics. We refer to the resulting theory, which is the central result of this work, as BO chiral effective field theory (\BOchiEFT).
As an illustrative application, we employ the framework to investigate the $J^{PC}=1^{++}$
sector of the bottomonium system.
All parameters are fixed by
\begin{itemize}
\item 
the experimentally measured masses of the three axial-vector bottomonium states listed in the RPP~\cite{ParticleDataGroup:2026mpi}, and 
\item lattice-QCD results for the unquenched static quark--antiquark potential with string breaking effects~\cite{Bulava:2024jpj}. 
\end{itemize}
Then the $\Xb$ and its properties emerge as predictions of the
approach.

We find that the experimentally established axial-vector bottomonium spectrum is best described by identifying the currently known state closest to the $B\bar{B}^*$ threshold, denoted as $\chi_{b1}(3P)$ in the RPP, as a predominantly molecular state, $\Xb$, providing EFT support for the phenomenological study of Ref.~\cite{Karliner:2014lta}. 
An additional, predominantly compact $\bar{b}b$ state is predicted at 10.544~GeV, about 60~MeV below the $B\bar{B}^*$ threshold. Thus, a molecular $\Xb$ may have already been observed, while an additional nearby $\chi_{b1}(3P)$ remains to be identified. Note, however, that a direct first-principle lattice-QCD study of the $\Xb$ is challenging. The large $b$-quark mass implies that $a m_b$ is sizable at currently accessible lattice spacings $a$, resulting in potentially large discretisation effects and necessitating either very fine lattices or dedicated heavy-quark formulations~\cite{Hashimoto:2004fv,Bicudo:2022cqi}. 

The paper is organised as follows.
The proposed framework is formulated in Sect.~\ref{sec:formulation}. This includes general remarks contained in Subsect.~\ref{sec:remarks}, power counting for the meson exchange and bottomonium exchange potentials introduced in Subsects.~\ref{sec:OMEcounting} and 
\ref{sec:bbcounting}, respectively, the full interaction potential built in Subsect.~\ref{sec:fullpotential}, lattice input explained in Subsect.~\ref{sec:lattice},
and BO degrees of freedom in \BOchiEFT addressed in detail in Subsect.~\ref{sec:BO}. In the two subsequent subsections of Sect.~\ref{sec:potential}, we explicitly construct the interaction potential for the $B^{(*)}\bar{B}^*$ system, including the contribution from exchanged quark--antiquark states, discussed in Subsect.~\ref{sec:VQM}, and the one-meson-exchange (OME) contribution, described in Subsect.~\ref{sec:OME}. 
The proposed formalism is applied to the 
$\Xb$ state in Sect.~\ref{sec:appXb}. In particular, the Lippmann--Schwinger equation (LSE) is formulated in Subsect.~\ref{sec:LSE}, the parameters of the framework are determined in Subsect.~\ref{sec:params}, and the probabilities of components are derived in Subsect.~\ref{sec:probabilities}. The results are presented and discussed in Subsect.~\ref{sec:Xb}, where we find evidence supporting the existence of the $\Xb$ and investigate its properties. 
Near-threshold scattering in the $B\bar{B}^*$ system is addressed in Subsect.~\ref{sec:threshold-scattering}. 
In Subsect.~\ref{sec:comparison}, we compare the results obtained within \BOchiEFT and BOEFT and, in Subsect.~\ref{sec:exp}, discuss the experimental situation around searches for the axial-vector bottomonium states. Finally, our conclusions are summarised in Sect.~\ref{sec:conclusions}.
Furthermore, Appendix~\ref{app:cornell} solves the Schr{\"o}dinger equation for the Cornell potential and discusses the calculation of transition form factors; Appendix~\ref{app:VSD} provides details of spin-dependent interactions in heavy quarkonia; Appendix~\ref{app:OPE} discusses one-pion exchange (OPE) in charmonium and bottomonium systems; Appendix~\ref{app:PWA} presents the partial-wave projection of the OME potentials; Appendix~\ref{app:tensor-off-reference} contains, for comparison, the results obtained without the OME tensor force; Appendix~\ref{app:probabilities} derives the expressions for the component probabilities; and Appendix~\ref{app:deep} addresses a spurious pole that appears in a strong attractive singular potential due to tensor forces.

\section{Towards a formulation of the \BOchiEFT}
\label{sec:formulation}

\subsection{General remarks}
\label{sec:remarks}

In Refs.~\cite{Hammer:2016prh,Hanhart:2022qxq}, a mechanism for the formation of near-threshold molecular states was proposed based on a nontrivial interplay between quark and mesonic degrees of freedom. In particular, strong unitarisation effects in a system consisting of a tower of conventional quark--antiquark states coupled to a 2-body open-flavour hadronic channel were shown to give rise to a ``collective'' state, whose pole is attracted towards a nearby open-flavour threshold.
In the present work, we put these exploratory studies on more systematic grounds by presenting a formalism that includes both heavy-quark--heavy-antiquark interactions as well as hadron--hadron interactions formulated in a way 
that lays the groundwork for the development of a full-fledged EFT. 
For illustration, we investigate the radially excited bottomonium states with quantum numbers $J^{PC}=1^{++}$ coupled to the open-bottom channels,
\be
\{B\bar{B}^*(^3S_1),B\bar{B}^*(^3D_1),B^*\bar{B}^*(^5D_1)\}
\label{BBchannels}
\ee
hereinafter labelled by Greek letter $\alpha=1,2,3$, respectively.
While at LO the quarkonium states couple directly only to the $S$-wave channel above, the OME induces couplings among all channels in Eq.~\eqref{BBchannels}.

Since isospin mass splittings in the $S$-wave bottom mesons are negligibly small, we do not distinguish between charged and neutral channels and work in the isospin limit, so the employed $B^{(*)}$ meson masses are
\be
m_B=5.279~\mbox{GeV},\quad m_{B^*}=5.325~\mbox{GeV},
\label{mBBst}
\ee
which give 
\be
\begin{split} 
\Delta m&=m_{B^*}-m_B=46~\mbox{MeV},\\
M_{\rm th}&=m_B+m_{B^*}=10.604~\mbox{GeV},\\
\mu_1&=\mu_2=\frac{m_Bm_{B^*}}{m_B+m_{B^*}}=2.651~\mbox{\rm GeV}\equiv\mu,\\
\mu_3&=\frac12 m_{B^*}=2.663~\mbox{\rm GeV},
\label{mthmu}
\end{split}
\ee
for the $B^*$-$B$ mass splitting, $B\bar{B}^*$ threshold position and the reduced masses in the two $B\bar{B}^*$ channels and the $B^*\bar{B}^*$ channel in Eq.~\eqref{BBchannels}, respectively.
Throughout this paper, we count the energy from the $B\bar{B}^*$ threshold,
\be
M=M_{\rm th}+E,
\label{energyth}
\ee
and for convenience introduce the on-shell centre-of-mass (c.m.) momentum $k$ such that
\be
E=\frac{k^2}{2\mu}.
\label{kdef}
\ee

At LO, the interaction potential between $B$ and $\bar{B}^*$ is constructed as the sum of three contributions: the exchange of the ground-state pseudoscalar mesons $\pi$ and $\eta$ with the masses \cite{ParticleDataGroup:2026mpi}
\be
m_\pi=138~\mbox{MeV},\quad m_\eta=548~\mbox{MeV},
\label{mphi}
\ee
denoted as $\VOME$, an $s$-channel interaction mediated by quark--antiquark states, $\VQQ$, and a short range contribution, $C$. In this formalism $\VQQ$ is represented by a sum in $n$ over the tower of radially excited bottomonium states
that will be denoted as $\bbnP$, where the subscript indicates that we focus on states with the heavy quark-antiquark spin 1, but consideration for the heavy quark-antiquark spin 0 is analogous. 
The wave functions and masses of the bottomonia are obtained by solving the Schr{\"o}dinger equation with the LO static colour-singlet potential. Within potential nonrelativistic QCD (pNRQCD), this potential is a nonperturbative quantity at distances relevant for the excited bottomonium states studied in this work, so we parameterise it using the Cornell form---see Eq.~\eqref{Cornell} below---consistent with the lattice study of Ref.~\cite{Bulava:2024jpj} underlying this work.

Denoting by $Q$ a generic momentum scale below the chiral symmetry breaking scale, 
\be
\Lambda_\chi \approx 1~\mbox{GeV},
\label{Lambdachi}
\ee
 we now discuss the power counting of the interactions involved.

\subsection{Power counting for $\VOME$}
\label{sec:OMEcounting}

For the light-pseudoscalar exchanges in $\VOME$, each LO chiral vertex counts as $\mathcal{O}(Q)$ [see the interaction Lagrangian in Eq.~\eqref{LPhi1} below], while the light-meson propagator counts as $\mathcal{O}(Q^{-2})$, yielding an $\mathcal{O}(Q^0)$ contribution to the potential. Within an EFT framework, the OME interaction is well defined only when supplemented by a short-range contact interaction, $C$, that explicitly depends on the hard scale of the EFT~\cite{Baru:2015nea}, which is $\Lambda_\chi$ in Eq.~\eqref{Lambdachi} when ${\chi}$EFT is involved. 
Conceptually this contact term also absorbs the effect of additional hard scales that are not included in the theory, like higher thresholds. For the case studied here this is most prominently the $B_s^{(*)}\bar{B}^{(*)}_s$ threshold, where the $B\bar{B}^*$ c.m. momentum is 0.97~GeV, numerically close to $\Lambda_\chi$. 

For $B\bar{B}^*$ scattering, the OME potential possesses a left-hand cut (lhc)
with the branch point at 
\be
p_{\mathrm{lhc}}^2= - \kappa_{\mathrm{lhc}}^2, \quad \kappa_{\mathrm{lhc}}= \frac{1}{2}\sqrt{m_\phi^2-\left(m_{B^*}-m_B\right)^2},
\label{eq:lhc-formula}
\ee
where $p$ is the momentum of the $B$ meson in the $B\bar{B}^*$ c.m. frame, and $m_\phi$ is the mass of the exchanged meson. For the masses quoted in Eqs.~\eqref{mthmu} and \eqref{mphi}, one finds 
\begin{align}
\kappa_{\mathrm{lhc}} = 65\quad\mbox{and} \quad 273~\mbox{MeV}
\label{eq:lhc-OME}
\end{align}
for the pion- and $\eta$-exchange, respectively.
These momenta are smaller than the characteristic binding momentum of the $\Xb$ obtained below, so 
the contribution of the pion- and $\eta$-exchange cannot be represented by a derivative expansion in local contact interactions over the momentum region relevant for the bound state. Thus, both these meson exchanges need to be included explicitly in $\VOME$. 

\subsection{Power counting for $\VQQ$}
\label{sec:bbcounting}

For the $s$-channel $\bbnP$ exchanges in $\VQQ$, each $S$-wave $[\bar{b}b]_1B\bar{B}^*$ vertex counts as $\mathcal{O}(Q^0)$.\footnote{There is no simple counting in powers of the heavy-quark mass and velocity for the quarkonium--heavy-meson-pair couplings; see Ref.~\cite{Guo:2010ak} for a model-based discussion.} The significance of each $s$-channel propagator is state dependent because the energy separation of different quarkonium states from the $B\bar{B}^*$ threshold varies considerably (see Table~\ref{tab:bottom} below). For a state whose energy counted from the threshold is of order $\mathcal{O}(Q^2/(2\mu))$, with $\mu$ the $B\bar{B}^*$ reduced mass in Eq.~\eqref{mthmu}, the propagator, and hence the corresponding exchange contribution, is kinematically enhanced and counts as $\mathcal{O}(Q^{-2})$. However, whether such an enhancement occurs is not known a priori.
For a state with the separation from the threshold of the order $\Lambda_\chi$ or larger, both the propagator and the exchange contribution count as $\mathcal{O}(Q^0)$. The latter is analytic in the momentum region $\sim Q$ and can, in principle, be absorbed order by order into contact meson--meson interactions. Related EFT-based discussions of shallow resonances, kinematic enhancements for near resonance poles, and explicit resonance fields can be found, for example, in Refs.~\cite{Bedaque:2003wa,Alhakami:2017ntb,Habashi:2020qgw,Habashi:2020ofb}. 
 
In the formalism to be developed here, 
quarkonium $\bbnP$ states with the principal quantum number $n~\leqslant~N$, irrespective of their energy separations from the $B\bar{B}^*$ threshold, are included in the sum in $\VQQ$ for a uniform treatment. Here $N$ should be sufficiently large to ensure that the states with a possible infrared kinematical enhancement are included explicitly. 
States with higher radial excitations do not induce any significant energy dependencies and can be absorbed by the local counterterm $C$ of the order $\mathcal{O}(Q^0)$ discussed above. 
In this way the parameter $C$ also acquires dependence on the truncation number $N$. 

\subsection{The full interaction potential}
\label{sec:fullpotential}

Summarising the discussions above, the full potential in the $B\bar{B}^*$ system at LO,
\be
V=\VOME+\sum_{n=1}^N \VQQ_n+C(\Lambda,N)\ ,
\label{Vdef}
\ee
contains the OME potential,
the $s$-channel exchanges of $\QQ$ states,
and the non-derivative contact interaction.
All these contributions are unitarised by solving a Schr{\"o}dinger equation or LSE. For distant states, retaining the full propagators amounts to a partial resummation of analytic higher-order terms. 
The truncation of the quarkonium tower and the numerical impact of this resummation can be examined by varying $N$.
Moreover, to render the scattering equation well defined, we introduce a momentum cutoff $\Lambda$ treated as the hard scale of the theory discussed above.
Results of physical observables in the fully renormalised theory should not (strongly) depend on this parameter. We checked by explicit calculations that variations of both $\Lambda$ and $N$ can indeed be compensated by variations in $C$, which will be discussed in more detail below.

Proceeding to higher orders will imply 
several modifications of the potential in Eq.~\eqref{Vdef}:
\begin{itemize}
\item Adding 
momentum-dependent counterterms as well as few-meson exchanges\footnote{Note that one-loop amplitudes can
be well approximated by polynomials~\cite{Chacko:2024cax}.} in line with the chiral expansion of the two-nucleon interaction; see Ref.~\cite{Epelbaum:2008ga} for a review.
Momentum-dependent counterterms can also generate short-distance contributions to higher partial waves as well as transitions between partial waves. In the channel considered here, the long-distance $S$-$D$ and $D$-$D$ interactions from OME are already present at LO, while derivative short-range operators will enter at higher orders.
\item Calculating the quark--antiquark states and wave functions
from an interquark potential of higher order in pNRQCD~\cite{Brambilla:1999xf,Brambilla:2000gk,Pineda:2000sz,Brambilla:2004jw}.
\end{itemize}
Clearly the two extensions above are controlled by different expansion parameters:
while the latter is organised as an expansion in momentum over the heavy quark mass, the
former is performed in powers of inter-meson momenta and pion masses over the typical hadronic scale in Eq.~\eqref{Lambdachi}, in line with the chiral expansion mentioned above.
A quantitative estimate of the associated theoretical uncertainty still needs to be developed, and the experience gained from studies of the nucleon--nucleon interaction may provide useful guidance~\cite{Epelbaum:2014efa,Epelbaum:2014sza,Furnstahl:2014xsa}. These and related issues will be investigated in subsequent work.

\subsection{Lattice input}
\label{sec:lattice}

\begin{table*}[t!]
\caption{Parameters of the potential in Eq.~\eqref{Hfit} as provided in Ref.~\cite{Bulava:2024jpj}, given in lattice units (upper row) and physical units (lower row). The conversion from lattice to physical units is performed using the scale $\sqrt{t_0}=0.1443(7)(13)$~fm.}
\label{tab:parametersalt}
\begin{ruledtabular}
\begin{tabular}{cccccccc}
 Quantity & $\hat{E}_1$ & $\hat{E}_2$ & $g_l$&$g_s$&$\sigma$& $\hat{V}_0$ &$\gamma$ \\
 \hline
In units of $t_0$ & $0.0034(3)/\sqrt{t_0}$ &$0.1005(7)/\sqrt{t_0}$&$0.0256(3)/\sqrt{t_0}$&$0.0134(4)/\sqrt{t_0}$&$0.1061(7)/t_0$&$-0.835(7)/\sqrt{t_0}$& $-0.434(11)$ \\
 Physical units &5 MeV&137 MeV&35 MeV&18 MeV &0.198 GeV$^2$& $-1.14$ GeV& $-0.434$
 \label{tab:parameters}
\end{tabular}
\end{ruledtabular}
\end{table*}

The leading input for the inter-quark potential and the coupling of quarkonia to open-flavour channels can be taken directly from unquenched lattice studies of a static quark--antiquark pair, corresponding to the exact heavy-quark limit, $m_Q\to\infty$. In Refs.~\cite{Bulava:2019iut,Bulava:2024jpj}, lattice data for the static potential between two bottom quarks---including string-breaking effects induced by light-quark $q\bar{q}$ pairs ($q=u,d,s$)---were analysed using a matrix potential of the form
\be
\begin{pmatrix}
V(r)&\sqrt{2}g_l&g_s\\
\sqrt{2}g_l&\hat{E}_1&0\\
g_s&0&\hat{E}_2
\end{pmatrix},
\label{Hfit}
\ee
defined in the basis $\{\bar{b}b, B \bar{B}, B_s\bar{B}_s\}$.
Here, $V(r)$ is the static $\bar{b}b$ potential, parameterised in the Cornell form,
\be
V(r)=V_0+\sigma r+\frac{\gamma}{r},\quad \gamma\equiv-\frac43\alpha_s<0,
\label{Cornell}
\ee
while the off-diagonal matrix elements $\sqrt{2}g_l$ and $g_s$ govern the transitions between the quark--antiquark states and the open-bottom channels,
\be
|B\bar{B}\rangle=\frac{1}{\sqrt{2}}\left(|B_u\bar{B}_u\rangle+|B_d\bar{B}_d\rangle\right)\quad\mbox{and}\quad|B_s\bar{B}_s\rangle,
\label{BBstates}
\ee
respectively. For convenience, we tagged the charged and neutral $B$ mesons by their light-quark content.

The numerical values of the parameters defining the potential in Eq.~\eqref{Hfit}, as extracted in Ref.~\cite{Bulava:2024jpj}, are summarised in Table~\ref{tab:parameters}. Not all of these parameters are used in our analysis. In particular,
\begin{enumerate}[label=(\roman*)]
\item because the $B\bar{B}^*$ c.m.\ momentum at the $B_s\bar{B}_s^*$ threshold is 0.97~GeV, a value comparable to the cutoff scale $\Lambda_\chi$ in Eq.~\eqref{Lambdachi}, we omit the $B_s\bar{B}_s^*$ channel and thus discard the parameters $\hat{E}_2$ and $g_s$;
\item we do not fix $V_0$ in Eq.~\eqref{Cornell} to the lattice value $\hat{V}_0$ given in Table~\ref{tab:parametersalt}; instead, we treat it as a free parameter;
\item rather than using a constant $\hat{E}_1$, we explicitly incorporate OME interactions supplemented by a contact term in the $B^{(*)}\bar{B}^{(*)}$ channel.
\end{enumerate}

The quark--antiquark potential in Eq.~\eqref{Cornell} can be used directly to construct the spectrum of bottomonium states with $J^{PC}=1^{++}$. What remains is to establish the relation between their coupling to an $S$-wave $B\bar{B}^*$ pair (hereafter denoted as $g$) and the coupling $g_l$ in Eq.~\eqref{Hfit}, extracted in lattice calculation in Ref.~\cite{Bulava:2024jpj}.

On the one hand, the two-meson operator---constructed following Ref.~\cite{Bali:2005fu} and employed in Refs.~\cite{Bulava:2019iut,Bulava:2024jpj}---corresponds to a pair of \emph{pseudoscalar} heavy mesons, as defined in Eq.~\eqref{BBstates}.
By writing the spin wave function of each meson as $(|{\uparrow_b\downarrow_\ell}\rangle-|{\downarrow_b\uparrow_\ell}\rangle)/\sqrt2$ (where the subscripts $b$ and $\ell$ indicate the $b$ quark and the light degree of freedom, respectively) and projecting onto the bases $\ket{j_\ell, \lambda_\ell}$ 
and $\ket{s_{\bar{b}b},m_{\bar{b}b}}$ ( $j_\ell$ denotes the angular momentum of the light degrees of freedom and $s_{\bar{b}b}$ is the total spin of the $\bar{b}b$ pair), we obtain
\be
\begin{split}
\ket{B\bar{B}}={}&\frac12\ket{0_\ell,0}\ket{0_{\bar{b}b},0}\\
&+\frac12\sum_{\lambda_\ell=\pm1,0}(-1)^{\lambda_\ell+1}
\ket{1_\ell,\lambda_\ell}\ket{1_{\bar{b}b},-\lambda_\ell}.
\label{eq:BBdecomp}
\end{split}
\ee

On the other hand, light degrees of freedom are absent in compact bottomonia, so their quantum numbers are those of the vacuum, $j_{\ell}^{PC}=0^{++}$. 
To decipher the multiplet structure dictated by heavy quark spin symmetry (HQSS) for $S$-wave interactions between hadron pairs, we observe that both $j_{\ell}^{PC}$ and $s_{\bar{b}b}^{PC}$ in the final state can be either $0^{-+}$ or $1^{--}$, with $PC$ referring to those of the light and heavy degrees of freedom correspondingly. Consequently, the spin multiplet with $j_{\ell}=0$ contains two states~\cite{Bondar:2011ev,Guo:2017jvc},
\be
0_{\ell}^{-+} \otimes 0_{\bar{b}b}^{-+}=0^{++}, \quad 0_{\ell}^{-+} \otimes 1_{\bar{b}b}^{--}=1^{+-},
\label{eq:qn0}
\ee
while the multiplet with $j_{\ell}=1$ contains four states,
\be
1_{\ell}^{--} \otimes\, 0_{\bar{b}b}^{-+}=1^{+-}, \quad 1_{\ell}^{--} \otimes 1_{\bar{b}b}^{--}=0^{++} \oplus 1^{++} \oplus 2^{++}.
\label{eq:qn}
\ee

The $J^{PC}=1^{++}$ system investigated here emerges from the $1_{\ell}^{--} \otimes 1_{\bar{b}b}^{--}$ combination in Eq.~\eqref{eq:qn}.
The static $\bar{b}b$ potential is extracted on the lattice~\cite{Bulava:2019iut, Bulava:2024jpj} from an interpolating operator corresponding to a symmetric spin state of the $\bar{b}b$ pair, specifically $\ket{1_{\bar{b}b},0}$. Selecting
the corresponding term on the right-hand side in Eq.~\eqref{eq:BBdecomp}, we find that the coupling of the state $\ket{1_\ell,0}\ket{1_{\bar{b}b},0}$ to the $J^{PC}=1^{++}$ bottomonium acquires an additional factor of 2 compared to the coupling $g_l\sqrt{2}$ from Eq.~\eqref{Hfit}; see also Sect.~III of Ref.~\cite{Bicudo:2019ymo} for a detailed discussion.

Finally, the total transition strength extracted from the static potential in Eq.~\eqref{Hfit} is distributed among the $B^{(*)}\bar{B}^*$ states in different partial waves, enumerated in Eq.~\eqref{BBchannels}, as \cite{Baru:2016iwj,Bruschini:2023zkb}\footnote{Here the sign for the $^5D_1$ term follows the sign convention in Eq.~\eqref{projectors} and is opposite to that in Ref.~\cite{Bruschini:2023zkb}.}
\be
\begin{split}
\ket{\bbnP}&\to\frac{1}{\sqrt{3}}\ket{B\bar{B}^*(^3S_1)}
+\frac{1}{\sqrt{6}}\ket{B\bar{B}^*(^3D_1)}\\&\hspace*{0.32\columnwidth}-
\frac{1}{\sqrt{2}}\ket{B^*\bar{B}^*(^5D_1)}.
\end{split}
\label{eq:PWdecomp}
\ee
Thus, the $\bbnP$ coupling to the $S$-wave $B\bar{B}^*$ pair acquires an additional factor of $1/\sqrt{3}$.

Collecting all the factors derived above, we arrive at the relation between the coupling of the $J^{PC}=1^{++}$ bottomonium states to an $S$-wave $B\bar{B}^*$ pair and the lattice coupling $g_l$ from Eq.~\eqref{Hfit},
\be
g=\frac{2}{\sqrt{3}}\sqrt{2}\, g_l= 57~\mbox{MeV},
\label{g}
\ee
where the numerical value of $g_l$ is taken from the central value listed in Table~\ref{tab:parameters}.

The value of the string--meson-pair mixing coupling provided in Eq.~\eqref{g} warrants a clarifying comment. The pseudoscalar--pseudoscalar operator employed in the lattice studies of Refs.~\cite{Bali:2005fu,Bulava:2019iut,Bulava:2024jpj} contains the normalised $\Sigma_g^+$ (see Subsect.~\ref{sec:relBOEFT} for explanation of the notation) two-meson component with a coefficient of $-1/2$, as detailed in Eq.~\eqref{eq:BBdecomp} (see also Ref.~\cite{Bruschini:2023zkb}). This factor was omitted in the seminal lattice calculation of string breaking in Ref.~\cite{Bali:2005fu}, but was later explicitly accounted for in a reanalysis of those data in Ref.~\cite{Bicudo:2019ymo}. Comparing Table~IV of Ref.~\cite{Bicudo:2019ymo} and Table~I of Ref.~\cite{Bali:2005fu} at $r\approx1.25$~fm or equivalently $r/a\approx 15$, with $a$ the lattice spacing, where string breaking occurs~\cite{Bali:2005fu}, shows that the re-extracted mixing strength is roughly doubled, increasing from about 25 to 50~MeV.
Although this latter value is numerically rather close to the chirally extrapolated result, $\sqrt{2}g_l\approx 50$~MeV, extracted in Ref.~\cite{Bulava:2019iut},\footnote{This value was used in a series of BOEFT calculations in Refs.~\cite{Brambilla:2024imu,Brambilla:2026zfw,Brambilla:2026jqz,Alasiri:2026vfz,Brambilla:2026ujo}.} the pion mass $m_\pi=654$~MeV used in Refs.~\cite{Bali:2005fu,Bicudo:2019ymo} is very large, lying closer to the physical kaon mass than to the physical pion mass. Furthermore, a strong dependence of this coupling constant on the pion mass (or, equivalently, the light-quark mass) was observed in Ref.~\cite{Bulava:2019iut} (see Fig.~6 therein), as evidenced by the relation $g_l\approx 2g_s$ that holds numerically for the chirally extrapolated values quoted in Table~\ref{tab:parameters}.
Therefore, pending a reanalysis of the lattice data from Ref.~\cite{Bulava:2024jpj} along the lines of Ref.~\cite{Bicudo:2019ymo}, applying a correction factor of 2 in Eq.~\eqref{g} serves as a sound approximation, as can be inferred from comparing the results of Refs.~\cite{Bicudo:2019ymo} and \cite{Bali:2005fu} (note that the sign for the coupling is immaterial here).

As a concluding remark, we note that the lattice parameterisation in Ref.~\cite{Bulava:2024jpj} treats the transition potentials as constants, whereas their explicit $r$-dependence has been investigated within pNRQCD in Refs.~\cite{TarrusCastella:2022rxb,Berwein:2024ztx,Brambilla:2024imu}. 
After Fourier transformation, the mentioned $r$-dependence translates to a $Q^2$-dependence that is of subleading order in our power counting.
Thus, in this LO study we use the constant coupling $g$ in Eq.~\eqref{g}, in line
with the lattice parameterisation.

\subsection{BO degrees of freedom in \BOchiEFT}
\label{sec:BO}

\subsubsection{BO quantum numbers and potentials}
\label{sec:relBOEFT}

The static $b$ and $\bar{b}$ sources fix an axis $\hat{\ver}$ in space, leading to cylindrical symmetry, $PC$ symmetry, and reflection symmetry with respect to a plane containing that axis. The irreducible representations (irreps) for these symmetries are denoted as $\Lambda_\eta^\sigma$, as adopted in the BOEFT literature, with the quantum numbers
\begin{itemize}
\item $\Lambda=|\lambda_\ell|$ for the modulus of the projection of $\bm{j}_\ell$ onto the axis along the static $\bar{b}b$ sources, with $\Sigma,\Pi,\dots$ representing $\Lambda=0,1,\dots$, respectively;
\item $\eta=g,u$ for $PC$ (even and odd, respectively) of the light degrees of freedom;
\item $\sigma=P(-1)^{j_\ell}$, which is quoted only for $\Lambda=0$.
\end{itemize}

In this notation, the interaction potential between $b$ and $\bar{b}$, with quantum numbers of the light degrees of freedom $j_{\ell}^{PC}=0^{++}$, corresponds to the string state from the Wilson line connecting the heavy quark--antiquark pair and is tagged as $\Sigma_g^+$. The situation becomes more intricate when the static two-meson channels couple to $\bar{b}b$ states, so further options open up. For $S$-wave meson pairs with $j_\ell^{PC}$ in Eqs.~\eqref{eq:qn0} and \eqref{eq:qn}, one has
\begin{itemize}
\item the longitudinal component of $1_\ell^{--}$ with $\lambda_\ell=0$ corresponds to $\Sigma_g^{+\prime}$;
\item the transverse components of $1_\ell^{--}$ with $\lambda_\ell=\pm1$ form $\Pi_g$;
\item the term $0_\ell^{-+}$ gives rise to $\Sigma_u^-$.
\end{itemize}
Thus, in addition to the $\bar{b}b$ potential $\Sigma_g^+$ from the spatial Wilson line, when considering $S$-wave meson pairs, one has two additional energy levels, $\Sigma_g^{+\prime}$ and $\Pi_g$~\cite{Berwein:2024ztx,Brambilla:2026jqz}, which account for the coupling of $\bbnP$ to the $S$-wave $B\bar{B}^*$ system as well as for the coupling between the meson pairs. Accordingly, the
light-quark states
that appear, for example,
in the decomposition Eq.~\eqref{eq:BBdecomp}, namely
$\ket{1_\ell,0}$, $\ket{0_\ell,0}$, $\ket{1_\ell,\pm 1}$, can be identified with the components in the $\Sigma_g^{+\prime}$, $\Sigma_u^{-}$, and $\Pi_g$ irreps, respectively.

The situation becomes more complicated, when $D$-wave meson pairs are taken into account. They
can be analysed by considering angular-momentum recoupling at a fixed direction $\hat{\ver}$ of the axis between the static quarks. From Eq.~\eqref{eq:qn} we have several spin components: 
the heavy spin $s_{\bar{b}b} = 1$, the angular momentum of the light degrees of freedom ${j}_\ell=1$,\footnote{It is also the total spin of the light quark--antiquark pair since, in the heavy-quark limit, the positions of the heavy hadrons coincide with those of the heavy quarks, so that their relative orbital angular momentum is identical to that of the two heavy quarks.} and the orbital angular momentum $L$ in the $B^{(*)}\bar{B}^*$ pair. Then we have two alternative equivalent descriptions of the system:
\begin{enumerate}
\item Coupling first the $\bm{s}_{\bar{b}b}$ and $\bm{j}_\ell$ spin states gives $S=1$ for the $C$-even $B\bar{B}^*$ pair and $S=2$ for $B^*\bar{B}^*$, noting that the $S=0$ states cannot couple to the $J^{PC}=1^{++}$ quantum numbers of interest. Combining these spin states with the possible orbital angular momenta $L=0,2$ yields the three partial-wave channels in Eq.~\eqref{BBchannels}.
\item Alternatively, one can first 
couple $\bm{j}_\ell$ to the orbital angular momentum to get $\bm{J}_\ell\equiv\bm{L}+\bm{j}_\ell$,\footnote{In the notation of Ref.~\cite{Berwein:2024ztx}, $\bm j_\ell$, $\bm L$, $\bm J_\ell$, and $\lambda_\ell$ are $\bm K$, $\bm L_Q$, $\bm L$, and $\lambda$, respectively, so that our $J_\ell$ is $l$ in Ref.~\cite{Berwein:2024ztx}.} which is a good quantum number in the heavy quark limit. 
For $L=0$, one has $J_\ell =1$, while,
for $L=2$, one has $J_\ell =1,2,3$. However, $J_\ell=3$ does not contribute as it cannot couple with $s_{\bar{b}b} = 1$ to form a $1^{++}$ state. In addition, 
the state $|J_\ell=2,\lambda_\ell=0\rangle$ does not contribute either. To see this, notice that $\bm{L}\cdot\hat{\ver}=0$, so the relevant Clebsch-Gordan coefficient (in the Condon--Shortley convention) for $\lambda_\ell=0$ reads
\be
\begin{split}
\langle L,&0,1,\lambda_\ell=0|J_\ell,\lambda_\ell=0\rangle\\
&=\sqrt{\frac{L+1}{2L+1}}\delta_{J_\ell,L+1}-\sqrt{\frac{L}{2L+1}}\delta_{J_\ell,L-1},
\end{split}
\ee
and, therefore, the case of $J_\ell=L=2$ is excluded. Hence, there are three $J^{PC}=1^{++}$ BO configurations~\cite{Brambilla:2026zfw}: 
$\ket{\Sigma_g^{+\prime}}$ for $J_\ell=1$ plus $\ket{\Pi_g;1}$ and $\ket{\Pi_g;2}$ for $J_\ell=1$ and 2, respectively; the number is the same as that of the partial-wave channels in Eq.~\eqref{BBchannels}. The configuration with $j_\ell^{PC}=0^{-+}$ ($\Sigma_u^-$) cannot give $J^{PC}=1^{++}$ for any $L$~\cite{Berwein:2024ztx,Brambilla:2026zfw}.
\end{enumerate}

\subsubsection{Matching BO to partial-wave potentials}
\label{sec:BOWPmatch}

The consideration above implies that there exists an orthogonal basis transformation between the three partial-wave channels in Eq.~\eqref{BBchannels} and the three relevant BO configurations, $\{\ket{\Sigma_g^{+\prime}},\ket{\Pi_g;1},\ket{\Pi_g;2}\}$. To establish this connection, let us first couple $\bm{j}_\ell$ to $\bm{L}$ in the $D$ wave and use the Wigner $6j$ symbols $ \{L\,j_\ell\,J_\ell;s_{\bar{b} b}\,J\,S\} = \{2\,1\,J_\ell;1\,1\,S\}$. This gives the states with definite $J_\ell$'s~\cite{Brambilla:2026zfw},
\be
\begin{split}
\ket{(L=2,j_\ell=1)J_\ell=1}={}&-\frac12\ket{B\bar{B}^*(^3D_1)}\\
&+\frac{\sqrt3}{2}\ket{B^*\bar{B}^*(^5D_1)},\\
\ket{(L=2,j_\ell=1)J_\ell=2}={}&\frac{\sqrt3}{2}\ket{B\bar{B}^*(^3D_1)}\\
&+\frac12\ket{B^*\bar{B}^*(^5D_1)},
\end{split}
\label{eq:l12}
\ee
Within the states with $J_\ell=1$, the projection with $\lambda_\ell=0$ contains $L=0$ and 2 as mentioned above, and those with $\lambda_\ell=\pm1$ contain $L=0$, 1, and 2, with $L=1$ excluded, since the positive parity of the system
needs to be conserved. Then, using the basis transformation formula~\cite{Jacob:1959at},
\begin{align}
&\left|J_{\ell} M_\ell; \lambda_{\ell}\right\rangle \notag\\
={}&\sum_L \sqrt{\frac{2 L+1}{2 J_{\ell}+1}}\left\langle L\, 0\, j_{\ell}\, \lambda_{\ell} \mid J_{\ell}\, \lambda_{\ell}\right\rangle\left|(L\,j_\ell) J_{\ell}\, M_\ell\right\rangle,
\end{align}
where $M_\ell$ denotes the projection of $J_\ell$ on a fixed quantisation axis (the coefficients are independent of it),
to relate the states of definite projection $\lambda_\ell$ on $\hat{\ver}$ to those of definite $L$~\cite{Berwein:2024ztx}, one finally arrives at
\be
\begin{pmatrix}\ket{\Sigma_g^{+\prime}}\\[0.5mm] \ket{\Pi_g;1}\\[0.5mm] \ket{\Pi_g;2}\end{pmatrix}
=\begin{pmatrix}
\frac{1}{\sqrt3} & \frac{1}{\sqrt6} & -\frac{1}{\sqrt2}\\[1mm]
\sqrt{\frac23} & -\frac{1}{2\sqrt3} & \frac12\\[1mm]
0 & \frac{\sqrt3}{2} & \frac12
\end{pmatrix}
\begin{pmatrix}\ket{B\bar{B}^*(^3S_1)}\\[0.5mm] \ket{B\bar{B}^*(^3D_1)}\\[0.5mm] \ket{B^*\bar{B}^*(^5D_1)}\end{pmatrix},
\label{eq:BOchannels}
\ee
with the same partial-wave coefficients in the first row as in Eq.~\eqref{eq:PWdecomp}. 

In the heavy quark limit of $m_b{\to}\infty$, (i) the kinetic energies of the $b$ and $\bar{b}$ sources vanish, (ii) there is no hyperfine splitting between $B$ and $B^*$ mesons, and (iii) the recoil terms from OME (see Subsect.~\ref{sec:OME}) vanish. Then, at each fixed $b\bar{b}$ separation $r$, the eigenenergies of the light degrees of freedom define the $r$-dependent, energy-independent BO potentials~\cite{Juge:1997nc,Braaten:2014qka}.
Furthermore, for fixed $\bar{b}b$ sources, $\ket{\Pi_g;1}$ and $\ket{\Pi_g;2}$ share the same static potential $V_{\Pi_g}(r)$. 
Then, since 
\be
\operatorname{diag}(V_{\Sigma_g^{+\prime}},V_{\Pi_g},V_{\Pi_g})=V_{\Pi_g}\bm{1}+(V_{\Sigma_g^{+\prime}}-V_{\Pi_g})\operatorname{diag}(1,0,0),
\ee
the elements of the radial potential matrix, 
\be
V(r)=U^{T}{\rm diag}(V_{\Sigma_g^{+\prime}},V_{\Pi_g},V_{\Pi_g})U,
\ee
with $U$ the orthogonal matrix in Eq.~\eqref{eq:BOchannels}, obtained by projecting on the orthonormal spin-angular basis states $\mathcal{Y}^{J=1}_{L_\alpha,S_\alpha}$ of the channels in Eq.~\eqref{BBchannels}, with $(L_\alpha,S_\alpha)=(0,1)$, $(2,1)$, $(2,2)$ for $\alpha=1,2,3$, respectively,
read
\be
\begin{split}
V_{\alpha'\alpha}(r)={}&V_{\Pi_g}(r)\,\delta_{\alpha'\alpha}\\
&+\big[V_{\Sigma_g^{+\prime}}(r)-V_{\Pi_g}(r)\big]\braket{\alpha'|\Sigma_g^{+\prime}}\braket{\Sigma_g^{+\prime}|\alpha},
\end{split}
\label{eq:Vstatic}
\ee
where, as per Eq.~\eqref{eq:BOchannels}, $\braket{\alpha|\Sigma_g^{+\prime}}=1/\sqrt3$, $1/\sqrt6$, and $-1/\sqrt2$ for $\alpha=1,2,3$, respectively.
Then, explicitly we have
\be
V(r)
=
V_C(r)\,\mathbbm{1}
+
V_T(r)
\begin{pmatrix}
0 & -\sqrt2 & \sqrt6\\
-\sqrt2 & 1 & \sqrt3\\
\sqrt6 & \sqrt3 & -1
\end{pmatrix},
\label{eq:cylindricalstuff}
\ee
with the central and tensor radial potentials,
\be
\begin{split}
V_C(r) &= \frac13 \left[V_{\Sigma_g^{+\prime}}(r)+2V_{\Pi_g}(r)\right], \\
V_T(r) &= \frac16\left[V_{\Pi_g}(r) - V_{\Sigma_g^{+\prime}}(r) \right],
\end{split}
\label{eq:VCT-BO}
\ee
respectively.

Note that the potential in Eq.~\eqref{eq:Vstatic} enters the radial Schr{\"o}dinger equation [see Eq.~\eqref{mthmu} for the definition of the reduced masses $\mu_\alpha$ and the threshold splitting $\Delta m$],
\be
\begin{split} 
{\left[-\frac{1}{2 \mu_\alpha} \frac{d^2}{d r^2}+\frac{L_\alpha\left(L_\alpha+1\right)}{2 \mu_\alpha r^2}+\delta_{\alpha 3} \Delta m\right] u_\alpha(r)} & \\ +\sum_{\alpha'} V_{\alpha \alpha'}(r)\, u_{\alpha'}(r)=E \, u_\alpha(r)&,
\end{split}
\ee
with $\left(L_1, L_2,L_3\right)=(0,2,2)$, so it
does not include the centrifugal barrier---the corresponding information sits in the wave functions. 
Then, to obtain momentum-space matrix elements of the potential $\hat{V}$, 
\be 
V_{\alpha'\alpha}(p',p)=\bra{p',\alpha'}\hat{V}\ket{p,\alpha},
\ee
one needs to take into account the free incoming and outgoing $B^{(*)}\bar{B}^*$ radial partial-wave states $|p,\alpha\rangle$,
normalised as 
\be
\langle p',\alpha'|p,\alpha\rangle
=\delta_{\alpha'\alpha}\frac{2\pi^2}{p^2}\delta(p'-p).
\label{eq:pLket}
\ee
We further employ their $r$-space representation,\footnote{For future convenience, we include a factor of $1/\sqrt{4\pi}$ into the definition of the angular part, so that $\int d\Omega |{\mathcal Y}|^2=1/(4\pi)$; see Subsect.~\ref{sec:schannel} for more details.}
\be
\braket{\ver|p,\alpha}
=4\pi\,i^{L_\alpha}j_{L_\alpha}(pr)
\mathcal{Y}^{J=1}_{L_\alpha,S_\alpha}(\hat{\ver}),
\label{eq:pw-free-coordinate}
\ee
where $j_{L}(x)$ is a spherical Bessel function, to perform a double Fourier-Bessel transform of the potential and get
\begin{align}\nonumber
V_{\alpha' \alpha}(p', p) ={}&4\pi i^{L_\alpha - L_{\alpha'}} \\
&\times\!\! \int_0^\infty \!\!\!\! dr r^2 \, j_{L_{\alpha'}}\!(p'r)\, V_{\alpha' \alpha}(r) \, j_{L_\alpha}\!(pr).
\label{eq:Vlocal-Bessel}
\end{align}

In the local static limit, the difference between $\Sigma_g^{+\prime}$ and $\Pi_g$ vanishes for $r\to 0$, and both potentials approach the same meson-pair threshold at large distances~\cite{Berwein:2024ztx},
while they may differ sizably at intermediate distances. In our formalism, 
the long-range parts of these two levels are given by the OME potential $\VOME$ evaluated in the basis of two-meson channels, Eq.~\eqref{BBchannels}. In the local static limit, $\VOME$ contributes to the radial potentials $V_{\alpha'\alpha}(r)$. The short-range parts of these levels are parameterised through a derivative expansion of local operators appearing in the two-hadron interaction terms.
This is necessary, since we use the two-hadron states explicitly as low-energy dynamical degrees of freedom, without resolving their internal structure at short distances.
Thus, at LO in \BOchiEFT,
the short range part of the hadron-hadron scattering potential in the $1^{++}$ channel is parametrised by
the constant contact term $C$, as detailed in Eq.~\eqref{Vdef}, and
additional contact terms
accompanied by powers of the incoming and outgoing momenta $p$ and $p'$ [each counted at $\mathcal{O}(Q)$] at higher orders. In coordinate space, these correspond to potentials $\propto\delta^{(3)}(\ver)$ and its derivatives.\footnote{Note that OME also contains short-range contributions, whose coefficients are fixed by the axial coupling terms in the chiral Lagrangian (here only the LO is of interest). Their separation from contact interactions is scheme and scale dependent.}
The equality of $V_{\Pi_g}(r)$ and $V_{\Sigma_g^{+\prime}(r)}$ at $r\to 0$ does not entail equality of the contributions of the $\Sigma_g^{+\prime}$ and $\Pi_g$ irreps to the contact terms at a given order.

Since the spherical Bessel functions in Eq.~\eqref{eq:Vlocal-Bessel} behave as
\be
j_L(x)=x^L \sum_{n=0}^{\infty} \frac{(-1)^n x^{2 n}}{2^n n!(2 L+2 n+1)!!}\mathop{\sim}_{x\to 0} \frac{x^L}{(2L+1)!!},
\label{jLx}
\ee
the contact terms, which come from integrating over the short-range piece, scale as $(p')^{L_{\alpha'}}p^{L_\alpha}$ for small momenta. Thus, the constant non-derivative contact term, which is the sole LO contact term $C$, contributes only to $V_{11}(p',p)$, with $L_1=0$.
At higher orders, two-derivative and four-derivative contact terms contain further short-range information in the
$S$-$S$, $S$-$D$, and $D$-$D$ channels.
This illustrates how the short-range BO potentials are encoded in the derivative expansion of \BOchiEFT.

Moreover, the off-diagonal short-distance contact terms carry additional suppression factors, which can be assigned to $\mathcal{O}(Q^2)$.
To demonstrate it, we perform the integration in Eq.~\eqref{eq:Vlocal-Bessel} in a short-distance region that is not resolved by the hadronic degrees of freedom, which should be $\mathcal{O}(\Lambda_\chi^{-1})$ in \BOchiEFT. We use the notation $V^{<}_{\alpha' \alpha}(p',p)$ to denote these short-distance pieces.\footnote{We note that the truncation prescription is scheme dependent.}
Since the static potentials associated with the same light-degrees-of-freedom multiplet share their leading short-distance terms, and their splitting starts at order $r^2$~\cite{Brambilla:2017uyf,Berwein:2024ztx}, we can write~\cite{Brambilla:2024imu} (see also Ref.~\cite{Bali:2003jq} for a discussion on hybrid potentials)
\begin{equation}
\begin{split}
V_T(r)&=b_T r^2+c_T r^4+\cdots,\\
b_T&=\frac16(b_\Pi-b_\Sigma),\quad
c_T=\frac16(c_\Pi-c_\Sigma),
\end{split}
\label{eq:bo-cutoff-splitting}
\end{equation}
where parity conservation has been considered to exclude terms with odd powers of $r$,  radiative corrections have been neglected, and $\{b_\Sigma,b_\Pi\}$ and $\{c_\Sigma,c_\Pi\}$ are the coefficients at the $r^2$ and $r^4$ terms in the multipole expansion of the BO static potentials, respectively. The mass dimensions of the coefficients $b$ and $c$ in Eq.~\eqref{eq:bo-cutoff-splitting} are three and five, so they naturally scale as $\mathcal{O}(\Lambda_{\rm QCD}^3)$ and $\mathcal{O}(\Lambda_{\rm QCD}^5)$, respectively~\cite{Brambilla:2017uyf, Berwein:2024ztx},
since these coefficients encode the nonperturbative dynamics of the light degrees of freedom. Alternatively one may use $\sqrt{\sigma}$ as the nonperturbative scale which is of a similar order of magnitude as $\Lambda_{\rm QCD}$.
Using the representation for the spherical Bessel functions in Eq.~\eqref{jLx} and defining short-distance coefficients,
\be 
\begin{split}
C^<_{T,n} &\equiv 4\pi \int _0^{1/\Lambda_\chi} dr\, r^n\, V_T(r) \\
&= \frac{4\pi}{\Lambda_\chi^n} \left[\frac{b_T}{(n+3) \Lambda_\chi^3}+\frac{c_T}{(n+5) \Lambda_\chi^5}\right],
\end{split}
\ee 
we arrive at 
\begin{align}
V_{12}^{<}\left(p^{\prime}, p\right) &=   \frac{\sqrt{2}}{15}\left[p^2 C_{T,4}^< -\left(\frac{p^{\prime 2} p^2}{6}+\frac{p^4}{14}\right) C_{T,6}^< \right]+\ldots\,, \notag\\
V_{23}^{<}\left(p^{\prime}, p\right) &= \frac{ \sqrt{3}}{225} p^{\prime 2} p^2 C_{T,6}^< +\ldots\,,\\
V_{13}^{<}\left(p^{\prime}, p\right) &= -\sqrt{3}
V_{12}^{<}\left(p^{\prime}, p\right),\notag
\end{align}
where the last relation is a consequence of HQSS, and the remaining off-diagonal elements follow from the symmetry $V_{ij}^<(p',p) = V_{ji}^<(p, p')$.

Therefore, each successive higher-order short-distance term in the multipole expansion carries an additional suppression factor of $(\Lambda_\text{QCD}/\Lambda_\chi)^2$. To maintain a uniform power counting,  we assign this factor as $\mathcal{O}(Q^2)$, although this assignment can be adjusted for specific systems. 
Combined with the aforementioned centrifugal barrier terms $(p')^{L_{\alpha'}}p^{L_\alpha}$, this counting implies that for the $1^{++}$ quantum numbers, higher-order BO potentials enter the $S$-wave $B\bar{B}^*$ channel [channel 1 in Eq.~\eqref{BBchannels}] at $\mathcal{O}(Q^2)$. In contrast, they begin contributing to the other diagonal interactions [$D$-wave channels 2 and 3 in Eq.~\eqref{BBchannels}] and to all off-diagonal interactions from at least $\mathcal{O}(Q^4)$. 
In this way, the cylindrical symmetry of the BO framework with two static sources is incorporated into \BOchiEFT. Note that while this observation yields nontrivial relations between different counterterms, it does not circumvent the fact that the counterterms of the hadronic theory are strongly cutoff-dependent and require proper renormalisation.

\section{Construction of the LO hadron--hadron potential}
\label{sec:potential}

\subsection{Interaction through bottomonium states}
\label{sec:VQM}

\begin{table*}[t!]
\caption{The reference, bare, and experimental (if available) masses and energies relative to the open-bottom threshold in Eq.~\eqref{energyth} for the seven low-lying states $\bbnP$. The parameters of the interaction are taken from Table~\ref{tab:parameters}, and the bottom quark mass in Eq.~\eqref{mb} is used.
}
\label{tab:bottom}
\begin{ruledtabular}
\begin{tabular}{lccccccc}
$n$ & 1 & 2 & 3 & 4 & 5 & 6 & 7 \\
\hline
Reference masses $M_n^{(0)}$ for $V_0=0$ (GeV)
& 9.523 & 9.890 & 10.190 & 10.454 & 10.696 & 10.920 &
11.131 \\
Reference energies $E_n^{(0)}=M_n^{(0)}-M_{\rm th}$ (GeV)
& $-1.081$ & $-0.715$ & $-0.414$ & $-0.150$ & 0.091 & 0.315 &
0.527\\
\hline
\vphantom{$\Bigl[$}Experimental masses $M_n^{\rm exp}$ (GeV) \cite{ParticleDataGroup:2026mpi}
& 9.893 & 10.255 & 10.513 & --- & --- & --- & --- \\
\hline
Bare masses $\Mbb_n$ for $V_0=V_0^{(\QQ)}$ in Eq.~\eqref{V0QM} (GeV)
& 9.876 & 10.242 & 10.543 & 10.807 & 11.048 & 11.272 &
11.484 \\
Bare energies $E_n^{(\QQ)}=\Mbb_n-M_{\rm th}$ (GeV)
& $-0.728$ & $-0.362$ & $-0.062$ & 0.203 & 0.444 & 0.668 &
0.879 \\
\end{tabular}
\end{ruledtabular}
\end{table*}

\subsubsection{Spectrum of the quark--antiquark states}
\label{sec:qmstates}

We start from evaluating the spectrum and wave functions of generic quark--antiquark states in the spectrum of bottomonium. We refer to them as ``bare'' states as they are not yet coupled to the open-bottom channels.
Since low-lying bottomonia are nonrelativistic systems, we start from the Schr{\"o}dinger equation in coordinate space,
\be
\hatH\psibb=\Mbb\psibb,
\label{Scheq}
\ee
where
\be
\hatH=2m_b-\frac{1}{m_b}\nabla^{2}+V(r) \ .
\label{H0}
\ee
To LO the potential
$V(r)$ is spin-independent.
It
 is provided in Eq.~\eqref{Cornell} and its parameters are listed in Table~\ref{tab:parameters}. They are taken from the unquenched lattice-QCD determination of the interaction potential between a static $b$ quark and a static $\bar{b}$ antiquark reported in Ref.~\cite{Bulava:2024jpj}. 
 At higher orders this potential can be improved by including the short-distance part provided by pNRQCD; see, for example, Ref.~\cite{Brambilla:2024imu}.
 
 Since the avoided level crossing of the energy levels that drives the dynamics of interest here occurs at distances larger than the inverse pion mass in the lattice calculation, we regard the values of $g_l$ and $g_s$ extracted in Ref.~\cite{Bulava:2024jpj} as a reasonable approximation, despite the fact that meson exchange was not included in that analysis. The impact of OME on these parameters, particularly at lighter pion masses, is left for further investigation.

The lattice data discussed so far are generated in the static limit, referring to $m_b\to\infty$. To connect them to experimental observables, the Schr{\"o}dinger equation in Eq.~\eqref{Scheq} requires the heavy-quark mass as input. 
We consider a relatively broad range of $m_b \in [4.3,5.3]$~GeV, which covers both the values employed in constituent quark models as well as
the kinetic mass employed in Refs.~\cite{Berwein:2024ztx,Brambilla:2024imu}. Meanwhile, we adopt
\be
m_b\simeq 4.534~\mbox{GeV},
\label{mb}
\ee
as the central value in our formalism with the full LO OME potential, which includes a tensor force, since this value yields the best agreement with the spectrum of experimentally observed bottomonium states in the $J^{PC}=1^{++}$ channel; see the discussion in Subsect.~\ref{sec:params} and Table~\ref{tab:parameters2} below.
Whenever $m_b$ is varied, the entire calculation is repeated from the beginning, while the lattice parameters in Table~\ref{tab:parameters} are kept fixed.

For the axial-vector bottomonium states $\bbnP$ with quantum numbers $J^{PC}=1^{++}$ that are in the spotlight of this investigation, $L=s_{\bar b b}=1$ for the angular momentum and total spin of the $\bar{b}b$ pair, respectively, and $n$ is the radial excitation number. Then the eigenstates of the Hamiltonian in Eq.~\eqref{H0} can be presented in the form
\be
\psibb_{n,L=1}(r)=Y_{1m}(\hat{\ver})R_{1n}(r),
\label{psi}
\ee
where $Y_{1m}(\hat{\ver})$ and $R_{1n}(r)$ are the spherical harmonics with $L=1$ and the radial wave function, respectively. We then proceed in steps.

First, in calculating the spectrum of the Hamiltonian in Eq.~\eqref{H0}, we temporarily set $V_0=0$ in the potential in Eq.~\eqref{Cornell}; this constant is restored explicitly in Eq.~\eqref{vs} below as a free parameter, which can absorb the variation of the bottom quark mass. For all other parameters, the values listed in Table~\ref{tab:parameters} are employed, and details of the numerical procedures are provided in Appendix~\ref{app:cornell}. 
The resulting masses, denoted by $M_n^{(0)}$, will be referred to as the ``reference'' masses. Correspondingly, the quantities $E_n^{(0)}$, defined relative to the 2-body threshold in Eq.~\eqref{energyth}, will be referred to as the ``reference'' energies. The results of this pure quark--antiquark calculation obtained using $m_b$ in Eq.~\eqref{mb} are presented in the upper part of Table~\ref{tab:bottom}.
It can also be verified explicitly that the variation of the subleading spin-dependent corrections to the mass $\Mbb$ in Eq.~\eqref{Scheq} (more specifically, the spin-orbital and spin-tensor ones since the hyperfine term vanishes for orbitally excited states) in the $1/m_b$ expansion \cite{Eichten:1980mw,Gromes:1984ma} remains within approximately $1$~MeV for all states considered in the analysis; see Appendix~\ref{app:VSD} for details. Therefore, these corrections can be absorbed by the parameter $V_0$.

Second, to improve the agreement between the predictions and the experimental spectrum, we fix $V_0^{(\QQ)}$ by minimising the sum of squared deviations,
\be
(\Delta M^{(\QQ)})^2=\frac13\sum_{n=1}^3\Bigl(M_n^{(0)}+V_0^{(\QQ)}-M_n^{\rm exp}\Bigr)^2,
\ee
with respect to $V_0^{(\QQ)}$. For $m_b$ in Eq.~\eqref{mb}, this procedure yields
\be
V_0^{(\QQ)}= 353~\mbox{MeV},\quad \Delta M^{(\QQ)}_{\rm min}= 21~\mbox{MeV},
\label{V0QM}
\ee
and the resulting shifted spectrum is presented in the last two rows of Table~\ref{tab:bottom}. We therefore summarise by highlighting several conclusions that follow from the results in this table and Eq.~\eqref{V0QM}:
\begin{itemize}[leftmargin=*]
\item As expected,
the LO calculation without $B\bar{B}^*$ contribution, which agrees with 
the nonrelativistic potential quark model defined by the Hamiltonian in Eq.~\eqref{H0} and the interaction in Eq.~\eqref{Cornell}, is capable of reproducing the gross features of the $1^{++}$ bottomonium spectrum.
\item Only the three lowest $\bbnP$ bottomonium states with $n=1,2,3$ are predicted to lie below the $B\bar{B}^*$ threshold, while the states with $n\geqslant 4$ appear above it.
\item The experimental masses of the three observed low-lying bottomonium states can be reproduced with a root-mean-square deviation of 21~MeV for the masses given in Table~\ref{tab:bottom}.\footnote{For the central value of $m_b$ in Eq.~\eqref{mb}, the experimental masses of the three established axial-vector bottomonia ($\chi_{b1}(1P)$, $\chi_{b1}(2P)$, and $\chi_{b1}(3P)$ in RPP \cite{ParticleDataGroup:2026mpi}) quoted in Table~\ref{tab:bottom}, can be reproduced as eigenvalues of the quark--antiquark Hamiltonian in Eq.~\eqref{Scheq} only by adopting an unphysically small string tension, $\sigma=0.120~\mbox{GeV}^{2}$, together with a Coulomb strength parameter $\gamma=-0.920$, corresponding to an unnaturally large value of the strong coupling constant, $\alpha_s=0.69$. The overall shift of the spectrum is $V_0=827$~MeV in this case.}
\end{itemize}

The origin of the discrepancy between the masses of the quark--antiquark states and the experimental spectrum can be identified by noting that the difference between the theoretical level spacings,
\be
\Bigl(M_2^{(\QQ)}-M_1^{(\QQ)}\Bigr)-\Bigl(M_3^{(\QQ)}-M_2^{(\QQ)}\Bigr)
\approx 66~\mbox{MeV},
\ee
is considerably smaller than the corresponding experimental value,
\be
\Bigl(M_2^{\rm exp}-M_1^{\rm exp}\Bigr)-\Bigl(M_3^{\rm exp}-M_2^{\rm exp}\Bigr)
\approx 104~\mbox{MeV}.
\ee
In other words, the pure quark--antiquark calculation fails to reproduce the reduction of the level spacing between successive bottomonium states as the open-bottom threshold is approached, where effects from couplings to the continuum are expected to play an essential role in the dynamics of the system. Such continuum-induced modifications of the heavy-quark spectra have a long history since the seminal paper on the Cornell model~\cite{Eichten:1978tg}. 
For some early works, see the review in Ref.~\cite{Tornqvist:1984xz} on the account for the quark loops in the unitarised quark model and Refs.~\cite{Brisudova:1998wq,Brisudova:1999ut} for their phenomenological account through the effective screening of the confining interaction in quarkonia. Later, such modifications were investigated in coupled-channel descriptions~\cite{Eichten:2004uh,Kalashnikova:2005ui,Liu:2011yp,Lu:2016mbb,Ni:2025gvx,Zhang:2026nmz}, including applications to the $ \X$, charmonium around 3.9~GeV, and the $\Xb$~\cite{Ortega:2009hj,Cincioglu:2016fkm,Ortega:2017qmg,Ortega:2021zgk}. Related formulations treat compact quarkonium and two-hadron degrees of freedom simultaneously at the hadronic level~\cite{Baru:2010ww,Hanhart:2011jz,Guo:2016bjq,Zhou:2017dwj,Xiao:2023lpv,Wang:2023ovj,Shi:2024llv}.
One of the main goals of the present work is to systematically account for such effects by incorporating the full complexity of the hadron--hadron interaction within a $\chi$EFT framework that also includes compact quark--antiquark states.

\subsubsection{Potential of $s$-channel exchanges through bottomonia}
\label{sec:schannel}

To construct the $s$-channel interaction, we first isolate the $S$-wave $B\bar{B}^*$ component in the set of states in Eq.~\eqref{BBchannels} that couples directly to the bottomonia ${\bbnP}$. The $D$-wave components and their OME couplings are restored in the full partial-wave equation; see the discussion in Subsects.~\ref{sec:OME} and \ref{sec:LSE} below. 
As before, all energies are measured relative to the two-meson threshold as detailed in Eq.~\eqref{energyth}. Within this $S$-wave construction, the physical state is a superposition of an infinite quarkonium tower and the $S$-wave continuum state,
\be
|\Psi\rangle=
\begin{pmatrix}
\lambda_1\ket{\bbarb(1P)}\\ \lambda_2\ket{\bbarb(2P)}\\ \cdots\\ \phi(q)\ket{B\bar{B}^*}
\end{pmatrix},
\label{state}
\ee
normalised as $\braket{\Psi|\Psi}=1$, which implies
\be
\sum_n|\lambda_n|^2+\int\frac{q^2dq}{(2\pi)^3}|\phi(q)|^2=1,
\label{eq:state-normalisation}
\ee
where, for future convenience, the spin-angular wave function of the $B\bar{B}^*$ pair is included in the ket vector $\ket{B\bar{B}^*}$.

\begin{figure}[t!]
\centerline{\includegraphics[width=0.8\columnwidth]{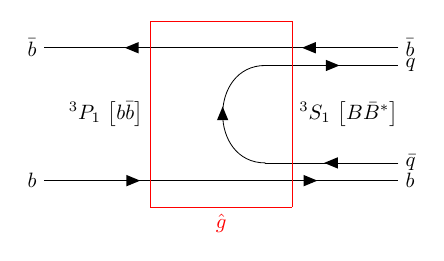}}
\caption{A sketch of the transition in the $J^{PC}=1^{++}$ channel between the $^3P_1$ bottomonium state and the $S$-wave $B\bar{B}^*$ system mediated by the operator $\hat{g}$ in the matrix Hamiltonian in Eq.~\eqref{H}.
}
\label{fig:transition}
\end{figure}

The multicomponent wave function in Eq.~\eqref{state} satisfies a Schr{\"o}dinger equation,
\be
\calH\ket{\Psi(E)}=E\ket{\Psi(E)},
\label{Sheq}
\ee
with the matrix Hamiltonian 
\be
\calH=
\begin{pmatrix} \hatH-M_{\rm th} & 0 & \cdots & 0 & \hat g \\ 0 & \hatH-M_{\rm th} & \cdots & 0 & \hat g \\ \vdots & \vdots & \ddots & \vdots & \vdots \\ 0 & 0 & \cdots & \hatH-M_{\rm th} & \hat g \\ \hat g & \hat g & \cdots & \hat g & \Hoo
\end{pmatrix},
\label{H}
\ee
where $\hatH$ is defined in Eq.~\eqref{H0}, $\Hoo$ denotes the meson-meson Hamiltonian,
and $\hat{g}$ is the transition operator between the $\bar{b}b$ and $B\bar{B}^*$ states, whose matrix elements are connected with the parameter $g_l$ introduced in Eq.~\eqref{Hfit}. Written in a component form, the Schr{\"o}dinger equation in Eq.~\eqref{Sheq} reads
\be
\begin{split}
\lambda_n \En+\int\frac{q^2dq}{(2\pi)^3}\phi(q)\braket{\bbnP|\hat{g}|B\bar{B}^*}&=\lambda_n E,\\ 
\sum_n\lambda_n\hat{g}\ket{\bbnP}+\Hoo\ket{B\bar{B}^*}&=E\ket{B\bar{B}^*},
\label{system}
\end{split}
\ee
where the first equation is obtained by projecting onto $\bra{\bbnP}$, which gives rise to the transition form factor $\braket{\bbnP|\hat{g}|B\bar{B}^*}$ between the bottomonium state with $J^{PC}=1^{++}$ and the $S$-wave $B\bar{B}^*$ system, as sketched in Fig.~\ref{fig:transition}. To evaluate this form factor, we employ the coordinate-space wave functions of the $S$-wave $B\bar{B}^*$ pair [\emph{c.f.} Eq.~\eqref{eq:pw-free-coordinate}],
\be
\braket{\ver|B\bar{B}^*}=4\pi j_0(pr)\mathcal{Y}_{L=0,S=1}^{J=1}(B\bar{B}^*;\hat{\ver}),
\label{BBr}
\ee
and the $n$-th bottomonium state,
\be
\braket{\ver|\bbnP} = R_{1n}(r)\mathcal{Y}_{L=1,S=1}^{J=1}(\bar{b}b;\hat{\ver}),
\label{chibr}
\ee
where, in both cases above, $\mathcal{Y}_{L,S}^{J=1}$ denotes the spin-angular wave function of the system with the total spin $J=1$.
The radial wave functions $R_{1n}(r)$ for the bottomonium states with
orbital angular momentum $L=1$, defined in Eq.~\eqref{psi}, were previously obtained as solutions of the Schr{\"o}dinger equation in Eq.~\eqref{Scheq}.

Then it is straightforward to find for the transition form factor,
\be
\braket{B\bar{B}^*|\hat{g}|\bbnP}=g\psi_n(p),
\label{psigMM}
\ee
where $p$ is the relative momentum in the $B\bar{B}^*$ system,
\be
\psi_n(p)=
4\pi \int_0^\infty r^2dr\, j_0(pr)R_{1n}(r),
\label{psinp}
\ee
and
\be
g=\int d\Omega \left[\mathcal{Y}_{L=0,S=1}^{J=1}(B\bar{B}^*;\hat{\ver})\right]^*\hat{g}\left[\mathcal{Y}_{L=1,S=1}^{J=1}(\bar{b}b;\hat{\ver})\right]
\label{gcoupling}
\ee
is the $n$-independent coupling constant that accounts for the recoupling of the spin and orbital degrees of freedom between the $\bar{b}b$ pair and the $S$-wave $B\bar{B}^*$ system, both in the state with $J^{PC}=1^{++}$. Its relation to the constant $g_l$ extracted in lattice-QCD studies in Ref.~\cite{Bulava:2024jpj} from the unquenched static potential (see its value quoted in Table~\ref{tab:parameters}) is provided in Eq.~\eqref{g}. Notice that the coordinate $r$ 
denotes the $\bar{b}b$ separation in the compact bottomonium channel [see Eq.~\eqref{chibr}] and at the same time the relative c.m. coordinate of the two heavy--light mesons in the open-bottom channel [see Eq.~\eqref{BBr}].
These two definitions agree at LO in the heavy mass expansion, since the heavy-quark positions are preserved during the transition. 
The relation in Eq.~\eqref{psigMM} can be viewed as a particular implementation of the Franck--Condon principle \cite{Franck:1926,Condon:1926,LlanesEstrada:2008fc,TorresRincon:2010fc}, according to which the heavy quarks act as spectators during the transition.

Then, extracting $\lambda_n$ from the first relation in Eq.~\eqref{system},
\be
\lambda_n=\frac{1}{E-\En}\intk \phi(q)\braket{\bbnP|\hat{g}|B\bar{B}^*},
\label{eq:lamn}
\ee
and employing it in the second relation, the latter can be brought to the form
\be
\Hoo\ket{B\bar{B}^*}+\int \frac{q^2dq}{(2\pi)^3} \phi(q)\hat{V}^{(\QQ)}\ket{B\bar{B}^*}=E\ket{B\bar{B}^*},
\label{BBsttransitions}
\ee
where the second term on the left-hand side admits a natural interpretation as the $B\bar{B}^*$ interaction mediated by the $s$-channel exchange of the bottomonium states, and the corresponding induced energy-dependent effective potential is
\be
\hat{V}^{(\QQ)}
=
\sum_{n=1}^{\infty}
\frac{\hat g\ket{\bbnP}
\bra{\bbnP}\hat g}{E-E_n} .
\ee
Then, for the potential in momentum space we find\footnote{If one needs to extend the analysis into the complex momentum plane beyond the bound-state poles on the physical Riemann sheet, which are the primary focus of this work, the bilinear form $\psi_n(p')\psi_n^*(p^*)$ should be used, which provides the analytic continuation
away from the real momentum axis.}
\be
\begin{split}
\VQQ&(\vep',\vep)=\braket{\vep'|\hat{V}^{(\QQ)}|\vep}=\sum_{n=1}^\infty \VQQ_n(\vep',\vep)\\
&=\frac{g^2}{4\pi}\sum_{n=1}^\infty\frac{\psi_n(p')\psi_n^*(p)}{E-\En},
\end{split}
\label{Vgamma}
\ee
where $\vep'$ and $\vep$ denote the outgoing and incoming momenta of the open-bottom mesons in their c.m. frame, respectively, the relation in Eq.~\eqref{psigMM} was used, and
we explicitly took into account $Y_{00}(\hat{\vep})=1/\sqrt{4\pi}$ for the $S$-wave angular wave function of the $B\bar{B}^*$ system. This restores the standard normalisation of the potential. Consequently, hereafter the integration measure in all momentum integrals is taken to be $d^3q/(2\pi)^3$ including the angular integration, rather than $q^2dq/(2\pi)^3$ employed above.
The formal expression provided in Eq.~(\ref{Vgamma}) to be used in Eq.~\eqref{Vdef} is the same also 
at higher orders of the inter-quark potential. What needs to be adapted then are the vertex functions derived from the potential, that is, $\{\psi_n(p)\}$ and the bare input energies $E_n$. Moreover, at higher orders, also the coupling of the $\bar{b}b$ states to the $D$-wave two-hadron states needs to be included, thus turning $V^{(\bar{b}b)}$ to a matrix in the channel space.

The $S$-wave projection of the potential in Eq.~\eqref{Vgamma} is trivial, since by construction it contains only $S$-waves
(below we use the subscript $S$ to indicate its $S$-wave projection). Within the given EFT framework,
the sum in Eq.~\eqref{Vgamma} is truncated at a finite $N$ and the
neglected part of the exchanges through high-lying states is absorbed into an $N$-dependent short-ranged contact term, so
\be
V^{(\QQ)+{\rm CT}}_S(p',p)=\frac{g^2}{4\pi}
\sum_{n=1}^N\frac{\psi_n(p')\psi^*_n(p)}{E-\En}+C(N),
\label{VQMab}
\ee
where the superscript $(\QQ)+\rm{CT}$ stresses that for convenience the contact term $C$ is included in the interaction potential through the quark--antiquark states.
Below we demonstrate by explicit calculations that variations in $N$ can be absorbed into variations of $C$,
as long as $N\geqslant 4$.

By construction
the potential in Eq.~\eqref{VQMab} takes a separable form,
\be
V^{(\QQ)+{\rm CT}}_S(p',p)=f^T(p')\hat{v}(E)f^*(p),
\label{VQMsep}
\ee
where
\be
f_n(p)=
\begin{cases}
\dfrac{1}{\sqrt{4\pi}}\,\psi_n(p), & n=1,..,N,\\[8pt]
\quad 1, & n=N+1,
\end{cases}
\label{eq:fn}
\ee
while
\be
\hat{v}=\mbox{diag}(v_1,..,v_N,v_{N+1}),
\label{eq:vhat}
\ee
with
\be
v_n=
\begin{cases}
\dfrac{g^2}{E-(E_n^{(0)}+V_0)}, & n=1,..,N,\\[8pt]
\quad C, & n=N+1.
\end{cases}
\label{vs}
\ee
In the last equation, we used the relation between the bare energy $E_n$
of the $n$-th bottomonium state and its reference energy $E_n^{(0)}$ obtained as the eigenenergy of the Hamiltonian in Eq.~\eqref{H0} with $V_0=0$ (see Table~\ref{tab:bottom}). This way we made the dependence of the potential in Eq.~\eqref{VQMsep} on the parameter $V_0$ explicit.

In this LO calculation we do not quantify uncertainties, so we refrain from keeping those of the input parameters in the analysis.
This completes the construction of the potential $\VQQ$ ($V^{(\QQ)+{\rm CT}}$), which contains only one (two) free parameter(s), the overall shift $V_0$ (and the contact term $C$). 

\subsection{Interaction through OME}
\label{sec:OME}

\begin{figure}[t]
\centerline{\includegraphics[width=0.8\columnwidth]{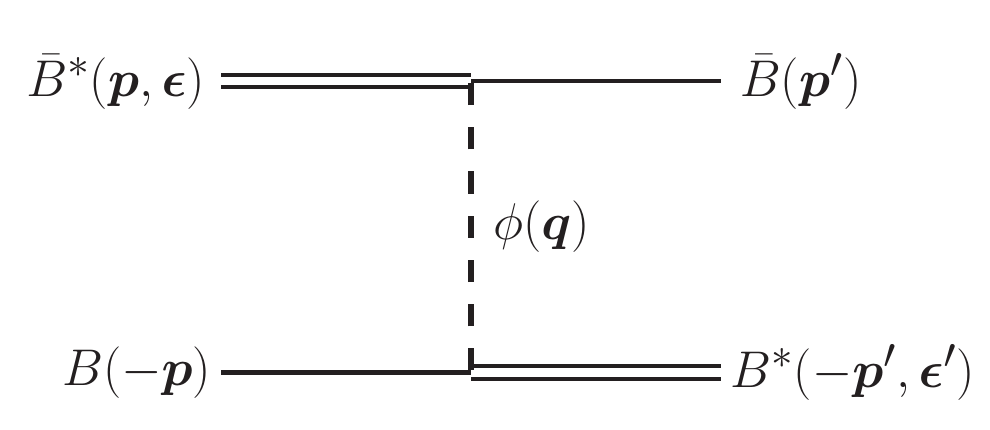}}
\caption{Feynman diagram for the OPE interaction in a $B\bar{B}^*$ system in the c.m. frame. In parentheses, we indicate the momentum and (if applicable) polarisation vector of the corresponding particle. Similar diagrams involving the $B^*B^*\phi$ and $\bar{B}^*\bar{B}^*\phi$ vertices can be straightforwardly obtained from the depicted one by replacing the corresponding $B$'s by $B^*$'s.}
\label{fig:ome}
\end{figure}

We choose the charge-conjugation phase convention such that $\mathcal C P_a\mathcal C^{-1}=\bar P_a$ and $\mathcal C V_a^i\mathcal C^{-1}=-\bar V_a^i$,\footnote{With this convention, the positive $C$-parity combination of $B\bar{B}^*$ and $B^*\bar{B} $ is $\left|B \bar{B}^* ; C=+\right\rangle=\left(\left|B \bar{B}^*\right\rangle-\left|B^* \bar{B}\right\rangle\right)/{\sqrt{2}}$.\label{footnoteC}} where $\mathcal C$ is the charge conjugation operator. With the charge conjugation $\mathcal C H_a\mathcal C^{-1}=\sigma_2\bar H_a^T\sigma_2$~\cite{Fleming:2008yn}, the heavy- and antiheavy-meson superfields read~\cite{Manohar:2000dt, Hu:2005gf}
\begin{equation}
H_a=P_a+\bm{V}_a\cdot\bm{\sigma},\quad
\bar{H}_a=\bar{P}_a+\bm{\bar{V}}_a\cdot \bm{\sigma},
\end{equation}
which annihilate the heavy-light $\bar{q}b$ and $\bar{b}q$ mesons, respectively, with $P_a=\{B^-,\bar{B}^0\}$, $V_a=\{B^{*-},\bar{B}^{*0}\}$, $\bar P_a=\{B^+,B^0\}$, and $\bar V_a=\{B^{*+},B^{*0}\}$.
Here $\bm{\sigma}$ denotes the spin Pauli matrices.

\begin{figure*}[t!]
\begin{tabular}{ccccccc}
\epsfig{file=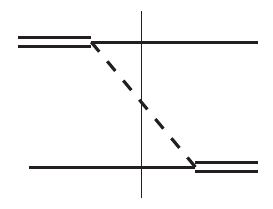, width=0.4\columnwidth}&\qquad&
\epsfig{file=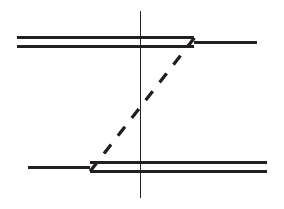, width=0.4\columnwidth}&\qquad&
\epsfig{file=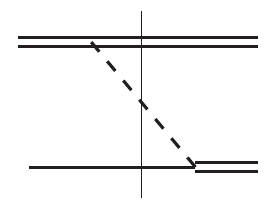,width=0.4\columnwidth}&\qquad&
\epsfig{file=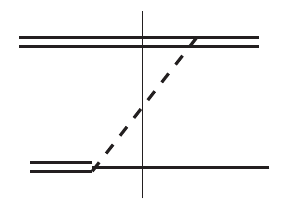,width=0.4\columnwidth}\\
(1)&&(2)&&(3)&&(4)
\end{tabular}
\caption{TOPT diagrams corresponding to the 3-body intermediate states (pinpointed by the thin vertical lines) enumerated in Eq.~\eqref{BBphichannels} and contributing to the OME potentials in Eqs.~\eqref{VBBstBBst}-\eqref{VBstBstBstBst}. The single and double solid lines represent the $B$ and $B^*$ mesons, respectively, and the dashed line represents the exchanged boson $\phi$.}
\label{fig:TOPT}
\end{figure*}

The LO chiral Lagrangian describing the axial coupling of heavy--light mesons with the light pseudoscalar bosons, which are the pseudo-Nambu-Goldstone bosons of the spontaneously broken chiral symmetry, reads~\cite{Wise:1992hn,Burdman:1992gh,Yan:1992gz} 
\be
\mathcal{L}_{\Phi}=-\frac{g_b}{4} \operatorname{Tr}\left(H_a^{\dagger} H_b \boldsymbol{\sigma} \cdot \boldsymbol{u}_{b a}\right)+\frac{g_b}{4} \operatorname{Tr}\left(\boldsymbol{\sigma} \cdot \boldsymbol{u}_{a b} \bar{H}_b \bar{H}_a^{\dagger}\right),
\label{LPhi1}
\ee
where the bottom-meson axial coupling, 
\be
g_b=0.56(3)_{\rm stat}(7)_{\rm sys},
\label{gb}
\ee
is determined at the physical bottom-quark mass employing the lattice-QCD techniques~\cite{Flynn:2015xna}. We will use its central value.

The axial current is given by $\vec{u}=-\vec{\nabla}\Phi/F_\pi+\mathcal{O}(\Phi^3)$, with $F_\pi=92.2$~MeV for the pion decay constant and the matrix of pseudoscalar fields reads
\be
\Phi=\pi_a\tau_a+\frac{\eta}{\sqrt{3}}\mathbbm{1}_2
=\begin{pmatrix}
\pi^0+\frac{1}{\sqrt{3}}\eta & \sqrt{2}\pi^+\\
\sqrt{2}\pi^- & -\pi^0+\frac{1}{\sqrt{3}}\eta
\end{pmatrix},
\label{eq:u-phi-def}
\ee
with $\tau_a$ $(a=1,2,3)$ the Pauli matrices in the flavour space.
The LO Lagrangian describing the coupling of a single $\Phi$ field to a pair of heavy--light mesons takes the form 
\begin{align}
\mathcal{L}_{\Phi}=\frac{g_b}{2F_\pi}\big[&
(V_a^{i\dagger}P_b+P_a^{\dagger}V_b^{i})\partial^{i}\Phi_{ba}\notag\\
&-(\bar{P}_a^{\dagger}\bar{V}_b^{i}+\bar{V}_a^{\dagger i}\bar{P}_b)
\partial^{i}\Phi_{ab} \label{eq:Lpi}\\
&+ i\varepsilon_{ijk}V_a^{i\dagger}V_b^j\partial^k\Phi_{ba}
+i\varepsilon_{ijk}\bar V_a^{i\dagger}\bar V_b^j\partial^k\Phi_{ab}\big].\notag
\end{align}
The Lagrangian in Eq.~\eqref{eq:Lpi} gives rise to two types of interaction vertices for $B^{(*)}$ mesons (and similarly for $\bar{B}^{(*)}$'s),
\begin{align}
v^a_{B^*B\phi}&=\frac{g_b}{2 F_\pi}\tau^a(\veep\cdot\veq),
\label{vBBst}\\
v^a_{B^*B^*\phi}&=-\frac{g_b}{\sqrt2 F_\pi}\tau^a(\veA \cdot\veq),
\label{vBstBst}
\end{align}
with $\veA=\frac{i}{\sqrt{2}}(\veep\times \veep^{\prime*})$; see, for example, Refs.~\cite{Nieves:2011zz,Baru:2016iwj}. Here $\veep\equiv\veep_{\lambda}(\vep)$ and $\veep^{\prime*}\equiv \veep_{\lambda'}^*(\vep')$ denote the polarisation vectors of the incoming and outgoing $B^*$ mesons, respectively (we have omitted the polarisation indices for brevity), and $\vep$ and $\vep'$ denote the incoming and outgoing momenta, and $\veq=\vep-\vep'$, as illustrated in Fig.~\ref{fig:ome}. 

In time-ordered perturbation theory (TOPT), OME in the $B^{(*)}\bar{B}^*$ system involves the 3-body intermediate states
\be
\{B\bar{B}\phi,B^*\bar{B}^*\phi,B\bar{B}^*\phi,\bar{B}^*B\phi\},
\label{BBphichannels}
\ee
as illustrated in Fig.~\ref{fig:TOPT}. 
In what follows we use the index $\beta=1,..,4$, respectively, to label them. Then, for the interaction potentials generated by the exchange of a generic light pseudoscalar boson, $\phi=\pi,\eta$, in the $B^{(*)}\bar{B}^*$ system, one obtains
\begin{align}
&\VOME_{B\bar{B}^*\to B\bar{B}^*}(\vep',\vep;m_\phi)\label{VBBstBBst}\\
&=-\frac{g_b^2}{4F_\pi^2}(\veep\cdot \veq)(\veepst\cdot \veq) \left[D_{1}(\vep',\vep;m_\phi)+D_{2}(\vep',\vep;m_\phi)\right],\notag
\\[3mm]
&\VOME_{B\bar{B}^*\to B^*\bar{B}^*}(\vep',\vep;m_\phi)\label{VBBstBstBst}\\
&=\frac{g_b^2}{4F_\pi^2}\Big[({\veA_1}\cdot \veq)({\veep_2}^{\prime *}\cdot \veq)+({\veA_2}\cdot \veq)({\veep_1}^{\prime *}\cdot \veq)\Big] \notag\\
&\hspace*{0.3\columnwidth}\times\left[D_{3}(\vep',\vep;m_\phi)+D_{2}(\vep',\vep;m_\phi)\right],\notag
\\[3mm]
&\VOME_{B^*\bar{B}^*\to B\bar{B}^*}(\vep',\vep;m_\phi)\label{VBstBstBBst}\\
&=\frac{g_b^2}{4F_\pi^2}\Big[({\veA_1}\cdot \veq)({\veep_2}\cdot \veq)+({\veA_2}\cdot \veq)({\veep_1}\cdot \veq)\Big] \notag\\
&\hspace*{0.3\columnwidth}\times\left[D_{4}(\vep',\vep;m_\phi)+D_{2}(\vep',\vep;m_\phi)\right],
\notag\\[3mm]
&\VOME_{B^*\bar{B}^*\to B^*\bar{B}^*}(\vep',\vep;m_\phi)\label{VBstBstBstBst}\\
&\hspace*{0.15\columnwidth}=-\frac{g_b^2}{2F_\pi^2}(\veA_1\cdot \veq)(\veA_2\cdot \veq) \left[ 2D_{2}(\vep',\vep;m_\phi)\right], \notag
\end{align}
where the TOPT propagators take the form
\be
D_\beta(\vep',\vep;\mG)=\frac1{2\EG(q)\,[\EG (q)+\mu_\beta]},
\label{eq:Dbeta}
\ee
with $\beta=1,..,4$ enumerating the 3-body intermediate states in Eq.~\eqref{BBphichannels}. Here, $\veq=\vep-\vep'$, $\EG(q)=\sqrt{q^2+\mG^2}$, and
nonrelativistically expanded $\mu_\beta$ functions read
\begin{align}
\mu_1&=E_B(p')+E_B(p)-M-i\epsilon \notag\\
&= -\Delta m+\frac{p^2+p^{\prime2}}{2m_B}-E-i\epsilon\ , \notag \\
\mu_2&=E_{B^*}(p)+E_{B^*}(p')-M-i\epsilon \notag\\ &= \Delta m+\frac{p^2+p^{\prime2}}{2m_{B^*}}-E-i\epsilon \ , \notag\\
\mu_3&=E_{B}(p)+E_{B^*}(p')-M-i\epsilon \notag \\ &=\frac{p^2}{2m_{B}}+
\frac{p^{\prime2}}{2m_{B^*}}-E-i\epsilon \ , \label{eq:alphaOME}
 \\
\mu_4&=E_{B}(p')+E_{B^*}(p)-M-i\epsilon \notag \\ &=\frac{p^{\prime2}}{2m_{B}}+
\frac{p^2}{2m_{B^*}}-E-i\epsilon \ . \notag 
\end{align}
Note that the isospin factor $\vetau\cdot\vetau^c$, with $\vetau^c=\tau_2\vetau^T\tau_2=-\vetau$ for the charge conjugated isospin operator, is not included in Eqs.~\eqref{VBBstBBst}-\eqref{VBstBstBstBst}. For the isoscalar channel the corresponding factors enter explicitly in Eq.~\eqref{VOME} below as $3$ for the pion exchange and $1/3$ for the $\eta$ exchange.
In Eqs.~\eqref{VBBstBstBst}--\eqref{VBstBstBstBst}, in which a $B^*\bar{B}^*$ pair appears, the subscripts $1$ and $2$ refer to the $B^*$ and the $\bar{B}^*$ of that pair: $\veep_1$, $\veep_2$ ($\veep_1^{\prime *}$, $\veep_2^{\prime *}$) are their incoming (outgoing) polarisations, and $\veA_a=\frac{i}{\sqrt2}(\veep_a\times\veep_a^{\prime *})$, with $a=1,2$, is the vertex structure in Eq.~\eqref{vBstBst} on the corresponding meson line. The two terms in Eqs.~\eqref{VBBstBstBst} and \eqref{VBstBstBBst} are the contributions of the two components, $B\bar{B}^*$ and $B^*\bar{B}$, of the $C$-even $B\bar{B}^*$ state; see footnote~\ref{footnoteC}.

Remarkably, since the potentials in Eqs.~\eqref{VBBstBBst}-\eqref{VBstBstBstBst} involve two $P$-wave vertices, they contain both $S$- and $D$-wave components. Let us comment on these components and the related effects employing the potential in Eq.~\eqref{VBBstBBst} as an example. The numerator structure can be decomposed as~\cite{Ericson:1988gk}:
\be
(\veep\cdot \veq)(\veq\cdot\veepst) = \frac13 q_iq_j\Bigl(
\delta_{ij}(\veep\cdot\veepst)+T_{ij} 
\Bigr),
\label{decomp}
\ee
where the first term, proportional to $\delta_{ij}$,
denotes the central force and the second, proportional to 
\be
T_{ij}= 3\epsilon_i \epsilon_j^{\prime*}-(\veep\cdot\veepst)
\delta_{ij},
\label{Sij}
\ee
denotes the tensor force (see also the discussion in Appendix~\ref{app:OPE}).
The latter term does not contribute, when only $S$-waves are considered, however, it is known to induce significant $S$-$D$
transitions. The tensor force in Eq.~\eqref{VBBstBBst} also induces mixing between the open-bottom channels enumerated in Eq.~\eqref{BBchannels}. Moreover, when transferred to $r$-space, in the local static limit this term generates a $1/r^3$ behaviour
(see, for example, Refs.~\cite{Ericson:1988gk,
Tornqvist:1993ng,Thomas:2008ja}) that calls for special care; see also the related discussion of a spurious pole in Sect.~\ref{sec:Xb}. 
The momentum space analogue of this was investigated in Ref.~\cite{Wang:2018jlv}
in a study of $Z_b(10610)$ and $Z_b(10650)$.
In the cited work, an additional counterterm needed to be promoted to LO
to absorb a significant regulator dependence of the results---at the
same time this term tamed
an otherwise very strong channel coupling in the $1^{+-}$ system of the $B^*\bar{B}^*$ 
channel to the lower lying
$B\bar{B}^*$ channel. In our study here, where there are no lower-lying channels, this additional counterterm is not necessary and the effects can be absorbed into the counterterm $C$ as is demonstrated by the cutoff independence of the results.

We also note that a systematic partial-wave projection 
of the OME potentials in Eqs.~\eqref{VBBstBBst}-\eqref{VBstBstBstBst} is conveniently performed as~\cite{Baru:2016iwj,Baru:2017gwo,Baru:2019xnh}
\begin{align}
&\VOME_{\alpha'\alpha}(p',p;m_\phi)= \frac{1}{2J+1}
\int\frac{d\Omega_{\ven'}}{4\pi}
\frac{d\Omega_{\ven}}{4\pi}
\label{VPWA}\\
&\times\mbox{Tr}\left[
P^\dagger(\alpha';\ven')
\VOME_{B^{(*)}\bar{B}^*\to B^{(*)}\bar{B}^*}
 (\vep',\vep;m_\phi)P(\alpha;\ven)\right],
\notag
\end{align}
where the index $\alpha^{(\prime)}=1,2,3$ enumerates the 2-body channels quoted in Eq.~\eqref{BBchannels}, the trace implies summation over all relevant polarisation indices,
and the projection operators satisfy the condition
\be
\frac1{2J+1}\int\frac{d\Omega_n}{4\pi}P(\alpha;\ven)P^\dagger(\alpha;\ven)=1.
\label{Pnorm}
\ee
We provide further details and quote explicit expressions for the partial-wave-projected potentials in Eqs.~\eqref{VBBstBBst}-\eqref{VBstBstBstBst} in Appendix~\ref{app:PWA}.

It should be stressed that the OME potential discussed here is essentially different compared to the one employed, for example, in the two-nucleon system:
\begin{enumerate}
\item It contains the $B^*$-$B$ mass difference $\Delta m$ [see Eq.~\eqref{mthmu}] that reduces the effective exchanged-meson mass in the first diagram of Fig.~\ref{fig:TOPT}, while increasing it in the second diagram; see also Appendix~\ref{app:OPE}. As a result, the former contribution becomes more important than the latter, particularly for pion exchange.
\item More intriguing are the kinetic-energy terms that scale as $1/m_{B^{(*)}}$, as they render the propagator nonlocal. These recoil terms are essential for obtaining the correct analytic structure when the intermediate state pinpointed by the thin vertical line in Fig.~\ref{fig:TOPT} can go on shell. For the physical $B^*$-$B$ mass difference ($\Delta m<m_\pi$), this condition is not met. In contrast, for pion exchange in the $D\bar{D}^*$ system relevant to the $ \X$, the corresponding effect is very important; see Appendix~\ref{app:OPE} for details and Sect.~3 of Ref.~\cite{Lensky:2005hb} for a discussion in a different context.
\end{enumerate}

When the recoil and $E$ terms are dropped in $\mu_1$ and $\mu_2$ in Eqs.~\eqref{eq:alphaOME}, the contributions (1) and (2) in Fig.~\ref{fig:TOPT} can be combined together to form a static OPE propagator, however, with the meson mass reduced by
$\Delta m$, $[\veq^2+(m_\phi^2-(\Delta m)^2)]^{-1}$; see Appendix~\ref{app:OPE} for details.
This form, and its Fourier transform into $r$-space is used by many authors; see, for example, Refs.~\cite{Tornqvist:1993ng,Swanson:2003tb,Suzuki:2005ha,Thomas:2008ja}.
The larger bottom-meson masses suppress the recoil energies relative to the charm system at fixed momenta. In the latter case, the OPE even develops an imaginary part when all the meson masses take their physical values. Since it is desirable to apply the same formalism to both systems, we keep the recoil corrections
and the explicit energy dependence already here. With the recoils included, the OME interaction does not 
take the simple structure of a static meson-exchange potential anymore. On the contrary, its Fourier transform depends on both the incoming and outgoing relative distances separately.

Combining one-pion and one-$\eta$ exchanges gives the complete OME matrix in the isospin limit,
\be
\begin{split}
\VOME_{\alpha'\alpha}(p',p)={}&\,
3\VOME_{\alpha'\alpha}(p',p;m_{\pi})\\
&+\frac13 \VOME_{\alpha'\alpha}(p',p;m_\eta),
\label{VOME}
\end{split}
\ee
where, as before, $\alpha^{(\prime)}=1,2,3$ enumerates the channels quoted in Eq.~\eqref{BBchannels} and the isospin coefficients are displayed explicitly. 

The $s$-channel quarkonium exchange and contact interaction in Eq.~\eqref{VQMsep} act only in the $^3S_1$ component, so that the total potential is
\be
V_{\alpha'\alpha}(p',p)=\delta_{\alpha' 1}\delta_{\alpha 1}
V_S^{(\QQ)+{\rm CT}}(p',p)+\VOME_{\alpha'\alpha}(p',p),
\label{Vcoupled}
\ee
and we note that, once the OME interaction is included and the scattering equation [see Eq.~\eqref{LSEfull2} below] is solved, the contact term $C$ introduced in Eq.~\eqref{VQMab} acquires dependence on the momentum cutoff $\Lambda$ as detailed in Eq.~\eqref{Vdef}. 

\section{Application of \BOchiEFT to $\Xb$}
\label{sec:appXb}

\subsection{Lippmann-Schwinger equation}
\label{sec:LSE}

Now we proceed to the scattering equation for the system of coupled channels $B^{(*)}\bar{B}^*$ quoted in Eq.~\eqref{BBchannels}. 
The scattering $T$-matrix is defined in the standard way through the relation
\be
\hat T \ket{p,\alpha} =\hat{V}\ket{\psi^{(+)}_{p\alpha}},
\ee
where $\ket{\psi^{(+)}_{p\alpha}}$ is the scattering state obeying the outgoing boundary conditions. Then the matrix elements of the $T$-matrix are defined as 
\be 
T_{\alpha'\alpha}(p',p) =\bra{p',\alpha'}\hat T\ket{p,\alpha}
\ee 
and obey an LSE with the full interaction potential specified in Eq.~\eqref{Vcoupled}, 
\begin{align}
&T_{\alpha'\alpha}(p',p)=V_{\alpha'\alpha}(p',p) \label{LSEfull2}\\
&\quad -\sum_{\alpha''}\int_0^\Lambda\frac{q^2dq}{2\pi^2}
V_{\alpha'\alpha''}(p',q)G_{\alpha''}(k,q) T_{\alpha''\alpha}(q,p),\notag
\end{align}
where the on-shell momentum $k$ was introduced in Eq.~\eqref{kdef}, and
\be
\begin{split}
&G_1(k,q)=G_2(k,q)=\frac{2\mu}{q^2-k^2-i\epsilon},\\
&G_3(k,q)=\left(\frac{q^2}{m_{B^*}}+\Delta m-\frac{k^2}{2\mu}-i\epsilon\right)^{-1}
\label{Gpropdef}
\end{split}
\ee
are the nonrelativistic 2-body Green's functions.

\begin{figure}[t!]
\centerline{\includegraphics[width=0.8\columnwidth]{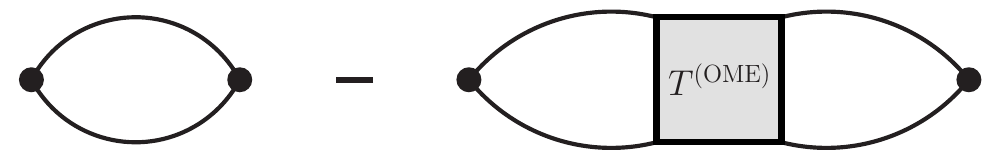}}
\caption{Schematic illustration of the expression for $\hat{J}(k)$ in Eq.~\eqref{Mdef}. The fat dots denote the channel-indexed vertices $f_{i,\alpha}$ and the shaded square denotes the coupled amplitude $T^{\rm (OME)}_{\alpha'\alpha}$, which is the solution of the LSE in Eq.~\eqref{LSEOME}. The loops correspond to the two-meson Green's functions $G$ in Eq.~\eqref{Gpropdef}.}
\label{fig:J}
\end{figure}

We solve Eq.~\eqref{LSEfull2} in two steps using the two-potential formalism~\cite{Gell-Mann:1953dcn,Nakano:1982bc}; see also Refs.~\cite{Doring:2009bi,Hanhart:2012wi} for applications close to the present problem. First, we solve the coupled LSE with the OME potential in Eq.~\eqref{VOME} only,
\begin{align}
&T^{\rm (OME)}_{\alpha'\alpha}(p',p)=\VOME_{\alpha'\alpha}(p',p)
\label{LSEOME}\\
&-\sum_{\alpha''}\int_0^\Lambda\frac{q^2dq}{2\pi^2}
\VOME_{\alpha'\alpha''}(p',q) G_{\alpha''}(k,q)
T^{\rm (OME)}_{\alpha''\alpha}(q,p),\notag
\end{align}
to obtain the corresponding $3\times3$ matrix amplitude $T^{\rm (OME)}_{\alpha'\alpha}(p',p)$.
Then, introducing $f_{i,\alpha}(p)=\delta_{\alpha 1}f_i(p)$ for the separable quarkonium and contact vertices (see Eq.~\eqref{eq:fn}), the full solution is built algebraically as
\begin{align}
T_{\alpha'\alpha}(p',p)= T^{\rm (OME)}_{\alpha'\alpha}(p',p)+\sum_{i,j}X^{{\rm out}}_{i,\alpha'}(p')\hat{\tau}_{ij}(k)
X^{{\rm in}}_{j,\alpha}(p).
\label{Tsol}
\end{align}
Here, the incoming and outgoing dressed vertices read, for $i=1,..,N+1$,
\be
\begin{split}
&X_{i,\alpha}^{{\rm in}}(p)=f_{i,\alpha}(p)\\
&\hspace*{0.05\columnwidth}-\sum_{\alpha'}\int_0^\Lambda\dqr f_{i,\alpha'}(q)G_{\alpha'}(k,q) T^{\rm (OME)}_{\alpha'\alpha}(q,p),\\
&X_{i,\alpha}^{{\rm out}}(p')=f_{i,\alpha}(p')\\
& \hspace*{0.05\columnwidth}-\sum_{\alpha'}\int_0^\Lambda\dqr T^{\rm (OME)}_{\alpha\alpha'}(p',q)G_{\alpha'}(k,q) f_{i,\alpha'}(q),
\end{split}
\label{Xs}
\ee
and $\tau_{ij}$ are the elements of
the OME-dressed reduced matrix amplitude $\hat{\tau}(k)$ in the $(N+1)$-dimensional separable space associated with the explicit $s$-channel quarkonium exchanges and the contact interaction. This matrix is built as
\be
\hat{\tau}=\left(1+
\hat{v}\cdot\hat{J}\right)^{-1}\cdot \hat{v}=\left(\hat{v}^{-1}+\hat{J}\right)^{-1},
\label{eq:tau}
\ee 
where the diagonal interaction matrix $\hat{v}$ was introduced in Eq.~\eqref{eq:vhat} and the matrix $\hat{J}$ has the components
\begin{align}
J_{ij}(k)
&=\sum_\alpha\int_0^\Lambda\dqr X_{i,\alpha}^{{\rm in}}(q)G_\alpha(k,q)f_{j,\alpha}(q)\notag\\
&=\sum_\alpha\int_0^\Lambda\dqr f_{i,\alpha}(q)G_\alpha(k,q)X_{j,\alpha}^{{\rm out}}(q)\notag\\
&=\sum_\alpha\int_0^\Lambda\dqr f_{i,\alpha}(q)G_\alpha(k,q)f_{j,\alpha}(q)\label{Mdef}\\
&\quad-\sum_{\alpha',\alpha}\int_0^\Lambda\dqr\int_0^\Lambda\dqrpr
f_{i,\alpha'}(q')G_{\alpha'}(k,q')\notag\\
&\hspace*{0.10\columnwidth}\times
T^{\rm (OME)}_{\alpha'\alpha}(q',q)G_\alpha(k,q)f_{j,\alpha}(q).\notag
\end{align}
Time-reversal invariance implies $J_{ij}=J_{ji}$ and therefore $\tau_{ij}=\tau_{ji}$.
The first term on the right-hand side of the last expression is the free two-meson loop connecting the separable vertices associated with quarkonium exchange and the contact interaction, while the second term contains the full coupled-channel OME amplitude inserted between the two free two-meson loops; see Fig.~\ref{fig:J}.

The poles of the full amplitude $T(p,p')$ correspond to the physical states generated within the formalism. They come as solutions of the secular equation,
\be
\mbox{det}(\hat{v}^{-1}+\hat{J})=0.
\label{det}
\ee

The decomposition of the full scattering amplitude in Eq.~\eqref{Tsol} is introduced for technical convenience. At fixed $m_b$, the numerically demanding quantities $T^{(\rm OME)}_{\alpha'\alpha}$, $X_{i,\alpha}^{\rm in,out}$, and $J_{ij}$, obtained by solving the integral equation or evaluating the corresponding loop integrals, are independent of the free parameters $C$ and $V_0$ and can therefore be precomputed. These parameters enter only through $\hat v$ and, consequently, $\hat\tau$ in the second term in Eq.~\eqref{Tsol}, which can then be evaluated algebraically by a matrix inversion.

\subsection{Fixing parameters}
\label{sec:params}

To complete the specification of the theoretical framework, the parameters need to be determined. As in the pure quark--antiquark calculations of Subsect.~\ref{sec:qmstates} (see, in particular, Eq.~\eqref{V0QM}), these include the constant shift parameter $V_0$ in the Cornell potential in Eq.~\eqref{Cornell}.
We note that, in the quark--antiquark calculation presented in Subsect.~\ref{sec:qmstates}, $V_0$ merely provides an overall shift of the bound-state spectrum generated by the Hamiltonian in Eq.~\eqref{H0}. In the full coupled-channel problem, however, the position of the $B\bar{B}^*$ threshold $M_{\rm th}$ in Eq.~\eqref{mthmu} does not depend on $V_0$. Consequently, this parameter controls the energy separations between the eigenstates of the Schr{\"o}dinger equation~\eqref{Scheq} and the open-bottom threshold. Since the interaction potential in Eq.~\eqref{Vdef} depends sensitively on these energy separations, $V_0$ becomes intertwined with the other parameters, and its role in determining the physical states extends beyond that of a simple overall shift of the spectrum.
The second parameter that needs to be adjusted is $C(\Lambda,N)$. Unless stated otherwise, we retain $N=4$ bottomonium states in Eq.~\eqref{Vdef} and use $\Lambda=4$~GeV to regularise the loop integrals entering the LSE. Then, 
for each fixed value of $m_b$ in the broad interval $[4.3,5.3]$~GeV, we vary both $C$ and $V_0$ of the complete LO calculation to minimise the root-mean-square deviation, 
\be
(\Delta M)^2=\frac13\sum_{n=1}^3\Bigl(M_n-M_n^{\rm exp}\Bigr)^2,
\label{dM}
\ee
where the theoretical bound-state masses are obtained as
\be
M_n=M_{\rm th}-\frac{(\kappa_n^{\rm fit})^2}{2\mu},
\ee
with the binding momenta $\kappa_n^{\rm fit}$ found as solutions of the secular equation in Eq.~\eqref{det},
\be 
\mbox{det}\!\left[\hat{v}^{-1}(i\kappa_n^{\rm fit})+\hat{J}(i\kappa_n^{\rm fit})\right]=0, \quad n=1,2,3. \label{mincond} 
\ee
The masses for the three lightest bottomonium states observed experimentally~\cite{ParticleDataGroup:2026mpi} are quoted in Table~\ref{tab:bottom}. 
The corresponding binding momenta, defined through the relation
\be
M_{\rm th}-M_n^{\rm exp}=\frac{\kappa_n^2}{2\mu}>0,\quad n=1,2,3,
\label{gammas}
\ee
are
\be
\begin{split}
&\kappa_1=1.942~\GeV,\\
&\kappa_2=1.360~\GeV,\\
&\kappa_3=0.695~\GeV.
\label{gammas123}
\end{split}
\ee
The results are collected in Table~\ref{tab:parameters2}, from which one can conclude that the best agreement with the measured masses is achieved for $m_b$ in Eq.~\eqref{mb}. Table~\ref{tab:parametersN} shows that changing the number $N$ of explicitly retained quark-model states in the sum in Eq.~\eqref{Vdef} from 4 to 7 is absorbed by the refitted contact interaction $C$. For comparison, we also collect results from the LO calculation with the tensor force neglected (for brevity referred to as the ``tensor-off'' case) in Appendix~\ref{app:tensor-off-reference}.

\begin{table}[t!]
\centering
\caption{Sets of $C$ and $V_0$ parameters obtained for several values of $m_b$, using $N=4$ states in Eq.~\eqref{Vdef} and $\Lambda=4$~GeV. The dagger marks the value yielding the best agreement with the measured masses and adopted as the central value of the $b$-quark mass in Eq.~\eqref{mb}; in the table, it is rounded to one decimal place. 
}
\label{tab:parameters2}
\begin{ruledtabular}
\begin{tabular}{cccc}
$m_b$\unit{GeV} & $C$\unit{GeV$^{-2}$} & $V_0$\unit{MeV} & $\Delta M$\unit{MeV} \\
\hline
4.3 & $-16.17$ & 820 & 1.8 \\
~$4.5^\dagger$ & $-16.27$ & 371 & 0.0 \\
4.6 & $-16.30$ & 245 & 0.5 \\
4.7 & $-16.34$ & 53 & 1.2 \\
4.8 & $-16.38$ & $-140$ & 1.8 \\
4.9 & $-16.42$ & $-332$ & 2.5 \\
5.1 & $-16.50$ & $-718$ & 3.7 \\
5.3 & $-16.56$ & $-1105$ & 4.9 \\
\end{tabular}
\end{ruledtabular}
\end{table}

\begin{table}[t!]
\centering
\caption{Dependence of the contact term $C$ on the number of explicitly retained quark-model states in the sum in Eq.~\eqref{Vdef}, evaluated at the central value of $m_b$ in Eq.~\eqref{mb} and $\Lambda=4$~GeV. In all cases, $V_0=371$~MeV and $\Delta M=0.0$~MeV.
}
\label{tab:parametersN}
\begin{ruledtabular}
\begin{tabular}{l|cccc}
$N$ & 4 & 5 & 6 & 7 \\
\hline
$C$\unit{GeV$^{-2}$} & $-16.27$ & $-16.26$ & $-16.26$ & $-16.26$
\end{tabular}
\end{ruledtabular}
\end{table}

\subsection{Component probabilities}
\label{sec:probabilities}

To gain an understanding of the resulting spectrum and the composition of the states it contains, we now decompose the full wave functions into their components. Since the formalism includes three different, mutually orthogonal sectors---generic $\bar{b}b$ states $\bbnP$, 2-body $B^{(*)}\bar{B}^*$ states, and 3-body $B^{(*)}\bar{B}^{(*)} \phi$ states---we perform this decomposition in an enlarged Hilbert space that contains all three sectors. Note, however, that this is merely a different bookkeeping of the various ingredients, which gives us access to the quantities of interest, while remaining fully equivalent to the formulation discussed above. 
Accordingly, the completeness condition reads
\be
\begin{split}
\mathbbm{1}={}&\sum_{n=1}^\infty\ket{\bbnP}\bra{\bbnP}\\
&+\sum_\alpha\sum_{a} \ket{B^{(*)}\bar{B}^*[\alpha;a]}\bra{B^{(*)}\bar{B}^*[\alpha;a]}\\
&+\sum_\beta\sum_{b}\ket{B^{(*)}\bar{B}^{(*)}\phi[\beta;b]}\bra{B^{(*)}\bar{B}^{(*)}\phi[\beta;b]},
\end{split}
\label{completeness}
\ee
where $\alpha$ and $\beta$ count the 2- and 3-body channels enumerated in Eqs.~\eqref{BBchannels} and \eqref{BBphichannels}, respectively, while $a$ and $b$ denote complete sets of 
quantum numbers identifying these channels, respectively. As before, we truncate the sum over the bottomonium states at a finite value $N$, and the properly normalised 2-body $B^{(*)}\bar{B}^*$ states ${\ket{p,\alpha}}$ were defined in Subsect.~\ref{sec:OME}; see the normalisation condition in Eq.~\eqref{eq:pLket} and the text above it. We do not need to specify explicitly the quantum numbers $\{b\}$ for the states $\{\beta\}$ and only note that they include both discrete (spins and angular momenta of all subsystems) and continuous (various relative momenta) degrees of freedom, corresponding to the phase space of the 3-body states.

In this formulation, the wave function of the system satisfies a Schr{\"o}dinger equation,
\be
\tilH\ket{\Psi(E)}=E\ket{\Psi(E)},
\label{eq:schroedinger-full}
\ee
where $\tilH$ acts in the mentioned enlarged Hilbert space.
Let $E_B>0$ denote the binding energy of a bound state, so the corresponding pole is located on the physical Riemann sheet at $E=-E_B$. 
Then, using the completeness relation in Eq.~\eqref{completeness}, we decompose the bound-state wave function as 
\begin{align}
\ket{\Psi(-E_B)}={}&
\sum_{n=1}^{N}\lambda_n
\ket{\bbnP}\notag\\
&+\sum_{\alpha}
\int_0^\Lambda\frac{p^2dp}{2\pi^2}\Psi_\alpha(p)
\ket{p,\alpha}\label{eq:bound-state-decomposition}\\
&+\sum_\beta\sum_b
\varphi_\beta(-E_B;b)
\ket{B^{(*)}\bar{B}^{(*)}\phi[\beta;b]}\notag \ ,
\end{align}
where the change of integral measure compared to Eq.~\eqref{eq:state-normalisation} for convenience has been explained below Eq.~\eqref{Vgamma}. 
The parameters $\lambda_n$ were introduced in Eq.~\eqref{state},
\be
\Psi_\alpha(p)\equiv \Psi_\alpha(-E_B;p)=\braket{p,\alpha|\Psi(-E_B)},
\label{psiL}
\ee
and
\be
\varphi_\beta(-E_B;b)=\braket{B^{(*)}\bar{B}^{(*)}\phi[\beta;b]|\Psi(-E_B)}.
\ee

The normalisation of the bound-state wave function implies the following decomposition of its component probabilities:
\be
1={\cal P}_{\bar{b}b} +\sum_{\alpha}{\cal P}_\alpha+{\cal P}_{\text{3-body}}=
{\cal P}_{\bar{b}b} +{\cal P}_S+{\cal P}_D+{\cal P}_{\text{3-body}},
\label{eq:norm_generalised}
\ee
where 
\be
{\cal P}_{\bar{b}b}
=\sum_{n=1}^N|\lambda_n|^2
\label{Pbb}
\ee
is the probability of finding the bound state in the compact bottomonium components. The sum in $\alpha$ in Eq.~\eqref{eq:norm_generalised} runs over the 2-body states in Eq.~\eqref{BBchannels}, with
\be
{\cal P}_\alpha=\int_0^\Lambda\dqr\,|\Psi_\alpha(q)|^2,
\label{eq:P-2body}
\ee
while ${\cal P}_S$ and ${\cal P}_D$ measure the $S$-wave 2-body component $B\bar{B}^*(^3S_1)$ and the sum of the $B\bar{B}^*(^3D_1)$ and $B^*\bar{B}^*(^5D_1)$ 2-body components, respectively. The last term ${\cal P}_{\text{3-body}}$ is the probability carried by the 3-body intermediate states
enumerated in Eq.~\eqref{BBphichannels} and illustrated in Fig.~\ref{fig:TOPT}.

In addition, we define the $S$-wave compositeness by summing the two- and 3-body $S$-wave probabilities,
\be
X_S={\cal P}_S+{\cal P}_{\text{3-body},S},
\label{eq:XS}
\ee
where both terms are evaluated from the $S$-wave component of the complete coupled-channel state, so that all iterated $S$-$D$-$S$ effects are retained. Further details and explicit expressions for the individual probabilities entering Eq.~\eqref{eq:norm_generalised} are provided in Appendix~\ref{app:probabilities}.

It is worth noting that, in a theory containing only contact interactions, the normalisation integral for the $S$-wave 2-body component of a shallow bound state is ultraviolet (UV) convergent, while those for higher partial waves are UV divergent. Accordingly, the probability of only the $S$-wave component in the bound-state wave function (the compositeness) is model independent at LO in the weak-binding expansion; see Refs.~\cite{Guo:2017jvc,Hyodo:2013nka} for reviews. In our LO calculation, by contrast, all 2- and 3-body contributions entering Eq.~\eqref{eq:norm_generalised}, including the $D$-wave 2-body component, are UV convergent. This convergence results from the compensation between the momentum factors generated by the derivative couplings at the two $B^*B\phi$ vertices and the UV falloff of the exchanged light-meson propagator; see Fig.~\ref{fig:ome}.

\subsection{Results and discussion}
\label{sec:Xb}

\begin{table*}[t!]
\centering
\caption{Predictions for the four lightest axial-vector states in the bottomonium spectrum in the complete LO calculation, obtained for the parameters listed in Tables~\ref{tab:parameters} and \ref{tab:parameters2} with $N=4$ and $\Lambda=4$~GeV; see the main text for a discussion of the interpretation of the $\chi_{b1}(3P)$ and $\Xb$ states. 
The values of the deviation $\Delta M$ for each $b$-quark mass $m_b$ are given in Table~\ref{tab:parameters2}. The dagger marks the value of $m_b$ defined in Eq.~\eqref{mb} (rounded here to one decimal place for presentation purposes), which provides the best agreement with the measured masses, yielding $\Delta M=0.0$~MeV. The probabilities are defined in Eq.~\eqref{eq:norm_generalised}, with their explicit expressions given in Appendix~\ref{app:probabilities}. The $S$-wave compositeness $X_S$ is defined in Eq.~\eqref{eq:XS}.
The $D$-wave probability ${\cal P}_D$ is a sum of the $B\bar{B}^*(^3D_1)$ and $B^*\bar{B}^*(^5D_1)$ 2-body components.
All percentages are rounded to full values. Accordingly, a 0 still allows for contributions of less than 0.5\%. 
}
\label{tab:results}
\begin{ruledtabular}
\setlength{\tabcolsep}{1.4pt}
\begin{tabular}{c|c|c|c|c|c|c|c}
$m_b$\unit{GeV} & Mass\unit{GeV} & $E_B$\unit{MeV}
& ${\cal P}_{\bar{b}b}$ (\%) & ${\cal P}_S$ (\%)
& ${\cal P}_D$ (\%) & ${\cal P}_{\text{3-body}}$ (\%)
& $X_S$ (\%) \\
\hline
\multicolumn{8}{c}{$\chi_{b1}(9893)=\chi_{b1}(1P)$\quad
$M^{\rm exp}[\chi_{b1}(1P)]=9.893$~GeV~\cite{ParticleDataGroup:2026mpi}}
\\
\hline
4.3 & 9.891 & 713 & 100 & 0 & 0 & 0 & 0 \\
~$4.5^\dagger$ & 9.893 & 711 & 100 & 0 & 0 & 0 & 0 \\
4.7 & 9.895 & 710 & 100 & 0 & 0 & 0 & 0 \\
4.9 & 9.896 & 708 & 100 & 0 & 0 & 0 & 0 \\
5.1 & 9.898 & 706 & 100 & 0 & 0 & 0 & 0 \\
5.3 & 9.899 & 705 & 100 & 0 & 0 & 0 & 0 \\
\hline
\multicolumn{8}{c}{$\chi_{b1}(10255)=\chi_{b1}(2P)$\quad
$M^{\rm exp}[\chi_{b1}(2P)]=10.255$~GeV~\cite{ParticleDataGroup:2026mpi}}
\\
\hline
4.3 & 10.257 & 347 & 98 & 1 & 0 & 0 & 1 \\
~$4.5^\dagger$ & 10.255 & 349 & 98 & 1 & 0 & 0 & 1 \\
4.7 & 10.254 & 350 & 98 & 1 & 0 & 0 & 1 \\
4.9 & 10.252 & 352 & 98 & 1 & 0 & 0 & 1 \\
5.1 & 10.251 & 354 & 98 & 1 & 0 & 0 & 1 \\
5.3 & 10.249 & 355 & 98 & 1 & 0 & 0 & 1 \\
\hline
\multicolumn{8}{c}{$\chi_{b1}(10513)=\Xb$
\quad
$M^{\rm exp}[\Xb]=10.513$~GeV~\cite{ParticleDataGroup:2026mpi}
} \\
\hline
4.3 & 10.513 & 91 & 2 & 41 & 36 & 21 & 42 \\
~$4.5^\dagger$ & 10.513 & 91 & 2 & 41 & 36 & 21 & 42 \\
4.7 & 10.513 & 91 & 2 & 41 & 36 & 21 & 42 \\
4.9 & 10.513 & 91 & 2 & 41 & 36 & 21 & 42 \\
5.1 & 10.513 & 91 & 2 & 41 & 36 & 21 & 42 \\
5.3 & 10.513 & 91 & 1 & 42 & 36 & 21 & 42 \\
\hline
\multicolumn{8}{c}{$\chi_{b1}(10544)=\chi_{b1}(3P)$} \\
\hline
4.3 & 10.549 & 55 & 83 & 16 & 1 & 0 & 16 \\
~$4.5^\dagger$ & 10.544 & 61 & 85 & 14 & 1 & 0 & 14 \\
4.7 & 10.540 & 64 & 86 & 13 & 0 & 0 & 13 \\
4.9 & 10.536 & 68 & 88 & 12 & 0 & 0 & 12 \\
5.1 & 10.532 & 72 & 89 & 10 & 0 & 0 & 10 \\
5.3 & 10.529 & 75 & 90 & 9 & 0 & 0 & 9
\end{tabular}
\end{ruledtabular}
\end{table*}
 
With all parameters listed in Tables~\ref{tab:parameters} and \ref{tab:parameters2} fixed either from lattice-QCD input or from the measured masses of the three lightest axial-vector states in the spectrum of bottomonium, we are now in a position to explore the LO results
of the \BOchiEFT for the observed $J^{PC}=1^{++}$ spectrum and in the region near the open-bottom threshold. We find that the amplitude in Eq.~\eqref{Tsol} possesses four below-threshold physical poles in the mass range of the observed $\chi_{b1}$ states on the first Riemann sheet of the complex energy plane, corresponding to four bound states. We recall that the calculation restricted to quark--antiquark states discussed in Subsect.~\ref{sec:VQM} predicts only three such states, which are conventionally identified 
as the generic bottomonia $\bbnP$, with $n=1,2,3$. 
The positions of the four poles obtained in the full calculation are listed in Table~\ref{tab:results}. Specifically, we quote the masses of the resulting physical states together with their binding energies relative to the $B\bar{B}^*$ threshold.

In Table~\ref{tab:results}, the probabilities ${\cal P}_{\bar{b}b}$, ${\cal P}_{S,D}$, and ${\cal P}_{\text{3-body}}$ defined in Eq.~\eqref{eq:norm_generalised} are listed for all found bound states. 
The $S$-wave compositeness $X_S$ is also given.
The results show that the hadronic components in the $\chi_{b1}(1P)$ and $\chi_{b1}(2P)$ are negligible. On the other hand, the third $\bar{b}b$-dominated state, denoted as $\chi_{b1}(3P)$ for convenience, is found with a mass in the range
10.529--10.549~GeV, depending on the value of the $b$-quark mass used in the calculation, and a hadronic compositeness of (10--17)\%. This mass is sizably higher than that quoted for $\chi_{b1}(3P)$ in the RPP, $(10513.4 \pm 0.7)$~MeV~\cite{ParticleDataGroup:2026mpi}.
Instead, the experimentally observed state with that mass is identified as a dynamically generated hadronic molecule with a hadronic compositeness of about 98\%, comprising approximately 41\% $S$-wave 2-body, 36\% combined $D$-wave 2-body (to be compared with the approximately 5\% $D$-wave admixture in the deuteron~\cite{Machleidt:1987hj,Epelbaum:2019kcf}), and 21\% three-body components. These results lead us to interpret this state as the exotic meson $\Xb$.
The cutoff dependence of the results is presented in Table~\ref{tab:cutoff}, with $\Lambda$ varying from 3 to 5~GeV. As expected, this dependence can be almost entirely absorbed into the strength of the short-range operator $C$, yielding the masses of the physical states essentially unchanged. 
The compact-bottomonium probability ${\cal P}_{\bar{b}b}$ of each state changes by at most 1\%.

Besides the four physical states in Table~\ref{tab:results}, the amplitude possesses a spurious pole far below the $\chi_{b1}(1P)$, between 8.958 and 9.399~GeV for $\Lambda=3$--5~GeV, with ${\cal P}_{\bar{b}b}<0.3\%$. Unlike the $\chi_{b1}(1P)$ and $\chi_{b1}(2P)$, whose energies are set by the Cornell potential, this state is bound by the singular tensor part of the OME potential and is spurious since the binding momentum $\sqrt{2\mu E_B}\simeq2.5$--3~GeV is much larger than the breakdown scale of the chiral force considered in this work. Besides, its position strongly depends on the regulator.
This is a phenomenon known from chiral nuclear forces: the singular attractive tensor part of OPE, renormalised by a single counterterm, leads to a limit-cycle-like running of that term~\cite{Wilson:1970ag,Bedaque:1998kg,Beane:2001bc}, and each time the counterterm $C(\Lambda)$ passes through infinity a deeply bound state enters the spectrum, which lies outside the range of the theory and decouples from the low-energy observables~\cite{Beane:2000wh,Nogga:2005hy,Epelbaum:2002ji,Birse:2005um}. 
Here, the spurious pole is absent for $\Lambda\leqslant1.25$~GeV. 
Although the spurious pole is present for $\Lambda=3$--5~GeV, its influence on the physical observables is small.
We have checked that explicitly projecting it out at $\Lambda=4$~GeV changes the masses in Table~\ref{tab:results} by less than 1~MeV and the scattering length $a$ and effective range $r$, given below in Eq.~\eqref{eq:aandr}, by less than $4\times10^{-3}$~fm each.
We further discuss this spurious pole and its removal by the projection method~\cite{Kukulin:1978he,Epelbaum:2002ji,Nogga:2005hy,Yang:2020pgi} in Appendix~\ref{app:deep}.

\begin{table*}[t!]
\centering
\caption{Cutoff dependence of the complete LO calculation with $N=4$. For each cutoff, the parameters are adjusted to reproduce the same three experimental masses, yielding $\Delta M=0.0$~MeV.}
\label{tab:cutoff}
\begin{ruledtabular}
\small
\setlength{\tabcolsep}{1.3pt}
\begin{tabular}{c|ccc|cc|cc|cc|cc}
& \multicolumn{3}{c|}{Parameters}
& \multicolumn{2}{c|}{$\chi_{b1}(1P)$}
& \multicolumn{2}{c|}{$\chi_{b1}(2P)$}
& \multicolumn{2}{c|}{$\Xb$}
& \multicolumn{2}{c}{$\chi_{b1}(3P)$} \\
$\Lambda$\unit{GeV} & $m_b$\unit{GeV} & $C$\unit{GeV$^{-2}$}
& $V_0$\unit{MeV}
& $E_B$\unit{MeV} & ${\cal P}_{\bar{b}b}$ (\%)
& $E_B$\unit{MeV} & ${\cal P}_{\bar{b}b}$ (\%)
& $E_B$\unit{MeV} & ${\cal P}_{\bar{b}b}$ (\%)
& $E_B$\unit{MeV} & ${\cal P}_{\bar{b}b}$ (\%) \\
\hline
3 & 4.578 & 10.55 & 287
& 711 & 100 & 349 & 98 & 91 & 1 & 61 & 86 \\
4 & 4.534 & $-16.27$ & 371
& 711 & 100 & 349 & 98 & 91 & 2 & 61 & 85 \\
5 & 4.517 & $-0.67$ & 405
& 711 & 100 & 349 & 98 & 91 & 2 & 60 & 85
\end{tabular}
\end{ruledtabular}
\end{table*}

\begin{figure}[t!]
\centerline{\includegraphics[width=\columnwidth]{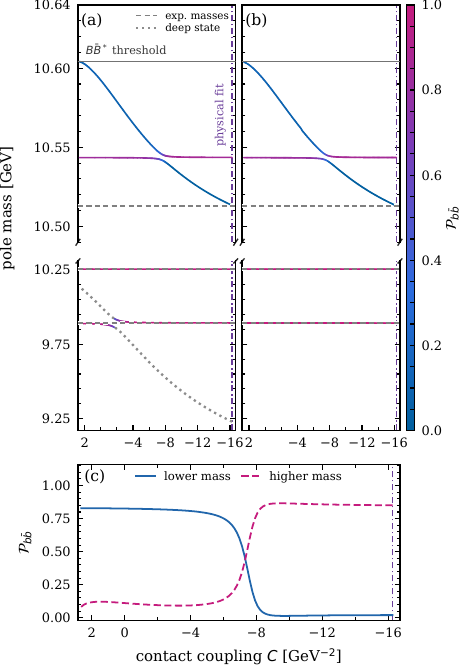}}
\caption{
Dependence of the four bound-state pole masses in the complete LO calculation for $N=4$, $\Lambda=4$~GeV, $m_b$ given in Eq.~\eqref{mb}, and $V_0=371$~MeV on the contact coupling $C$. 
Panels (a) and (b) are the results with the full OME potential and with the deep spurious state projected out, respectively; panel (c) shows ${\cal P}_{\bar{b}b}$ for the two near-threshold mass eigenstates of panel (b), illustrating the interchange of their compact-bottomonium content associated with the avoided level crossing. 
The corresponding parameter values, together with $C=-16.27$~GeV$^{-2}$, marked by the vertical dot-dashed line close to the right edge of each plot, are given in Table~\ref{tab:parameters2}. 
The colour of each trajectory denotes its compact-bottomonium probability ${\cal P}_{\bar{b}b}$ defined in Eq.~\eqref{Pbb}. The dashed horizontal lines mark the three experimental $\chi_{b1}$ masses, while the solid horizontal line marks the $B\bar{B}^*$ threshold. 
The grey dotted line in panel (a) corresponds to the predominantly hadronic member of the two lowest levels, that is, the regulator-dependent deep spurious pole.
}
\label{fig:levels}
\end{figure}

The results presented in Table~\ref{tab:results} also explain why the values of $\Delta M$ listed in Table~\ref{tab:parameters2} for all considered $b$-quark masses $m_b$ are substantially smaller than the minimum deviation that can be achieved using only quark--antiquark degrees of freedom, $\Delta M_{\rm min}^{(\QQ)}$, given in Eq.~\eqref{V0QM}. The improvement arises naturally from the inclusion of the two-hadron components discussed in Subsect.~\ref{sec:VQM}, which couple the $\bar{b}b$ states to the interacting hadronic $B\bar{B}^*$ channel. In particular, the appearance of the $\Xb$ pole between the dressed $\chi_{b1}(2P)$ and $\chi_{b1}(3P)$ states, together with its identification with the experimentally observed state at 10.513~GeV, provides an additional degree of freedom in the fit and leads to a considerable reduction of the mean deviation $\Delta M$ defined in Eq.~\eqref{dM}.
 
For illustration, we show in Fig.~\ref{fig:levels} the calculated bound-state pole masses [in panels (a) and (b)], as well as the compact-$\bar{b}b$ probabilities of the highest two poles [panel (c)], as functions of the contact term $C$ for $m_b$ given in Eq.~\eqref{mb}, keeping all other parameters at the values listed in the corresponding row of Table~\ref{tab:parameters2}.
At the left edge of the plot, $C=2.64$~GeV$^{-2}$, the upper near-threshold state is weakly bound, about 0.1~MeV below the $B\bar{B}^*$ threshold, and dominated by its hadronic components, while the lower one is predominantly compact. As $C$ decreases, the two poles approach one another, reach a minimum separation of 4~MeV at $C=-7.65$~GeV$^{-2}$, and exchange their compact-bottomonium content. After the avoided crossing, the upper level evolves into a predominantly compact state, which we identify with the $\chi_{b1}(3P)$ bottomonium, whereas the lower level follows the trajectory of the predominantly molecular $\Xb$ pole. 
Finally, at the fitted $C=-16.27$~GeV$^{-2}$ in Table~\ref{tab:parameters2}, the $\Xb$ pole approaches the experimental mass indicated by the highest dashed horizontal line. This mechanism allows the experimental spectrum to be reproduced.
In panel (a) of Fig.~\ref{fig:levels}, there is an additional avoided level crossing: the deeply bound spurious pole mentioned above approaches the $1P$ level as $C$ varies, where the dotted curve denotes the spurious pole. 
Panel (b) shows the results with the spurious pole projected out and, therefore, the avoided level crossing removed. The other poles are barely affected by the projection procedure, with the changes lying within 1~MeV.
The given example also illustrates the extent to which the experimental input drives the fitted parameters and, consequently, the interpretation of the resulting states.

\subsection{Near-threshold scattering}
\label{sec:threshold-scattering}

In this section, we discuss near-threshold $B\bar{B}^*$ scattering since the $S$-wave scattering length and effective range are closely related to the internal structure of near-threshold states, so their predictions provide valuable insight that can be checked in lattice-QCD calculations.

Below the $B^*\bar{B}^*$ threshold, only the $B\bar{B}^*$ channels are open.
In the fixed $\{^3S_1,^3D_1\}$ basis (we tag these basis states by the angular momentum $L=S,D$), we define the on-shell partial-wave amplitudes and the partial-wave $S$ matrix for isoscalar $J^{PC}=1^{++}$ $B\bar{B}^*\to B\bar{B}^*$ scattering as
\be
f_{LL'}(E)=-\frac{\mu}{2\pi}T_{LL'}(k,k;E),\quad
S(E)=\mathbbm{1}+2ik f(E).
\label{eq:SfromT}
\ee
The $S$- and $D$-wave coupled-channel $S$ matrix can be parameterised in the Blatt--Biedenharn form~\cite{Blatt:1952zza},
\be
\begin{split}
S&=O(\epsilon_{SD})
\begin{pmatrix}
e^{2i\delta_S} & 0\\
0& e^{2i\delta_D}
\end{pmatrix}
O^T(\epsilon_{SD}),\\
O(\epsilon_{SD})&=
\begin{pmatrix}
\cos(\epsilon_{SD})&-\sin(\epsilon_{SD})\\
\sin(\epsilon_{SD})&\phantom{-}\cos(\epsilon_{SD})
\end{pmatrix},
\end{split}
\label{eq:BBparam} 
\ee
where $\delta_S$ and $\delta_D$ are the eigenphase shifts, and $\epsilon_{SD}$ is the $S$-$D$ mixing angle. The eigenphases are defined modulo $\pi$. Since there is an additional bound state generated predominantly by the $S$-wave $B\bar{B}^*$ interaction compared with the pure quark--antiquark calculation, we adopt the convention in which, at threshold, $\epsilon_{SD}$ vanishes, while $\delta_S$ and $\delta_D$ approach $180^\circ$ and $0^\circ$, respectively. With this convention, the threshold behaviour is $\delta_S-\pi=\mathcal O(k)$, $\delta_D=\mathcal O(k^5)$, and $\epsilon_{SD}=\mathcal O(k^2)$.

\begin{figure}[t!]
\centering
\includegraphics[width=\columnwidth]{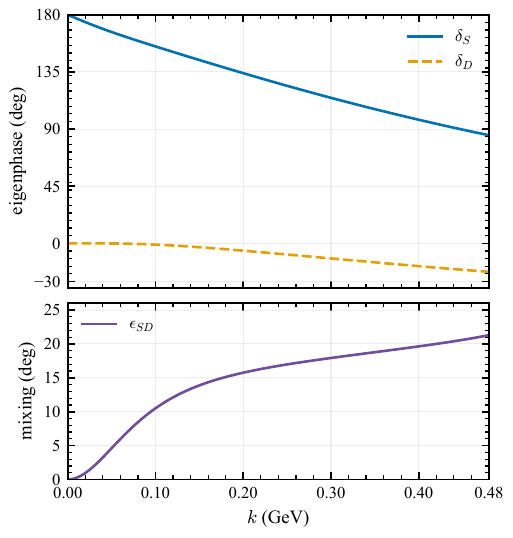}
\caption{The Blatt--Biedenharn eigenphase shifts $\delta_S$ and $\delta_D$ (upper panel) and the $S$-$D$ mixing angle $\epsilon_{SD}$ (lower panel) for $N=4$, $\Lambda=4$~GeV, and $m_b$ in Eq.~\eqref{mb}.}
\label{fig:phase-shift}
\end{figure}

The effective-range expansion (ERE)~\cite{Bethe:1949yr} for the predominantly $S$-wave eigenchannel reads
\be
k\cot\delta_S(k)=\frac{1}{a}+\frac{1}{2}rk^2
+\mathcal{O}(k^4),
\label{eq:ere}
\ee
where $a$ and $r$ are the $S$-wave scattering length and effective range, respectively. Note that we stick to the sign convention in which a positive scattering length signals an attractive interaction that is not sufficiently strong to produce a bound state, whereas a negative scattering length indicates either a repulsive interaction or the presence of a bound state.
The scattering parameters $a$ and $r$ of the ERE can also be extracted directly from the $SS$ element of the full $S$-$D$ coupled amplitude, 
\begin{align}
f_{SS}^{-1}(k) + ik = k\cot\delta_S(k) +\mathcal O(k^4),
\label{eq:projected}
\end{align}
since the $D$-wave coupled-channel effects in the $S$-wave-to-$S$-wave amplitude involve at least two $S$-$D$ transitions.

The calculation with $N=4$, $\Lambda=4$~GeV, and $m_b$ given in Eq.~\eqref{mb} yields
\be
a=-1.00~\mbox{fm},\quad r=-2.5~\mbox{fm}.
\label{eq:aandr}
\ee
The corresponding eigenphase shifts and the Blatt--Biedenharn mixing angle are shown in Fig.~\ref{fig:phase-shift}.

If, instead of using the results obtained in this work and reported in Table~\ref{tab:results}, where we tune the system such that the molecular $\Xb$ lies only 1~MeV below the threshold (see Table~\ref{tab:comp}), the scattering parameters become
\be
a=-3.01~\mbox{fm},\quad r=0.57~\mbox{fm},
\label{eq:aandr2}
\ee
in line with general expectations
for the effective range parameters in the presence of
near-threshold molecular states, as discussed in Ref.~\cite{Guo:2017jvc}.

These findings can be confronted with future lattice-QCD results.
However, extractions of the low-energy scattering parameters from lattice calculations have to face complications caused by the presence of the lhc branch points~\cite{Du:2023hlu} below but near the $B\bar{B}^*$ threshold, as given in Eq.~\eqref{eq:lhc-OME} for the physical meson masses. We provide additional details 
in Appendix~\ref{app:tensor-off-reference}.

\subsection{Comparison of \BOchiEFT and BOEFT}
\label{sec:comparison}

Here we compare the formalism and results of \BOchiEFT, introduced in this work, with 
selected BOEFT results. The BO approach and its EFT formulation have been developed in
Refs.~\cite{Berwein:2015vca,Brambilla:2017uyf,Oncala:2017hop,Soto:2020xpm, TarrusCastella:2022rxb,Berwein:2024ztx,Brambilla:2024imu,Brambilla:2026zfw,Brambilla:2026jqz} and, among others, relevant spectrum calculations for the system in focus here can be found in Refs.~\cite{Braaten:2024tbm,Brambilla:2024imu,Brambilla:2026zfw,Brambilla:2026jqz}. 
Both formalisms share the static $\bar{b}b$ potentials extracted from lattice QCD and provided in coordinate space as a key ingredient.
However, they differ in the treatment of short-distance physics, the long-range interactions, and the inputs used to fix the parameters.

To keep this chapter largely self-contained and ease reading,
we begin with a brief summary of the key features of \BOchiEFT
derived and presented above. In this formalism, 
at the first step, the $r$-space Schr{\"o}dinger equation is solved
for the lowest quark--antiquark potential, chosen in the Cornell form in Eq.~\eqref{Cornell}. The resulting $\bar{b}b$ spectrum and the corresponding wave functions are then used 
as input for the scattering kernel in momentum space.
UV-type contributions, such as those from higher-lying $\bar{b}b$ states and the high-momentum components of the hadronic loops, are absorbed into local operators---at LO this is the constant $C$ introduced in Eq.~\eqref{Vdef} and 
discussed further around Eq.~\eqref{VQMab} and below Eq.~\eqref{Vcoupled}. Both the strength of this operator and the energy scale $V_0$ that fixes the $\bar{b}b$ spectrum relative to the $B\bar{B}^*$ threshold need to be determined from a fit to data.
In particular, $C$ is not renormalisation-group invariant, as evidenced by its strong dependence on the regulator $\Lambda$ introduced to render the scattering equation well defined; see Table~\ref{tab:cutoff}. Consequently, $C$ cannot be directly related to lattice observables. Moreover, its dependence on the heavy-quark mass cannot be fixed within
the formalism at the present stage, which prevents a combined description of the $\bar{c}c$ and $\bar{b}b$ sectors. 
Whether this issue can be solved by further matching the contact-term operators to the quark-level potential at short distances will be investigated in the future.
Observables are determined by the full, renormalised amplitudes rather than by the short-range contact terms alone. In other words, since hadronic degrees of freedom are introduced explicitly in the formalism, short-range physics in the given sector is not explicit and instead is encoded in a series of local operators, with the constant $C$ at LO being supplemented with momentum-dependent terms at higher orders:
The BO structure is incorporated into the hadronic
sector by matching the partial-wave channels to the $\Sigma_g^{+\prime}$ and $\Pi_g$ configurations in the static limit. As shown in Sect.~\ref{sec:BO}, their short-distance degeneracy and the $\mathcal{O}(r^2)$ onset of their splitting constrain the higher-order contact interactions.

If the parameters of the LO calculation are fixed as detailed in 
Sect.~\ref{sec:params}, the experimentally observed state at 10.513~GeV appears as a dynamically generated molecule, $\Xb$. Furthermore, the calculation predicts an additional predominantly compact $\bar{b}b$ bottomonium at 10.544~GeV, about 60~MeV below the $B\bar{B}^*$ threshold. If this identification survives further refinements, it suggests that the state interpreted here as the $\Xb$ may have been observed already, while an additional compact bottomonium $\chi_{b1}$ should lie between it and the $B\bar{B}^*$ threshold.

We now turn to the BOEFT. 
The presently existing versions thereof 
solve the Schr\"odinger equations in $r$-space, with the short-distance potentials constrained by the QCD multipole expansion~\cite{Berwein:2024ztx}, and include the $\Sigma_g^{+\prime}$ and $\Pi_g$ tetraquark potentials explicitly. 
The lattice-QCD input constrains the $\Sigma_g^{+\prime}$ and mixing potentials near string breaking~\cite{Bulava:2024jpj}.
Since lattice data at short distances are not yet available,
in this regime both $\Sigma_g^{+\prime}$ and $\Pi_g$ are modeled using parametrisations of quenched hybrid potentials~\cite{Alasiri:2024nue,Brambilla:2026zfw}.
As soon as improved lattice data are available they can be
employed instead.
This makes a clear difference with the \BOchiEFT, where short-distance interactions in those channels are parametrised by a series of local operators and 
are therefore included only effectively.
On the other hand, the long-range parts of the potentials 
used in BOEFT
are parametrised phenomenologically, with their strengths and matching distances fixed by requiring continuity and smoothness of the resulting curves. One of the key parameters in those calculations is the adjoint meson mass (following Ref.~\cite{Brambilla:2024imu}, we use the notation $\Lambda_{1^{--}}$ for it), which can in principle be determined from gauge-invariant lattice correlators in a specified subtraction scheme, although its value has not yet been determined from unquenched lattice QCD. 

In the existing studies~\cite{Braaten:2024tbm,Brambilla:2024imu,Brambilla:2026zfw,Brambilla:2026jqz}, $\Lambda_{1^{--}}$ is tuned in the charm sector to produce a shallow bound state associated with the $\X$. Then predictions for the hidden-bottom sector are made without an additional fit to data in the bottom sector. 
The first quantitative study 
in this spirit is presented 
 in Ref.~\cite{Braaten:2024tbm}, where, however, the
effect of the avoided level crossing was not included.
The long-range part of the potential was taken in the form of $\exp(-r/d)$, with the range parameter $d$ set to the Sommer scale, $d=r_0=0.5~\mathrm{fm}$ (corresponding to an inverse range of approximately 400~MeV). The adjoint-meson mass was then fixed by requiring the hidden-charm state associated with the $\X$ to lie exactly at the spin-isospin averaged $D^{(*)}\bar{D}^{(*)}$ threshold. The spin-averaged $\Xb$ was subsequently predicted to lie 17~MeV below the corresponding bottom threshold---notice that the same binding energy remains relative to the $B\bar B^*$ threshold after including the first order correction of spin splittings, thus demonstrating a strong dependence of the binding energy on the mass of the heavy quark. 
Shortly after this work, the Munich group published a series of papers~\cite{Brambilla:2024imu,Brambilla:2026zfw,Brambilla:2026jqz},
employing the formalism of Ref.~\cite{Berwein:2024ztx}, that also contain results
for the $1^{++}$ sector within BOEFT.
The range of the long-range part
of the potential is modeled 
using either a form of $e^{-r/(2m_\pi)}/r^2$~\cite{Brambilla:2024imu} or a one-pion exchange tail. In each case, $\Lambda_{1^{--}}$
was tuned to reach values of 90--100~keV for the $\X$
binding energy relative to the spin-isospin averaged threshold,\footnote{The most precise determination of the $\X$ pole to date places it $160_{-57}^{+74}$~keV below the $D^0\bar{D}^{*0}$ threshold~\cite{Ji:2025hjw}.}
and the matching parameters were adapted accordingly.
The resulting spin-averaged binding energies of the $\Xb$, relative to the spin-averaged bottom threshold,
were in the range from 15~MeV to 233~keV, depending on the parametrisation employed for the long-range part and the details of the tuning of the parameters\footnote{In particular, when the pion range was changed from $1/(2m_\pi)$~\cite{Brambilla:2024imu} 
to $1/m_\pi$~\cite{Brambilla:2026jqz}
with $\Lambda_{1^{--}}$ fine tuned
to keep the $\X$ as a shallow bound state, the $\Xb$ binding energy dropped from 15~MeV to 1.1~MeV.}, thus demonstrating a 
significant sensitivity to both the
long-range part of the potential employed and the strength of the
short range potential controlled
by $\Lambda_{1^{--}}$.
It should be stressed that 
the $\Xb$, characterised by its prominent two-hadron composition, lies closer to the corresponding open-bottom threshold than the three predominantly compact $1^{++}$ states also found in those calculations. 

While \BOchiEFT and BOEFT treat the pion exchange as the leading long-range contribution to the $B\bar{B}^*$ interaction, 
its implementation differs between the theories in four important aspects:
\begin{enumerate}
\item The OPE potential derived in Subsect.~\ref{sec:OME} and employed in the present analysis explicitly accounts for the $\Delta m=m_{B^*}-m_B$ mass splitting, thereby yielding the correct lhc branch point given in Eq.~\eqref{eq:lhc-formula}, rather than the one at $\kappa_\text{lhc}=m_\pi/2$ obtained in the static OPE approximation with $\Delta m$ neglected in the pion propagator or Yukawa-like parameterisation.
\item Here, the tensor force is derived from the one-pion and one-$\eta$ exchanges. Note that in the local static limit, this contribution has a $1/r^3$ dependence in configuration space that calls for special care.
\item Both OPE and one-$\eta$-exchange contributions included here have no free or adjusted parameters, with their coupling strengths constrained by lattice data and SU(3) symmetry.
\item The OME potential in \BOchiEFT incorporates 3-body effects through the recoil terms in Eq.~\eqref{eq:alphaOME}. Although these terms are formally ${\cal O}(Q^2)$, they modify the analytic structure of the heavy-meson--heavy-meson--pion propagator and should therefore be resummed. Moreover, since the $\bar{D}D\pi$ intermediate state can go on shell for the $ \X$, whereas the corresponding $\bar{B}B\pi$ state cannot for the $\Xb$, pion-exchange effects are qualitatively different in the two systems. This distinction is preserved by retaining such recoil terms.
\end{enumerate}
Items 1 and 2 above can be straightforwardly incorporated in 
the BOEFT framework.
For item 3 it remains to be verified that the matching condition between the long- and short-range potentials can provide pion couplings consistent with its standard value extracted from the $D^*\to D\pi$ decay width or calculated on the lattice.
It also remains to be developed, how the 3-body effects mentioned in item 4 above are to be incorporated into BOEFT framework and what impact this has
on the quantitative results, especially when connecting the charm and bottom sectors. 

Within \BOchiEFT, we investigate the above effects in the $b$-sector in Table~\ref{tab:approximations}, where we collect the results of the \BOchiEFT approach obtained with the mass splitting in the first equation in Eq.~\eqref{mthmu}, the recoil terms, and the energy dependence in the 3-channel OME potential switched on and off in all possible combinations. In particular, the first row in the table corresponds to the full calculation while the last row quotes the results in the static limit discussed in detail in Appendix~\ref{app:OPE}.
Throughout this procedure, $m_b$ is fixed to the value given in Eq.~\eqref{mb}, as before, while $C$ and $V_0$ are refitted by minimising $(\Delta M)^2$.
We note that the contact term $C$ undergoes a substantial variation to compensate for the changes in the OME potential, thereby masking some effects that may become relevant when different sectors are investigated simultaneously. In particular, since the recoil and energy dependent terms introduce momentum dependence, switching them off modifies the UV behaviour of the OME potential,
which results in a change in the magnitude and sign of $C$. 
This is not problematic from the EFT perspective, since $C$ itself is not an observable, but stresses the need to assess the impact of the modifications of the meson-exchange potential at the level of observables. In the adopted scheme, the binding energy of the $\Xb$ remains unchanged by construction when the pion propagator is modified,
since the parameters are readjusted. However, its $S$-wave compositeness varies noticeably, ranging from 39\% to 53\% depending on the truncation scheme adopted for the propagators of the exchanged pseudoscalar mesons.
For the predicted $\chi_{b1}(3P)$-dominated bottomonium state, the mass varies within approximately 3~MeV and the compositeness remains essentially unchanged.
We reiterate that such corrections become substantial for near-threshold states in the charm sector, in particular for the $\X$ and $T_{cc}(3875)$~\cite{Baru:2011rs,Du:2021zzh,Ji:2025hjw}, and are therefore retained explicitly in the \BOchiEFT framework. This observation might also be of relevance for BOEFT studies, since the $\X$ has so far been used as an anchor for fixing the parameters of the formalism---how large the impact of the
three-body cut is within
the BOEFT framework can only be
seen from an explicit calculation, however.

\begin{table*}[t!]
\centering
\caption{Pole positions and compositeness fractions for the $\Xb$ candidate and the predicted $\chi_{b1}(3P)$-dominated state for $m_b$ in Eq.~\eqref{mb}, under different treatments for the meson-exchange potential, including switching on (Yes) or off (No) the $B^*$-$B$ mass splitting, recoil terms, and $E$-dependence. The first row quotes the complete calculation in Table~\ref{tab:results}, while the last row corrsponds to the static limit discussed in Appendix~\ref{app:OPE}.
Each row is obtained by refitting $C$ and $V_0$ to minimise $(\Delta M)^2$ while fixing the mass of the third state to 10.513~GeV.
To obtain the results with $m_{B^*}=m_{B}$, the masses of both $B^{(*)}$ mesons are set to their spin-averaged value $\bar{m}_B=(m_B+3m_{B^*})/4\approx5.313$~GeV solely in the nonrelativistically expanded $\mu$'s in Eq.~\eqref{eq:alphaOME}.} 
\label{tab:approximations}
\begin{ruledtabular}
\begin{tabular}{ccc|cc|ccc|ccc|c}
\multicolumn{3}{c|}{} &
\multicolumn{2}{c|}{Interaction parameters} &
\multicolumn{3}{c|}{$\Xb$} &
\multicolumn{3}{c|}{$\chi_{b1}(3P)$} &
\multicolumn{1}{c}{} \\
$m_{B^*}\neq m_B$ & recoil & $E$-dep.
& $C$\unit{GeV$^{-2}$} & $V_0$\unit{MeV}
& $M$ (GeV) & ${\cal P}_{\bar{b}b}$ (\%) & $X_S$ (\%)
& $M$ (GeV) & ${\cal P}_{\bar{b}b}$ (\%) & $X_S$ (\%)
& $\Delta M$\unit{MeV} \\
\hline
Yes & Yes & Yes & $-16.27$ & $371$
& 10.513 & 2 & 42
& 10.544 & 85 & 14 & 0.0 \\
Yes & No & Yes & 2.21 & $369$
& 10.513 & 2 & 40
& 10.541 & 84 & 14 & 2.1 \\
No & Yes & Yes & $-14.64$ & $371$
& 10.513 & 2 & 41
& 10.544 & 85 & 14 & 0.0 \\
No & No & Yes & 2.66 & $369$
& 10.513 & 2 & 39
& 10.541 & 85 & 14 & 2.5\\ 
Yes & Yes & No & $-9.97$ & 372 & 10.513 & 2 & 53 & 10.544 & 86 & 13 & 0.3 \\
Yes & No & No & 4.65 & 371 & 10.513 & 2 & 51 & 10.543 & 86 & 13 & 0.8 \\
No & Yes & No & $-9.01$ & 372 & 10.513 & 2 & 53 & 10.544 & 86 & 13 & 0.3 \\
No & No & No & 5.19 & 371 & 10.513 & 2 & 50 & 10.543 & 86 & 13 & 0.9 
\end{tabular}
\end{ruledtabular}
\end{table*}

To summarise, in the spin-averaged predictions of Refs.~\cite{Brambilla:2024imu,Brambilla:2026zfw,Brambilla:2026jqz}, the coupled Born--Oppenheimer potentials generate a shallow molecular $1^{++}$ bound state below the open-bottom threshold. This state is identified with the $\Xb$, the bottomonium counterpart of the $\X$. In addition, the approach predicts three deeper bound states that can be identified with the spin-averaged masses of the corresponding $\chi_{bJ}(nP)$ spin multiplets, with $n=1,2,3$, defined as
\be
\overline{M}[\chi_b(nP)]=\frac{1}{9}
\sum_{J=0}^2(2J+1)M[\chi_{bJ}(nP)].
\label{Mbar}
\ee

We note that this pattern can also be naturally reproduced within \BOchiEFT by varying the parameter $C$, 
as already shown in Fig.~\ref{fig:levels}, 
albeit at the cost of a sizable deviation in the masses of the three lowest-lying states from their experimental values. 
In this figure, it is demonstrated that $\Delta M$ remains nearly constant as the $\Xb$ binding energy is varied from near zero until the $\Xb$ level approaches the $\chi_{b1}(3P)$ state, where an avoided crossing occurs. 
Thus, schematically, the results consistent in spirit with those reported in Refs.~\cite{Brambilla:2024imu,Brambilla:2026zfw,Brambilla:2026jqz} correspond to the almost leftmost edge of Fig.~\ref{fig:levels}, whereas our \BOchiEFT result, with the two free parameters of the LO calculation fixed from experimental input, corresponds to the rightmost edge.
As an illustration, in Table~\ref{tab:comp}, we present the results of the \BOchiEFT approach,
when the binding energy of the $\Xb$ is artificially fixed to be as shallow as 1~MeV by adjusting the parameter $C$. 

\subsection{A fresh look at the experimental situation}
\label{sec:exp}

\begin{figure}[t!]
\centering
\includegraphics[width=\columnwidth]{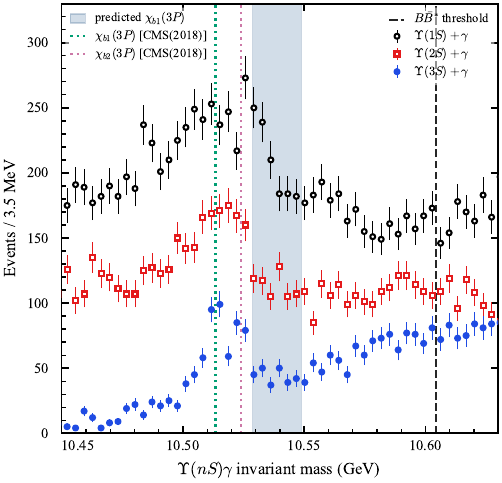}
\caption{The $\Upsilon(nS)\gamma$ invariant-mass spectra reported by CMS in the $\chi_{bJ}(3P)$ region~\cite{CMS:2018vgf}. The shaded band shows the predicted mass region of the $\chi_{b1}(3P)$-dominated state in Table~\ref{tab:results} for $m_b\in [4.3,5.3]$~GeV. The dotted vertical lines mark the masses of $\chi_{b1}(3P)$ and $\chi_{b2}(3P)$ reported by CMS~\cite{CMS:2018vgf}, and the dashed vertical line marks the isospin-limit $B\bar{B}^*$ threshold used in the present calculation.}
\label{fig:cms-chib3p}
\end{figure}

The most precise measurement of the $\chi_{b1}(3P)$ state was reported by CMS~\cite{CMS:2018vgf}, which is also the sole source of the $\chi_{b2}(3P)$ mass in the RPP~\cite{ParticleDataGroup:2026mpi}. 
Figure~\ref{fig:cms-chib3p} shows the $\Upsilon(nS)\gamma$ invariant-mass spectra in the $\chi_{bJ}(3P)$ region measured by CMS~\cite{CMS:2018vgf}. The CMS simulation study shows that the $\Upsilon(nS)\gamma$ mass resolution is approximately linear in the difference between the parent $\chi_{bJ}$ mass and the daughter $\Upsilon(nS)$ mass. Thus, CMS used only the $\Upsilon(3S)\gamma$ channel to extract the $\chi_{bJ}$ masses. 
In that channel, there is a single bin
lower than the others
in the peak region around the green and magenta dotted vertical lines in Fig.~\ref{fig:cms-chib3p}
showing the masses of the $\chi_{b1}(3P)$ and $\chi_{b2}(3P)$ states reported in Ref.~\cite{CMS:2018vgf}.
The mass resolutions are $(2.18\pm0.32)$, 7 and 15~MeV, respectively, for the $\Upsilon(3S)\gamma$, $\Upsilon(2S)\gamma$, and $\Upsilon(1S)\gamma$ spectra.
Interestingly, only a single peak is observed around 10.51~GeV in the $\Upsilon(2S)\gamma$ distribution, even though its energy resolution is smaller than the mass splitting of $(10.60\pm0.66)$~MeV between the $\chi_{b1}(3P)$ and $\chi_{b2}(3P)$ states reported in Ref.~\cite{CMS:2018vgf}.
Our results here suggest assigning the signal in all three final states to the same $\Xb$ state, while the $\chi_{b1}(3P)$ dominated one is yet to be discovered.
The vertical grey band in Fig.~\ref{fig:cms-chib3p} shows
the mass range we find for the $\chi_{b1}(3P)$ when varying the $b$-quark mass. Thus, either
the very steep drop off of the 
data in this mass range is driven not by a single state but by the interplay of two states with the same quantum numbers, namely
the $\Xb$ and the $\chi_{b1}(3P)$,
or one 
would need a mechanism to explain, why the predominantly compact $\chi_{b1}(3P)$ state in Table~\ref{tab:results} does not show up in the $\gamma\Upsilon(nS)$ spectra, in contrast to both the lower-lying states and the composite $\Xb$. A plausible explanation is that cancellations occur in the decay amplitudes, given that $\chi_{b1}(3P)$ is the only state with sizable contributions from both the $\bar{b}b$ and $B\bar{B}^*$ components. At this stage, however, this remains speculative, and further studies are required to clarify the situation.
How the $\chi_{b0}(nP)$ and $\chi_{b2}(nP)$ states are affected by the coupling to $B^{(*)}\bar{B}^{(*)}$ channels also needs to be investigated in the future. 
In closing we note that 
the D0 Collaboration observed a resonant peak with unknown spin at $(10.551\pm0.014\pm0.017)$~GeV~\cite{D0:2012pig}, which is compatible with the predicted $\chi_{b1}(3P)$ mass in Table~\ref{tab:results}.

\begin{table*}[t]
\centering
\caption{Predictions of \BOchiEFT for the bottomonium states $\chi_{b1}(nP)$, with $n=1,2,3$, when the counterterm $C$ is artificially tuned to fix the binding energy of the $\Xb$ equal to 1~MeV (marked by an asterisk) and $V_0$ is varied to minimise $(\Delta M)^2$. 
The results for the deeply bound spurious pole projected out (not projected out) are quoted in the first (second) row---see Fig.~\ref{fig:levels} and the discussion in Sect.~\ref{sec:Xb} and Appendix~\ref{app:deep} for further details. The $b$-quark mass is fixed to the value in Eq.~\eqref{mb} and 
the probabilities $\mathcal P_{\bar{b}b}$ and $X_S$ are given in per cent.}
\label{tab:comp}
\begin{ruledtabular}
\begin{tabular}{l|cc|ccc|ccc|ccc|ccc|c}
& & &
\multicolumn{3}{c|}{$\chi_{b1}(1P)$}
& \multicolumn{3}{c|}{$\chi_{b1}(2P)$}
& \multicolumn{3}{c|}{$\chi_{b1}(3P)$}
& \multicolumn{3}{c|}{$\Xb$}
& \\
Scheme
& $C$ (GeV$^{-2}$)
& $V_0$ (MeV)
& $M$ (GeV) & $\mathcal P_{\bar{b}b}$ & $X_S$
& $M$ (GeV) & $\mathcal P_{\bar{b}b}$ & $X_S$
& $M$ (GeV) & $\mathcal P_{\bar{b}b}$ & $X_S$
& $E_b$ (MeV) & $\mathcal P_{\bar{b}b}$ & $X_S$
& $\Delta M$ (MeV) \\
\midrule
Projected
& 2.66
& 362
& 9.884 & 100 & $<1$
& 10.246 & 99 & 1
& 10.535 & 86 & 13
& $1^*$ & 8 & 66
& 15.0 \\
Not projected 
& 2.52
& 363
& 9.881 & 100 & $<1$
& 10.246 & 99 & 1
& 10.535 & 86 & 13
& $1^*$ & 8 & 66
& 15.9
\end{tabular}
\end{ruledtabular}
\end{table*}

\section{Conclusions}
\label{sec:conclusions}

The existence and properties of the $\Xb$, the hidden-bottom counterpart of the celebrated
$\chi_{c1}(3872)$, also known as $ \X$, have been under active discussion for the past two decades, and a wide variety of predictions for it have been put forward. Within the hadronic EFT approach to exotic hadrons, two major obstacles hinder reliable predictions for this hypothetical state: (i) the lack of a controlled relation between the short-range dynamics in two-hadron systems across different heavy-quark sectors and (ii) insufficient information on the couplings of generic $\bar{b}b$ quark--antiquark states lying below the open-bottom threshold to the corresponding two-hadron channels. Recent unquenched lattice-QCD calculations have largely resolved the second issue above
and opened the avenue to systematic studies of the interplay of hadronic and quark degrees of freedom. In this spirit the BOEFT was developed and applied to exotic hadrons.
Presently existing versions of BOEFT are formulated in $r$-space, which allows for a straightforward implementation
of various short-distance constraints derived for interquark interactions 
within (p)NRQCD. 

Here we present an alternative approach by formulating an EFT in momentum space that builds on the same input of static quark-antiquark potentials with string breaking from lattice, but allows for the full inclusion of the chiral hadron-hadron interactions, dubbed \BOchiEFT. The new formalism allows us
to incorporate the unitarisation effects induced by the open-bottom $B\bar{B}^*$
and $B^*\bar{B}^*$
channels at a qualitatively higher level than is customary in unquenched quark models. In particular, we not only include the coupling between the bottomonium states and the open-bottom channels but also account for the relevant one-meson-exchange interactions, including the tensor force, within the latter; moreover, the hadron--hadron interactions can be improved systematically following the chiral EFT developed for the two-nucleon system. In the meantime, while this overcomes the problem with the hadron--hadron interaction, the suggested approach comes with the need to absorb the short-range physics into local counterterms to be determined either from experiment or via matching to other theoretical calculations.
In particular, in the new formalism, constraints from the cylindrical symmetry underlying a system of two static quarks enter only at subleading orders. It remains to be seen to what extent the BOEFT and \BOchiEFT formulations can be merged to provide an overarching approach to hidden-heavy-flavour hadrons.

To exemplify \BOchiEFT at work, we (i) exploit the fact that recent lattice-QCD calculations allow us to fix 
the parameters for both the interquark interaction inside bottomonia and the coupling between compact quark states and open-bottom channels and (ii) determine the remaining short-range interaction, that appears in the \BOchiEFT at LO, by requiring the experimental spectrum of the axial-vector bottomonia with $J^{PC}=1^{++}$ to be reproduced as accurately as possible. We find that, in the presented framework, the best description of the observed $1^{++}$ spectrum in the bottomonium mass region identifies the observed state at 10.513~GeV as a predominantly molecular one, $\Xb$, and predicts an additional predominantly compact axial-vector bottomonium at about 10.544~GeV.
Meanwhile, calculations in which the $\Xb$ lies between the generic $\chi_{b1}(3P)$ bottomonium state and the $B\bar{B}^*$ threshold, typically exhibit deviations between theory and experiment of more than 10~MeV for the experimentally known states. 
Although this scenario is also accessible in \BOchiEFT, it is not preferred when the two free parameters of the LO calculation are tuned to reproduce the existing data.

Both theories BOEFT and \BOchiEFT find four poles
in the $1^{++}$ sector 
below the open-bottom threshold $B\bar{B}^*$. The three lowest of those can be identified with the currently observed states. Meanwhile, the compositions of the two heaviest states differ substantially among the calculations, which we traced to how the parameters of \BOchiEFT were determined.
In any case, observable consequences of these different scenarios beyond the scattering parameters presented here should therefore be further explored, as they may eventually provide a way to distinguish the two interpretations experimentally.

The formalism presented in this work can be applied to the charm sector and, in particular, to the $ \X$. There are, however, three reasons why we do not perform such an analysis here.
\begin{enumerate}
\item In the $1^{++}$ charm sector, there is only a single established bound $\chi_{c1}$ state, which is insufficient to fix even the LO parameters of \BOchiEFT when considering the axial-vector sector alone, if the $\X$ is excluded from the fitting procedure. Note also that, within the EFT proposed here, the same contact terms cannot be employed in the $\bar{b}b$ and $c\bar{c}$ sectors at the present stage, since the heavy-quark-mass dependence of the short-range operators is not under theoretical control~\cite{Baru:2018qkb}. 
Whether this problem can be overcome by matching the contact-term operators to the quark-level potential at short distances will be investigated in the future.
\item The explicit inclusion of the $\pi D\bar{D}$ cut is numerically demanding.
\item The mass of the charm quark is substantially smaller than the mass of the bottom quark, so sizable corrections to the static potentials underlying the formalism need to be included.
\end{enumerate}
Once the higher-lying meson $\chi_{c1}(4010)$ \cite{LHCb:2024vfz}, which has been proposed to be a $\chi_{c1}(2P)$-dominated state \cite{Wang:2023ovj,Li:2024pfg,Shi:2024llv}, is established, the situation will change, putting us in a position to perform an analogous study, as the one presented here, also in the charm sector, albeit with the inclusion of $1/m_c$ corrections.

\acknowledgments
We acknowledge useful discussions with Eric Braaten, Roberto Bruschini, Francesco Knechtli, Ulf-G. Mei\ss ner, Juan Nieves, Andreas Nogga, Mike Peardon, Tommaso Scirpa and Antonio Vairo.
Special thanks go to Nora Brambilla for comments on the manuscript and on 
details of the calculations. 
This work is supported in part by the National Key R\&D Program of China under Grant No. 2023YFA1606703; by the National Natural Science Foundation of China (NSFC) under Grants No. 12125507, No. 12361141819, and No. 12447101, by the Chinese Academy of Sciences (CAS) under Grant No.~YSBR-101, and by
Deutsche Forschungsgemeinschaft (DFG) under Grant No. 576422137
and under Germany's Excellence Strategy -- EXC 3107 -- Project-ID~533766364. In addition, C.H., and A.N. thank the CAS President's International Fellowship Initiative (PIFI) for partial support under Grant Nos. 2025PD0087 and 2024PVA0004\_Y1, respectively.
Various AI agents were used for language polishing, cross-checking, code generation, and literature verifications, under the \texttt{Nullius} harness framework~\cite{nullius}. The authors take full responsibility for the work.

\appendix

\section{Coordinate-space Cornell wave functions and transition form factors}
\label{app:cornell}

In this appendix, we provide technical details on solving the Schr{\"o}dinger equation in Eq.~\eqref{H0} with the Cornell potential given in Eq.~\eqref{Cornell}. For angular momentum $L=1$ and setting $V_0=0$, as in the definition of the reference spectrum in Table~\ref{tab:bottom}, we obtain a radial eigenvalue equation for the radial wave function $u_n(r)=rR_{1n}(r)$ [see the definition of $R_{1n}(r)$ in Eq.~\eqref{psi}],
\be
\left[-\frac{1}{m_b}\frac{d^2}{dr^2}
+\frac{2}{m_b r^2}+\sigma r+\frac{\gamma}{r}\right]u_n(r)
=\epsilon_n u_n(r),
\label{eq:radial}
\ee
or, equivalently,
\be
\begin{split}
u_n''(r)&=Q_n(r)u_n(r),\\
Q_n(r)&=\frac{2}{r^2}
+m_b\left(\sigma r+\frac{\gamma}{r}-\epsilon_n\right),
\end{split}
\label{eq:radialQ}
\ee
where $\epsilon_n=M_n^{(0)}-2m_b$. The wave function scales as $u_n(r)\propto r^2$ near the origin and is normalised according to
\be
\int_0^\infty dr\,|u_n(r)|^2=1.
\label{unorm}
\ee

The numerical solution in Eq.~\eqref{eq:radial} is obtained by first discretising the equation on a uniform radial grid, $r_j=j\Delta r$ ($j=1,..,N_r$), where $\Delta r$ is the spacing between adjacent radial points and $r_{\rm box}=N_r\Delta r$ defines the size of the finite-difference grid. The boundary conditions imposed on the radial wave function are 
\be
u_n(0)=0,\quad u_n(r_{\rm box}+\Delta r)=0.
\ee
In this work we use
\be
\begin{split}
N_r&=1.2\times10^4,\\
r_{\rm box}&=30~{\rm GeV}^{-1}\simeq5.92~{\rm fm},\\
\Delta r&=2.5\times10^{-3}~{\rm GeV}^{-1}
\simeq4.93\times10^{-4}~{\rm fm}.
\end{split}
\ee
The resulting eigenvalue problem provides the eigenenergies $\epsilon_n$ and the corresponding wave functions in the box.

To construct the physical wave function tail beyond the finite-difference grid, we use the same radial Schr{\"o}dinger equation in Eq.~\eqref{eq:radialQ}. The outer classical turning point $r^{(0)}_n$ is defined as the largest solution of the equation $Q_n(r)=0$. Then, in the classically forbidden region at $r>r^{(0)}_n$, one has $Q_n(r)>0$, and, sufficiently far from the turning point, the regular quasiclassical solution is given by
\be
u_n(r)\propto Q_n(r)^{-1/4}
\exp\left[-\int_0^r d\rho\,\sqrt{Q_n(\rho)}\right],
\label{wkb}
\ee
with the logarithmic derivative from the Wentzel--Kramers--Brillouin (WKB) approximation,
\be
\bigl[\log u_n(r)\bigr]'
=-\sqrt{Q_n}-\frac{Q_n'}{4Q_n}+\cdots.
\label{eq:wkb}
\ee
The decaying boundary condition,
\be
u_n(r\to\infty)\to 0,
\ee
is implemented by taking
\be
r_{\rm far}=36~{\rm GeV}^{-1}\simeq7.10~{\rm fm},
\label{uboundary}
\ee
as the starting point for the inward Numerov integration, and fixing the logarithmic derivative there to the WKB form in Eq.~\eqref{eq:wkb}.
The overall coefficient of the inward Numerov solution is fixed by matching it to the finite-difference wave function in a least-squares fit over a common $1~{\rm GeV}^{-1}$ interval located beyond the outer turning point. The particular position of this interval depends on $m_b$, and for the central value of $m_b$ in Eq.~\eqref{mb}, it is
$10.30\leqslant r\leqslant11.30~{\rm GeV}^{-1}$
($2.03\leqslant r\leqslant2.23~{\rm fm}$).
The finite-difference solution is used below the matching interval, while the matched Numerov solution provides the wave function tail above it. 
The resulting piecewise wave functions are then normalised according to Eq.~\eqref{unorm}.

Finally, using Eqs.~\eqref{psinp} and \eqref{eq:fn}, the form factor $f_n(p)$ entering the separable potential in Eq.~\eqref{VQMsep} is evaluated as
\be
\begin{split}
f_n(p)=\frac{\psi_n(p)}{\sqrt{4\pi}}
&=\sqrt{4\pi}\int_0^\infty dr\,r^2j_0(pr)R_{1n}(r)\\
&=\frac{\sqrt{4\pi}}{p}\int_0^\infty dr\,u_n(r)\sin(pr),
\label{fneval}
\end{split}
\ee
where the momentum $p$ is taken real and 
the overall phase of the wave function is chosen such that $f_n(0)>0$.
Indeed, for a bound-state pole at $E=-\kappa^2/(2\mu)$ lying on the physical Riemann sheet, the loop momentum in Eqs.~\eqref{LSEOME} and \eqref{Mdef} takes values on the real interval $0\leqslant q\leqslant\Lambda$, so the form factors $f_n(q)$ are evaluated according to Eq.~\eqref{fneval}. 
By contrast, the on-shell subthreshold amplitude, used in Appendix~\ref{app:tensor-off-reference} below, contains the form factors $f_n$ evaluated at imaginary momenta, $p=k=i\kappa$. Then Eq.~\eqref{fneval} turns to 
\be
f_n(i\kappa)=\frac{\sqrt{4\pi}}{\kappa}
\int_0^\infty dr\,u_n(r)\sinh(\kappa r).
\label{fnkappa}
\ee
At large $r$ and up to an algebraic prefactor, Eq.~\eqref{wkb} gives an exponential behaviour,
\be
u_n(r)\sim\exp\left(-\frac23\sqrt{m_b\sigma r^3}
+\mathcal{O}(\sqrt{r})\right),
\ee
which decreases with $r$ faster than any exponent of the form $\exp(-\kappa r)$. It therefore dominates over the exponential growth of
$\sinh(\kappa r)\propto\exp(\kappa r)$ for any finite $\kappa$, ensuring convergence of the integral in Eq.~\eqref{fnkappa}.

To evaluate the subthreshold LSE amplitude, the external on-shell momentum entering the OME kernel is analytically continued from real values to $p=i\kappa$ along the path
\be
p=p_1+ip_2,\quad p_1\to 0^+,\quad 0\leqslant p_2\leqslant\kappa,
\ee
with the loop momentum kept real throughout. This prescription approaches the negative-energy region from the physical edge of the unitarity cut, $E+i\epsilon$, and thereby fixes the physical-sheet branch of the OME kernel. The $i\epsilon$ prescription in Eq.~\eqref{eq:alphaOME} is retained, and all square roots and logarithms in Eq.~\eqref{eq:deltaanalytic} are continued continuously along this path, including their physical boundary values when an OME cut is encountered. The compact-state wave functions at $p=i\kappa$ are evaluated according to Eq.~\eqref{fnkappa}.

\section{Spin-dependent interactions in bottomonia}
\label{app:VSD}

The spin-dependent interaction in a bottomonium system takes a well-known form ~\cite{Eichten:1980mw,Gromes:1984ma},
\be
\begin{split}
V_{\rm SD}(r)=V_{\rm SS}(r)(2\veS^2-3)+V_{\rm SL}(r)\left(\veS\cdot\veL\right)
+V_{\rm ST}(r)S_{12},
\label{VSD0}
\end{split}
\ee
where $S_{12}=3(\veS\cdot\hat \ver)^2-\veS^2$ denotes the spin-tensor structure, $\hat \ver$ is the unit vector for the coordinate $\ver$, $\veS$ is the total spin of the $\bar{b}b$ pair, and the hyperfine, spin--orbit, and spin-tensor potentials in Eq.~\eqref{VSD0}, derived from the general expressions in Refs.~\cite{Eichten:1980mw,Gromes:1984ma} for the static potential in Eq.~\eqref{Cornell}, read
\be
\begin{split}
&V_{\rm SS}(r)=-\frac{2\pi\gamma}{3m_b^2}\left|\Psi(0)\right|^2,\\
&V_{\rm SL}(r)=-\frac{\sigma}{2m_b^2r}-\frac{3\gamma}{2m_b^2r^3},\\
&V_{\rm ST}(r)=-\frac{\gamma}{4m_b^2r^3}.
\end{split}
\label{VSD}
\ee
Since $\gamma<0$ [see Eq.~\eqref{Cornell}], a well-known cancellation occurs in $V_{\rm SL}(r)$ between the contributions from the confining and color-Coulomb interactions; see, for example, Ref.~\cite{Schnitzer:1983tt} for an early phenomenological observation and Ref.~\cite{Isgur:1999rf} for a detailed discussion. In this work, we address $P$-level bottomonia with $\Psi(0)=0$ and, therefore, the hyperfine term $V_{\rm SS}(r)$ in Eq.~\eqref{VSD} vanishes. It is also straightforward to find that
\be
\begin{split}
\braket{\veS\cdot\veL} &=-2+\frac12J(J+1),\\
\braket{S_{12}} &=1-\frac{9}{2J+1}\left(\braket{1010|J0}\right)^2,
\end{split}
\ee
so the resulting mass of the bottomonium state is given by a sum of $\Mbb$, the eigenvalue of the spinless Hamiltonian in Eq.~\eqref{H0}, and the correction $\Delta M_{\rm SD}[\chi_{bJ}]$ from the spin-dependent interactions in Eq.~\eqref{VSD0},
\be
\begin{split}
\Delta M_{\rm SD}[\chi_{b0}] &= -2\braket{V_{\rm SL}} - 2\braket{V_{\rm ST}}, \\[2mm]
\Delta M_{\rm SD}[\chi_{b1}] &= -\braket{V_{\rm SL}} + \braket{V_{\rm ST}}, \\
\Delta M_{\rm SD}[\chi_{b2}] &= \braket{V_{\rm SL}} - \frac15\braket{V_{\rm ST}},
\end{split}
\label{DeltaEn}
\ee
where the symbol $\braket{\ldots}$ denotes an average over the radial wave function. It is easy to check that 
\be
\sum_{J=0}^2(2J+1)\Delta M[\chi_{bJ}]=0,
\label{chis2}
\ee
so that $\Mbb$ in Eq.~\eqref{Scheq} defines the centre of gravity of the spin multiplet $\{\chi_{bJ}(nP)\}$,
\be
\bar{M}[\chi_b]=\frac{1}{9}
\sum_{J=0}^2(2J+1)M[\chi_{bJ}]=\Mbb.
\label{chis}
\ee 

It can be verified by an explicit calculation that, for the parameters of the interaction in Table~\ref{tab:parameters} and the $b$-quark mass in Eq.~\eqref{mb}, the variation of the spin-dependent correction $\Delta M_{\rm SD}[\chi_{b1}]$ in Eq.~\eqref{DeltaEn} remains within approximately $1$~MeV for the $\bbnP$ states considered in the analysis. 

\section{OPE in charmonium and bottomonium systems}
\label{app:OPE}

In this appendix, we collect information on the OPE interaction relevant to the present investigation, with particular emphasis on the essential difference between its treatment in the charmonium and bottomonium sectors for shallow near-threshold states such as $\X$, $T_{cc}^+$, $\Xb$, and others. Thus, for a general discussion, we consider a generic two-meson system with the heavy quark denoted by $Q$ ($Q=c,b$) and use the notation $m$ and $m_*$ for the corresponding meson masses.

We start from the generic OME potential in the $B\bar{B}^*$ system and set $\phi=\pi$, so
\be
V^{\rm (OPE)}(\vep',\vep;m_\phi)
=\left(\frac{g_{\scriptscriptstyle Q}}{2F_\pi}
\right)^2
\frac{(\veep\cdot \veq)(\veq\cdot {\veep^{\prime *}})}{u-m_\pi^2},
\label{VOMEpi}
\ee
where the denominator can be written in the form
\be
u-m_\pi^2=q^2-m_\pi^2=-\veq^2-\left(m_\pi^2-q_0^2\right),
\ee
with $\veq=\vep'-\vep$. Expressing the temporal component of the transferred momentum, $q_0$, in terms of the energies of the mesons exchanging the pion leads to the expressions provided in Subsect.~\ref{sec:OME} and Appendix~\ref{app:PWA}, in particular to the denominators in Eq.~\eqref{eq:Dbeta} with the corresponding $\mu$'s specified in Eq.~\eqref{eq:alphaOME}. Because of the recoil terms ${\cal O}(p^{\prime 2},p^2)$ appearing in the latter equation, the OPE potential is nonlocal and cannot be expressed solely as a function of the transferred momentum $\veq$. To assess the physical significance of these recoil terms, we first neglect them, such that
\be
q_0=m_*-m=\Delta m,
\ee
and therefore
\be
u-m_\pi^2= -(\veq^2+\mu_\pi^2).
\ee
The OPE potential then becomes local, with the effective pion mass parameter \cite{Tornqvist:1993ng,Swanson:2003tb,Suzuki:2005ha,Thomas:2008ja},
\be
\mu_\pi^2=m_\pi^2-(\Delta m)^2,
\label{mupi}
\ee
controlling the long-distance behaviour of OPE. To make this feature explicit, we decompose the structure in the numerator in Eq.~(\ref{VOMEpi}) according to Eq.~\eqref{decomp}~\cite{Ericson:1988gk}\footnote{See also Eq.~\eqref{VPWA} and Appendix~\ref{app:PWA} for a generic way to perform partial wave decomposition of the potential in Eq.~\eqref{VOMEpi}.}
and for simplicity focus on the central contribution given by the first term in Eq.~(\ref{decomp}). Specifying the isoscalar channel for the OPE contribution, whose isospin factor is 3, we obtain
\be
3V^{\rm (OPE)}(\vep,\vep';m_\pi)\Longrightarrow
(\veep\cdot \veep^{\prime *})V^{\rm (OPE)}_{I=0;{\rm cent}}(\veq;m_\pi),
\ee
with
\be
\begin{split}
V^{\rm (OPE)}_{I=0;{\rm cent}}(\veq;m_\pi)&=\left(\frac{g_{\scriptscriptstyle Q}}{2F_\pi}\right)^2
\frac{\veq^2}{u-m_\pi^2}\\
&=\frac{g_{\scriptscriptstyle Q}^2}{4F_\pi^2}\left(-1+\frac{\mu_\pi^2}{\veq^2+\mu_\pi^2}\right).
\end{split}
\label{VOPE0}
\ee
In the last equation, the first term in parentheses on the right-hand side represents the short-range contribution to OPE and is proportional to $\delta^{(3)}(\ver)$ in coordinate space. The second term governs the long-distance tail, which is fully specified by the coupling $g_{\scriptscriptstyle Q}$ and the effective mass parameter in Eq.~(\ref{mupi})
leading to a modified
$r$ dependence proportional to $g_{\scriptscriptstyle Q}^2e^{-\mu_\pi r}/r$. 

In the TOPT approach, the static OPE propagator arises from 
the two contributions, schematically depicted as the first two diagrams in Fig.~\ref{fig:TOPT}, after the recoil and $E$ terms are dropped, so
in Eq.~\eqref{eq:alphaOME} $\mu_1\approx-\Delta m$, $\mu_2\approx\Delta m$, and, therefore,
\be
\begin{split}
D_1+D_2&=
\frac{1}{2E_\pi}\left[\frac{1}{E_\pi-\Delta m}+\frac{1}{E_\pi+\Delta m}\right]\\
&=\frac{1}{E_\pi^2-(\Delta m)^2}=\frac{1}{\veq^2+\mu_\pi^2},
\end{split}
\ee
where the parameter $\mu_\pi^2$ defined in Eq.~\eqref{mupi} naturally appears.

It is instructive to contrast pion exchange in the $D\bar{D}^*$ and $B\bar{B}^*$ systems relevant for the $\X$ and $\Xb$, respectively. A crucial difference between the two cases arises from the fact that, for the physical meson masses,
\be
\mu_\pi^2[D]=m_\pi^2-(m_{D^*}-m_D)^2<0,
\label{mupiD}
\ee
whereas
\be
\mu_\pi^2[B]=m_\pi^2-(m_{B^*}-m_B)^2>0.
\label{mupiB}
\ee
This distinction has several important consequences for OPE in the $\X$ and $\Xb$. The OPE interaction in the $\Xb$ case is closer in spirit to the conventional Yukawa potential, but with the pion mass effectively shifted due to the mass difference $m_{B^*}-m_B$, as given in Eq.~\eqref{mupiB}. The situation for the $\X$ is more subtle because $\mu_\pi^2[D]$ in Eq.~\eqref{mupiD} is negative, resulting in a purely imaginary $\mu_\pi[D]$. Consequently, the long-distance term in Eq.~\eqref{VOPE0} not only changes sign relative to the case of real $\mu_\pi[B]$, but in addition its spatial dependence in $r$-space changes from an exponential falloff in the $B$ sector to an oscillatory behaviour in the $D$ sector.
In the latter case, the intermediate 3-body state $\pi D\bar{D}$ (see the first diagram in Fig.~\ref{fig:TOPT}) can go on shell \cite{Suzuki:2005ha}, causing the potential to acquire an imaginary part. Under these circumstances, the recoil terms become essential for reproducing the correct analytic structure of the amplitude. Consequently, a simplified treatment of OPE for the $\X$ and similar shallow near-threshold states in the charm sector is not justified and may lead to unphysical results~\cite{Baru:2011rs,Du:2021zzh}. The full nonlocal potentials in Eqs.~\eqref{VBBstBBst}--\eqref{VBstBstBstBst} must therefore be retained.

\section{Partial wave projection of OME}
\label{app:PWA}

In this appendix, we provide explicit expressions for the partial-wave-projected potentials in Eqs.~\eqref{VBBstBBst}-\eqref{VBstBstBstBst}. As mentioned in the main text, this is straightforwardly gained by employing the projectors technique, as detailed in Eq.~\eqref{VPWA}. Those relevant to the studied system read
\begin{align}
P_i(\alpha=1;\ven)&=P_i\Bigl(B\bar{B}^*(^3S_1);\ven\Bigr)=\epsilon_i,\notag\\
P_i(\alpha=2;\ven)&=P_i\Bigl(B\bar{B}^*(^3D_1);\ven\Bigr)\notag\\
&=-\frac{3}{\sqrt{2}}\epsilon_j \left( n_i n_j-\frac13\delta_{ij}\right),
\label{projectors}\\
P_i(\alpha=3;\ven)&=P_i\Bigl(B^*\bar{B}^*(^5D_1);\ven\Bigr)=-\frac{\sqrt{3}}{2} i \varepsilon_{ijk}\notag\\
\times\Bigl(\epsilon_{1j}\epsilon_{2m}+\epsilon_{1m}&\epsilon_{2j}-\frac23\delta_{jm}(\veep_1\cdot\veep_2)\Bigr)
\left(n_k n_m-\frac13\delta_{km}\right),\notag
\end{align}
and are normalised as given in Eq.~\eqref{Pnorm}.

Then it is easy to find by an explicit calculation that 
\begin{align}
V_{11}^{\rm (OME)}(p',p;\mG)={}&\frac{g_b^2}{24 F_\pi^2}
\Bigl[2pp^{\prime}\Delta^{(12)}_1(p',p;\mG)\notag\\
&-\left(p^2+p^{\prime2}\right)\Delta^{(12)}_0(p',p;\mG)\Bigr],
\label{eq:OPES}
\\[2mm]
V_{12}^{\rm (OME)}(p',p;\mG)={}&\frac{g_b^2}{24\sqrt{2} F_{\pi}^2}
\Bigl[3 p^{\prime2}\Delta^{(12)}_2(p',p;\mG)\notag\\
&-4pp^{\prime}\Delta^{(12)}_1(p',p;\mG)
\label{eq:OPESD}\\
&+\left(
2p^2-p^{\prime2}\right)\Delta^{(12)}_0(p',p;\mG)\Bigr],
\notag\\[2mm]
V_{13}^{\rm (OME)}(p',p;\mG)={}&-\frac{g_b^2}{8\sqrt{6} F_{\pi}^2}
\Bigl[3 p^{\prime2}\Delta^{(24)}_2(p',p;\mG)
\notag\\
&-4pp^{\prime}\Delta^{(24)}_1(p',p;\mG)
\label{eq:OPESDstar}\\
&+\left(2p^{2}-p^{\prime2}\right)\Delta^{(24)}_0(p',p;\mG)\Bigr],
\notag\\[2mm]
V_{21}^{\rm (OME)}(p',p;\mG)={}&V_{12}^{\rm (OME)}(p,p';\mG),\\[2mm]
V_{22}^{\rm (OME)}(p',p;\mG)={}&\frac{g_b^2}{24 F_{\pi}^2}
\Bigl[\frac92pp^{\prime}\Delta^{(12)}_3(p',p;\mG)\notag\\
&-3\left(p^2+p^{\prime2}\right)\Delta^{(12)}_2(p',p;\mG)\notag\\
&-\frac12pp^{\prime}\Delta^{(12)}_1(p',p;\mG)\label{eq:OPED}\\
&+\left(p^2+p^{\prime2}\right)\Delta^{(12)}_0(p',p;\mG)\Bigr],
\notag\\
V_{23}^{\rm (OME)}(p',p;\mG)={}&\frac{g_b^2}{16\sqrt{3} F_{\pi}^2}
\Bigl[3pp^{\prime}\Delta^{(24)}_3(p',p;\mG)\notag\\
&-3\left(p^2+p^{\prime2}\right)\Delta^{(24)}_2(p',p;\mG)\notag\\
&+pp^{\prime}\Delta^{(24)}_1(p',p;\mG)\label{eq:OPEDDstar}\\
&+\left(p^2+p^{\prime2}\right)\Delta^{(24)}_0(p',p;\mG)\Bigr],
\notag\\[2mm]
V_{31}^{\rm (OME)}(p',p;\mG)={}&V_{13}^{\rm (OME)}(p,p';\mG) \\
V_{32}^{\rm (OME)}(p',p;\mG)={}&V_{23}^{\rm (OME)}(p,p';\mG)\\
V_{33}^{\rm (OME)}(p',p;\mG)={}&-\frac{g_b^2}{16 F_{\pi}^2}\,pp^{\prime}
\Bigl[\Delta^{(22)}_1(p',p;\mG)\notag\\
&\,-\Delta^{(22)}_3(p',p;\mG)\Bigr],
\label{eq:OPEDstarDstar}
\end{align}
where the auxiliary functions $\Delta_m^{(\beta_1\beta_2)}$ ($m=0,1,2,3$) are defined as angular integrals, 
\begin{align}
\Delta^{(\beta_1\beta_2)}_m(p',p;\mG) ={}& \int_{-1}^1 x^mdx\Bigl[
D_{\beta_1}(p',p,x;\mG)\notag\\
&+D_{\beta_2}(p',p,x;\mG)\Bigr],
\label{eq:delta}
\end{align}
with the intermediate states in Eqs.~\eqref{VBBstBBst}, \eqref{VBstBstBBst} and \eqref{VBstBstBstBst}, respectively. Here, the $\beta=3$ channel in Eq.~\eqref{VBBstBstBst} enters $V_{31}$ and $V_{32}$ through the interchange of the arguments, since $D_4(p,p',x;\mG)=D_3(p',p,x;\mG)$; see the last two diagrams in Fig.~\ref{fig:TOPT}. 
The TOPT propagators introduced in Eq.~\eqref{eq:Dbeta} can be presented in the form
\be
D_\beta(p',p,x;\mG)=\frac1{2\EG(q)\,[\EG (q)+\mu_\beta]},\quad \beta=1,..,4,
\label{eq:Dabapp}
\ee
where $x$ is the cosine of the angle between the momenta $\vep$ and $\vep'$, $q=\sqrt{p^2+p'^2-2pp'x}$, and the exchanged meson energy reads
\be
\EG(q)=\sqrt{q^2+\mG^2}=\sqrt{{\cal A}_\phi-{\cal B}x},
\label{EGdef}
\ee
with
\be
{\cal A}_\phi=p^2+p^{\prime2}+m_\phi^2,\quad
{\cal B}=2pp'.
\ee
For $m_B=m_{B^*}$, all functions $D_\beta$ coincide with each other and so do all $\Delta$'s in Eq.~\eqref{eq:delta}. In this limit, $V_{13}=-\sqrt3\,V_{12}$ and $2V_{23}=\sqrt3\,(V_{22}-V_{33})$, as dictated by HQSS.

The angular integral in Eq.~\eqref{eq:delta} can be taken analytically, through a variable change $x\to\EG=\sqrt{{\cal A}_\phi-{\cal B}x}$, yielding (we consider only the case of the potential in Eq.~\eqref{eq:OPES} as example)
\be
\begin{split}
\Delta_0={}&\frac{1}{{\cal B}}\sum_{\beta=1}^2 L_\beta,\\
\Delta_1={}&\frac{1}{{\cal B}^2}\sum_{\beta=1}^2
\Bigl[({\cal A}_\phi-\mu_\beta^2)L_\beta\\
&+\mu_\beta
(\sqrt{{\cal A}_\phi+{\cal B}}-\sqrt{{\cal A}_\phi-{\cal B}})-{\cal B}\Bigr],
\end{split}
\label{eq:deltaanalytic}
\ee
where
\be
L_\beta=\operatorname{Log}\left(\frac{\sqrt{{\cal A}_\phi+{\cal B}}+\mu_\beta}{\sqrt{{\cal A}_\phi-{\cal B}}+\mu_\beta}\right).
\label{eq:OMEaux}
\ee
We note that the original integral in Eq.~\eqref{eq:delta} remains finite in the limit ${\cal B}\to 0$, and therefore the expressions in Eq.~\eqref{eq:deltaanalytic} are finite as well. Indeed, a straightforward calculation yields
\be
\left.\Delta_0\right|_{{\cal B}=0}
=\sum_{\beta=1}^2\frac{1}{\sqrt{{\cal A}_\phi}(\sqrt{{\cal A}_\phi}+\mu_\beta)},\quad
\left.\Delta_1\right|_{{\cal B}=0}=0.
\label{eq:deltazeroB}
\ee

\section{Results without the OME tensor force}
\label{app:tensor-off-reference}

For comparison with the full results presented in the main text, which we refer to as ``tensor-on'' results throughout this appendix, we consider a calculation in which only the $V_{11}^{\rm (OME)}$ potential in Eq.~\eqref{eq:OPES}, mediated by the one-pion and one-$\eta$ exchanges, is retained, while the other entries of the OME matrix Eqs.~\eqref{eq:OPESD}-\eqref{eq:OPEDstarDstar}, which involve the $D$ waves, are omitted. We refer to this truncated calculation as ``tensor-off'' for brevity. We emphasise that $V_{11}^{\rm (OME)}$ in Eq.~\eqref{eq:OPES} is kept unchanged in this calculation. The same three experimental masses and the root-mean-square deviation defined in Eq.~\eqref{dM} are used in the refit.

The dependence of the fitted tensor-off parameters on the heavy-quark mass is summarised in Table~\ref{tab:tensor-off-parameters}. The dagger marks the value
\be
m_b=4.646~\mbox{GeV},
\label{mbalt}
\ee
which plays the same role in the tensor-off calculation as the central value in Eq.~\eqref{mb} does in the full tensor-on calculation presented in the main text. Namely, it corresponds to a description of the masses of the three observed $\chi_{b1}$ states with a vanishing $\Delta M$. Removing the $D$-wave contribution changes drastically the UV behaviour of the interaction. Consequently, the fitted contact term $C$ differs from that of the tensor-on calculation given in Table~\ref{tab:parameters2}. As in the full tensor-on calculation, $V_0$ compensates for the corresponding shift of the reference spectrum and takes similar values. 
The tensor-off spectrum evaluated at the $b$-quark mass given in Eq.~\eqref{mbalt} is presented in Table~\ref{tab:results-tensor-off}. In this single-channel limit, the compositeness $X_S$ as defined in Eq.~\eqref{eq:XS} equals to $1-{\cal P}_{\bar{b}b}$.

The regulator dependence of the results is illustrated in Table~\ref{tab:cutoff-tensor-off}. As in the full calculation (see Table~\ref{tab:cutoff} in the main text), and in accordance with general EFT principles, variations of the regulator are absorbed predominantly by the contact term $C$, while the physical observables remain essentially unchanged. Increasing the number of explicit bare quark-model states from $N=4$ to $7$ at fixed $\Lambda=4$~GeV changes $C$ only from $6.159$ to $6.160$~GeV$^{-2}$, with the spectrum of physical states remaining unchanged.
As mentioned above, removing the tensor force changes the UV behaviour, so the cutoff dependence of $C$ differs from that in the complete tensor-on calculation.

\begin{table}[tb]
\centering
\caption{Tensor-off results obtained for $N=4$ and $\Lambda=4$~GeV. The second row uses $m_b$ in Eq.~\eqref{mb} rounded to one decimal place for presentation purposes. The dagger indicates the similarly rounded value of $m_b$ in Eq.~\eqref{mbalt}.}
\label{tab:tensor-off-parameters} 
\begin{ruledtabular}
\begin{tabular}{cccc}
$m_b$\unit{GeV} & $C$\unit{GeV$^{-2}$} & $V_0$\unit{MeV} & $\Delta M$\unit{MeV} \\
\hline
4.3 & 6.155 & 822 & 2.4 \\
4.5 & 6.157 & 373 & 0.8 \\
~$4.6^\dagger$ & 6.159 & 158 & 0.0 \\
4.7 & 6.160 & 54 & 0.4 \\
4.9 & 6.163 & $-331$ & 1.6 \\
5.1 & 6.167 & $-717$ & 2.8 \\
5.3 & 6.173 & $-1103$ & 3.8
\end{tabular}
\end{ruledtabular}
\end{table}

\begin{table}[tb]
\centering
\caption{Tensor-off spectrum and compositeness of the corresponding states obtained for $N=4$, $\Lambda=4$~GeV, and $m_b$ in Eq.~\eqref{mbalt}.}
\label{tab:results-tensor-off}
\begin{ruledtabular}
\begin{tabular}{lccc}
State & Mass\unit{GeV} & $E_B$\unit{MeV} & $X_S$ \\
\hline
$\chi_{b1}(1P)$ & 9.893 & 711 & $<1\%$ \\
$\chi_{b1}(2P)$ & 10.255 & 349 & 1\% \\
$\Xb$& 10.513 & 91 & 90\% \\
$\chi_{b1}(3P)$ & 10.548 & 57 & 23\%
\end{tabular}
\end{ruledtabular}
\end{table}

\begin{table}[t!]
\centering
\caption{Cutoff dependence of the tensor-off results with $N=4$ and $m_b$ in Eq.~\eqref{mbalt}. The other parameters are refitted to the same three experimental masses quoted in Table~\ref{tab:bottom} at each cutoff.}
\label{tab:cutoff-tensor-off}
\begin{ruledtabular}
\setlength{\tabcolsep}{3pt}
\begin{tabular}{ccccc}
$\Lambda$\unit{GeV} & $C$\unit{GeV$^{-2}$} & $\Delta M$\unit{MeV}
& $X_S [\Xb]$ & $X_S[\chi_{b1}(3P)]$ \\
\hline
3 & 5.887 & 0.0 & 88\% & 25\% \\
4 & 6.159 & 0.0 & 90\% & 23\% \\
5 & 6.216 & 0.0 & 90\% & 22\%
\end{tabular}
\end{ruledtabular}
\end{table}

Finally, we discuss ERE in the presence of the lhc for the simplified tensor-off case studied in this appendix. Switching off the OME-induced tensor interaction and readjusting the parameters corresponding to $\Delta M=0.0$~MeV gives
\be
a=-1.07~\mbox{fm},\quad r=-2.0~\mbox{fm},
\label{eq:aandr-tensoroff}
\ee
which can be compared with the scattering length and effective range found in the full, tensor-on calculation and quoted in Eq.~\eqref{eq:aandr}. 

\begin{figure}[tb]
\centering
\includegraphics[width=\columnwidth]{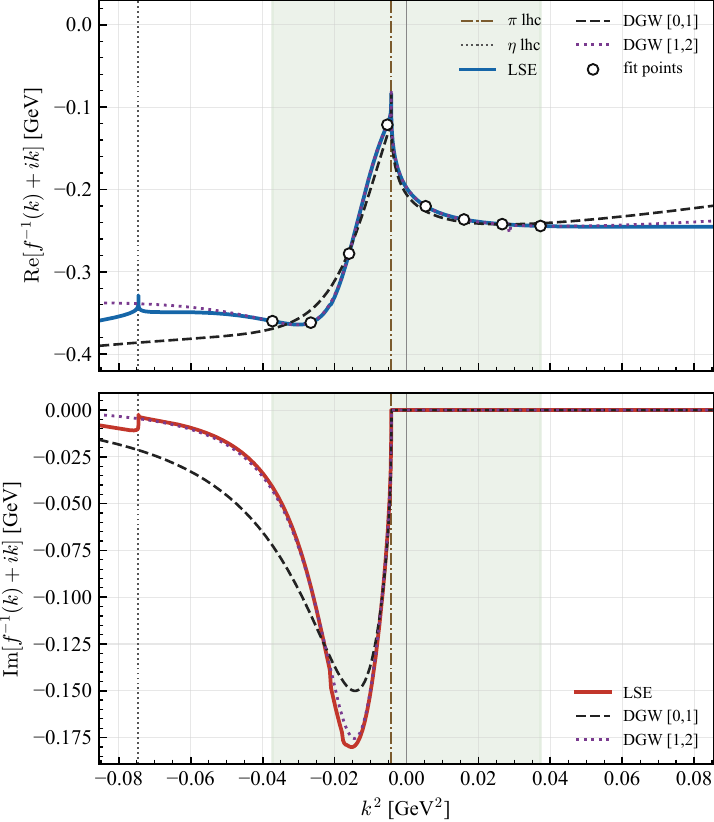}
\caption{Effects of lhc due to the one-pion and one-$\eta$ exchanges in $f^{-1}(k)+ik$ in the pure $S$-wave calculation with tensor force switched off. Upper and lower panels show the real and imaginary parts, respectively.
The solid lines represent the results from the tensor-off LSE obtained for $N=4$, $\Lambda=4$~GeV, and $m_b$ in Eq.~\eqref{mbalt}. 
The vertical dash-dotted and dotted lines mark the positions of the lhc branch points for the 
pion- and $\eta$-exchange, and the solid vertical line marks the threshold.
The dashed and dotted curves denote the best fit to the LSE real part using the $[0,1]$ and $[1,2]$ approximants of the parameterisation in Ref.~\cite{Du:2024gzw}, respectively, with the fitted points denoted by the open circles. }
\label{fig:dgw-ere}
\end{figure}

In Fig.~\ref{fig:dgw-ere}, we plot the real and imaginary parts of $f^{-1}(k)+ik$ obtained from the solution of the tensor-off LSE. One clearly sees that the branch point of the OPE-induced lhc lies rather close to the $B\bar{B}^*$ threshold and induces a significant effect. In particular, it limits the range of applicability of the traditional ERE to $|k|^2\lesssim 0.004$~GeV$^2$. A way to overcome this difficulty is to employ the Pad\'e-like parametrisation proposed by Du, Guo and Wu (DGW in Fig.~\ref{fig:dgw-ere}) in Ref.~\cite{Du:2024gzw}, which for one-particle exchange and single-channel scattering reads 
\be
f^{-1}_{[m, n]\left(k^2\right)}+ik=\frac{\sum_{i=0}^n \tilde{d}_i k^{2 i}-\tilde{g} d^{\mathrm{R}}(k^2)}{1+\sum_{j=1}^m \tilde{n}_j k^{2 j}+\tilde{g}\left[L(k^2)-L_0\right]},
\label{eq:dgw}
\ee
where $\tilde d_i$ and $\tilde n_j$ are parameters, $\tilde g$ is related to the pionic coupling constant and controls the strength of the lhc contribution, $L(k^2)$ and $d^R(k^2)$ are logarithmic functions (see Refs.~\cite{Du:2024gzw,Wang:2026ups} for explicit expressions), and $L_0 \equiv L(k^2=0)$. The parameters of the function in Eq.~\eqref{eq:dgw} are fitted to the real part of the LSE solution at several selected points, shown as open circles in Fig.~\ref{fig:dgw-ere}. The resulting curves for the real and imaginary parts of $f^{-1}(k)+ik$ obtained with the $[0,1]$ and $[1,2]$ approximants are compared in Fig.~\ref{fig:dgw-ere} with the exact result. One sees that the $[m,n]=[1,2]$ approximant describes the pure $S$-wave LSE amplitude fairly well up to the region where the lhc due to one-$\eta$ exchange starts to become relevant, whose correction yet remains small.

\begin{table}[t]
 \caption{Parameters of the low-energy expansion near the threshold extracted from Eq.~\eqref{eq:dgw}.}
 \label{tab:argb}
 \begin{ruledtabular}
 \begin{tabular}{cccc}
 &$a$ (fm) & $r$ (fm) & $g_b$\\ 
 \hline
 $f_{[0,1]}$ & $-1.04$ & $-1.8$ & 0.49 \\
 $f_{[1,2]}$ & $-1.08$ & $-2.0$& 0.56
 \end{tabular}
 \end{ruledtabular}
\end{table}

The scattering length and effective range extracted from the $[0,1]$ and $[1,2]$ approximants are listed in Table~\ref{tab:argb}. The parameters obtained with the $[1,2]$ approximant agree well with those from the LSE solution given in Eq.~\eqref{eq:aandr-tensoroff}. The table also lists the values of the pion coupling $g_b$ inferred from the strength of the lhc contribution for both cases. Again, the $[1,2]$ approximant reproduces well the value of $g_b$ quoted in Eq.~\eqref{gb} and used in the LSE. We stress that these results are presented here solely as a diagnostic of the OPE lhc structure.

\section{Component probabilities}
\label{app:probabilities}

In this appendix, we derive explicit expressions for the component probabilities defined in Sect.~\ref{sec:probabilities}. As before, we consider a bound state with the binding energy $E_B$ and binding momentum $\kappa_B=\sqrt{2\mu E_B}$, which corresponds to a pole on the complex momentum plane at $k=i\kappa_B$.

To determine the common normalisation of the components in Eq.~\eqref{eq:bound-state-decomposition}, we define the effective couplings $c_i$ $(i=1,..,N+1)$ via the
residues of the pole term in the Laurent expansion of the reduced amplitude $\hat\tau(k)$ defined in Eq.~\eqref{eq:tau} around $E=-E_B$, in the separable basis introduced below Eq.~\eqref{VQMsep},
\be
\hat\tau(k)
=
\frac{\bm{c}\bm{c}^T}{E+E_B}
+\hat{\tau}_R+\cdots.
\label{eq:residue}
\ee
Here, $\bm{c}$ denotes the vector of effective couplings, $\hat{\tau}_R$ is a constant matrix, and the ellipsis denotes terms that vanish as $E\to -E_B$. The inverse matrix $\hat{\tau}^{-1}(k)$ admits a Taylor expansion around $E=-E_B$,
\be
\hat{\tau}^{-1}(k)
=
\tauEB
+(E+E_B)\dtauinvEB
+\cdots,
\label{eq:tauinv}
\ee
where
\be
\tauEB \equiv\lim_{E\to -E_B}\hat{\tau}^{-1}(k),
\ee
$\partial_E$ denotes differentiation with respect to the energy $E$, and the ellipsis denotes higher-order terms. Then, from Eqs.~\eqref{eq:residue} and \eqref{eq:tauinv}, we find
\be
\begin{split}
\mathbbm{1}=
\hat{\tau}^{-1}\hat{\tau}
={}&
\frac{\tauEB\bm{c}\bm{c}^T}{E+E_B}
\\
&+\tauEB\hat{\tau}_R
+\dtauinvEB\bm{c}\bm{c}^T+\cdots\,.
\end{split}
\label{tausprod}
\ee
Since the leftmost side of this equation is a constant,
all terms on the rightmost side, except the constant ones, must vanish.
The condition that the pole term vanishes implies that $\bm{c}$ is an eigenvector of the matrix $\tauEB$ with a zero eigenvalue,
\be
\tauEB\bm{c}=0.
\label{eq:homogeneous-pole}
\ee
Consequently, $\bm{c}^T\tauEB=[\tauEB\bm{c}]^T=0$, where we have used the fact that the matrix $\hat{\tau}$, and thus $\tauEB$, is symmetric due to time-reversal invariance. Multiplying Eq.~\eqref{tausprod} by $\bm{c}^T$ from the left, we therefore obtain
\be
\bm{c}^T \dtauinvEB \bm{c}=1.
\ee
Since $\hat\tau^{-1}=\hat v^{-1}+\hat J$, this relation takes the form\footnote{This relation may be regarded as a generalisation of the ones between the couplings and wave functions in coupled-channel systems derived in Refs.~\cite{Gamermann:2009uq, Garcia-Recio:2015jsa}.}
\be
\sum_{i,j=1}^{N+1}c_i
\left[\partial_E\left(\hat v^{-1}+\hat J\right)\right]_{ij}
c_j
\Bigg|_{E=-E_B}
=1.
\label{eq:residue-normalisation}
\ee
Thus, the couplings in Eq.~\eqref{eq:residue} encode the normalisation of the complete bound-state wave function, which we decompose according to
\be
\begin{split}
\bm{c}^T \Bigl[\partial_E \hat{v}^{-1}\Bigr]_{E=-E_B}\bm{c}&=
\sum_{n=1}^N \frac{c_n^2}{g^2}={\cal P}_{\bar{b}b},\\
\bm{c}^T \Bigl[\partial_E \hat{J}\Bigr]_{E=-E_B} \bm{c} &=\sum_{
\alpha}\mathcal{P}_\alpha+\mathcal{P}_{3\text{-body}},
\end{split}
\label{eq:ceqs}
\ee
where the sum in $\alpha$ runs over the 2-body states enumerated in Eq.~\eqref{BBchannels}, the explicit expressions for $\hat v$ in Eqs.~\eqref{eq:vhat} and \eqref{vs} were used in the first equation above, and the condition $\partial_E C=0$ was taken into account, explicitly removing the $(N+1)$-th term in the sum in $n$.
The decomposition of the second relation in Eq.~\eqref{eq:ceqs} into its different components is given below.

The relation between the pole residue of the scattering amplitude and the two-meson wave function follows from the bound-state pole of the full Green's operator projected onto the two-meson channel,
\begin{align}
&\Psi_{\alpha'}(p')\Psi_\alpha^*(p)\notag\\
={}&
\lim_{E\to-E_B}(E+E_B)
\Big\langle p',\alpha'\left|
\bigl(E-\tilH+i\epsilon\bigr)^{-1}
\right|p,\alpha\Big\rangle
\notag\\
={}&
\lim_{E\to-E_B}(E+E_B)
G_{\alpha'}(k,p')T_{\alpha'\alpha}(p',p)G_\alpha(k,p),
\end{align}
where the LSE for the Green's function was used.
Substituting Eq.~\eqref{eq:residue} into Eq.~\eqref{Tsol}, we obtain for the wave function defined in Eq.~\eqref{psiL},
\be
\Psi_\alpha(p)=-G_\alpha(i\kappa_B,p)
\sum_{i=1}^{N+1}X^{\rm out}_{i,\alpha}(p)c_i,
\label{eq:wf}
\ee
where the dressed vertices on the right-hand side are evaluated at the pole.

Next, we express the probabilities in Eq.~\eqref{eq:norm_generalised} in terms of the two-meson wave functions.
We start with the first relation in Eq.~\eqref{eq:ceqs} and note that,
since the compact bottomonium states couple directly only to the $S$-wave $B\bar{B}^*$ pair at LO, the coefficient $\lambda_n$ for each compact component in the bound-state wave function is given by Eq.~\eqref{eq:lamn}. For $n=1,..,N$, from Eq.~\eqref{eq:ceqs} we have
\be
\lambda_n=\frac{c_n}{g}.
\ee
Evaluating Eq.~\eqref{eq:lamn} at the same pole and using the Hermiticity of $\hat g$, together with the relations $f_n(p)=\psi_n(p)/\sqrt{4\pi}$ in Eq.~\eqref{eq:fn} and $\phi(p)=\sqrt{4\pi}\,\Psi_{\alpha=1}(p)\equiv\sqrt{4\pi}\,\Psi_S(p)$, gives
\be
\lambda_n =-\frac{g}{E_B+E_n}
\int_0^\Lambda\dqr\,f_n^*(q)\Psi_S(q).
\label{eq:compact-amplitudes}
\ee
Then, Eq.~\eqref{Pbb} becomes
\be
\begin{split}
{\cal P}_{\bar{b}b}
&=\sum_{n=1}^N|\lambda_n|^2\\
&=\sum_{n=1}^N\frac{g^2}{(E_B+E_n)^2}
\left|\int_0^\Lambda\dqr\,f_n^*(q)\Psi_S(q)\right|^2\\
&=-\left.
\braket{\Psi_S|\partial_E V_S^{(\bar{b} b)}(E)|\Psi_S}
\right|_{E=-E_B}.
\end{split}
\label{eq:P-bbbar}
\ee

To derive the formula for $\mathcal{P}_{3\text{-body}}$ in Eq.~\eqref{eq:ceqs}, we project Eq.~\eqref{eq:schroedinger-full} onto $\ket{B^{(*)}\bar{B}^{(*)}\phi[\beta;b]}$,
\be
(E-E_\beta(b))\varphi_\beta(E;b)={\mathcal I}_\beta(E;b),
\label{projected}
\ee
where we used that
\be
\tilH=\tilH_0+\tilH_{\mathrm{int}},
\ee 
with $\tilH_0$ for the free Hamiltonian in the extended
Hilbert space such that
\be
\tilH_0\ket{B^{(*)}\bar{B}^{(*)}\phi[\beta;b]}
=E_\beta(b) \ket{B^{(*)}\bar{B}^{(*)}\phi[\beta;b]}.
\ee
As before, $\beta$'s count the 3-body channels enumerated in Eq.~\eqref{BBphichannels} and $b$'s stand for their discrete quantum numbers and momenta together. The matrix element on the right-hand side in Eq.~\eqref{projected} reads
\be
\begin{split}
{\mathcal I}_\beta(E;b)
&=\braket{B^{(*)}\bar{B}^{(*)}\phi[\beta;b]|\tilH_{\mathrm{int}}|\Psi(E)}\\
&=\sum_{\alpha}\int_0^\Lambda\dqr\,
W_{\beta \alpha}(q;b)\Psi_\alpha(E;q),
\end{split}
\ee
where (\emph{cf} Eq.~\eqref{psiL})
\be
\Psi_\alpha(E;q)=\braket{q,\alpha|\Psi(E)},
\label{psiLE}
\ee
and we introduced the $B^{(*)}\bar{B}^{(*)}\phi$ vertices embedded
in the two- and 3-body space,
\be
W_{\beta \alpha}(q;b)=\braket{\beta;b|\tilH_{\mathrm{int}}|q,\alpha}.
\ee

Thus, the 3-body probability is given by
\be
\begin{split}
{\cal P}_{\text{3-body}}
&=\sum_\beta\sum_b|\varphi_\beta(-E_B;b)|^2\\
&=\sum_\beta\sum_b\frac{|{\mathcal I}_\beta(-E_B;b)|^2}
{(E_B+E_\beta(b))^2}\\
&=-\sum_{\alpha,\alpha'}
\braket{\Psi_{\alpha'}|\partial_E
V^{\rm OME}_{\alpha'\alpha}(E)|\Psi_\alpha} \Bigg|_{E =-E_B},
\label{eq:P-3body}
\end{split}
\ee
where the spectral form of the OME kernel reads
\be
V^{\rm OME}_{\alpha'\alpha}(p',p;E)
=\sum_\beta\sum_b
\frac{W_{\beta \alpha'}^*(p';b)W_{\beta \alpha}(p;b)}
{E-E_\beta(b)+i\epsilon},
\label{eq:topt-spectral}
\ee
in agreement with the TOPT potential provided in Eq.~\eqref{VOME}.
The angular dependence of $E_\beta(b)$ is implicitly included in the sum over $b$, and the energy denominator remains inside the partial-wave projection in Eq.~\eqref{VPWA}.

Finally, we derive the formula for the contributions $\mathcal{P}_\alpha$ in Eq.~\eqref{eq:ceqs}.
We start with the definition of the matrix $\hat{J}$, with the matrix elements in Eq.~\eqref{Mdef}, and employ convenient matrix notation in which all the involved quantities become matrices with each row and column referring to the channel $i$ and the momentum-partial-wave space $(p, \alpha)$, respectively, and the momentum integrals become matrix products.
Then, the matrix $\hat{J}$ can be written as
\be
\hat{J}=\hat{f}\,\hat{G}\,\hat{f}^T - \hat{f}\,\hat{G}\,\hat{T}^{\mathrm{OME}}\,\hat{G}\,\hat{f}^T,
\ee
so that
\be
\begin{split}
\partial_E \hat{J} &=\hat{f}\left(1-\hat G\hat T^{\mathrm{OME}}\right) \hat{G}^2
\left(1-\hat{T}^{\mathrm{OME}}\hat{G}\right)\hat{f}^T \\
&\quad - \hat{f}\,\hat{G}\left(\partial_E \hat{T}^{\mathrm{OME}}\right)\hat{G}\,\hat{f}^T \\
&\quad
- \hat{f}\,\hat{G}\,\hat{T}^{\mathrm{OME}}\,\hat{G}^2\,
\hat{T}^{\mathrm{OME}}\,\hat{G}\,\hat{f}^T,
\end{split}
\label{Jprime}
\ee
where the relations $\partial_E\hat{f}=0$ and $\partial_E \hat{G}=\hat{G}^2$
were used. 

In the adopted matrix notation, the solution of the LSE, Eq.~\eqref{LSEOME}, takes the form
\be
\left[\hat{T}^{\mathrm{OME}}\right]^{-1}=
\left[\hat{V}^{\mathrm{OME}}\right]^{-1}+\hat{G}.
\ee
Then, on the one hand, we straightforwardly have
\be
\left(1-\hat{T}^{\mathrm{OME}}\hat{G}\right)\hat{V}^{\mathrm{OME}}=\hat{T}^{\mathrm{OME}},
\label{rel1}
\ee
and, on the other hand, we can write
\begin{align}
1&=1-\hat{T}^{\mathrm{OME}}\left(
\left[\hat{V}^{\mathrm{OME}}\right]^{-1}+\hat{G}-
\left[\hat{T}^{\mathrm{OME}}\right]^{-1}\right)\hat{V}^{\mathrm{OME}}\,\hat{G}\notag\\
&=\left(1-\hat{T}^{\mathrm{OME}}\,\hat{G}\right)\left(1+\hat{V}^{\mathrm{OME}}\,\hat{G}\right),
\end{align}
so that
\be
\left(1+\hat{V}^{\mathrm{OME}}\,\hat{G}\right)^{-1}=
1-\hat{T}^{\mathrm{OME}}\,\hat{G}.
\label{rel2}
\ee
Therefore, by applying the derivative $\partial_E$ to both sides of the LSE in Eq.~\eqref{LSEOME} and employing the relations in Eqs.~\eqref{rel1} and \eqref{rel2}, we arrive at
\begin{align}
\partial_E \hat{T}^{\mathrm{OME}} ={}& \left(1-\hat{T}^{\mathrm{OME}}\,\hat{G}\right)
\partial_E \hat{V}^{\mathrm{OME}} \left(1-\hat{G}\,\hat{T}^{\mathrm{OME}}\right)\notag\\
& -\hat{T}^{\mathrm{OME}}\,\hat{G}^2\,\hat{T}^{\mathrm{OME}}.\label{Tprime}
\end{align}

Using the relation in Eq.~\eqref{Tprime} in Eq.~\eqref{Jprime}
yields
\be
\partial_E \hat{J}=\hat{X}^{\text {in}}\,\hat{G}\Bigl[1
-\partial_E \hat{V}^{\mathrm{OME}}\Bigr]\hat{G}\bigl(\hat{X}^{\text {out}}\bigr)^T,
\label{Jprime2}
\ee
where the definitions of the dressed incoming and outgoing vertices in Eq.~\eqref{Xs} were also used, which read in matrix form
\be
\begin{split}
\hat{X}^{\text {in}} &=\hat{f}\left(1-\hat{G}\,\hat{T}^{\text {OME}}\right) , \\[1mm]
\bigl(\hat{X}^{\text {out}}\bigr)^T&=\left(1-\hat{T}^{\text {OME}}\,\hat{G}\right)\hat{f}^T.
\label{eq:Xmatrix}
\end{split}
\ee

Then, it is easy to see that, upon substitution of the relation in Eq.~\eqref{Jprime2} into Eq.~\eqref{eq:ceqs} and noting that, according to Eq.~\eqref{eq:wf},
\be
\begin{split}
\left[ \hat G\, \bigl(\hat X^\text{out} \bigr)^T \bm{c} \right]_{(p,\alpha)}\Big|_{E=-E_B} &=-\Psi_\alpha(p),\\
\left[ \bm{c}^T \hat X^\text{in} \,\hat G \right]_{(p,\alpha)}\Big|_{E=-E_B} &=-\Psi_\alpha^*(p),
\end{split}
\ee
the term with $\partial_E \hat{V}^{\mathrm{OME}}$ in Eq.~\eqref{Jprime2} reproduces the probability ${\cal P}_{\text{3-body}}$ in Eq.~\eqref{eq:P-3body}. Therefore, as per the first relation in Eq.~\eqref{eq:ceqs}, what remains is naturally identified with the probability
\be
\mathcal{P}_\alpha=\braket{\Psi_\alpha|\Psi_\alpha}=\int_0^\Lambda\dqr\,|\Psi_\alpha(q)|^2,
\ee
in agreement with Eq.~\eqref{eq:P-2body}.

\section{Spurious deeply bound state}
\label{app:deep}

\begin{table}[t!] 
\centering 
\caption{The spurious deep bound state in the calculation with $N=4$ for several cutoffs: mass, binding energy relative to the $B\bar{B}^*$ threshold and the probabilities in Eq.~\eqref{eq:norm_generalised} (in per cent). $\Lambda$ and $M$ are given in GeV, $C$ in GeV$^{-2}$ and $E_B$ in MeV. The parameters of Table~\ref{tab:cutoff} for each $\Lambda$ are used.}
\label{tab:deep}
\begin{ruledtabular}
\small
\setlength{\tabcolsep}{1.3pt}
\begin{tabular}{cc|cc|cccc}
$\Lambda$ & $C$ & $M$ & $E_B$ & ${\cal P}_{\bar{b}b}$ & ${\cal P}_S$ & ${\cal P}_D$ & ${\cal P}_{\text{3-body}}$ \\
\hline
3 & 10.55 & 8.958 & 1646 & 0.1 & 24 & 36 & 40 \\
4 & $-16.27$ & 9.230 & 1375 & 0.2 & 38 & 28 & 34 \\
5 & $-0.67$ & 9.399 & 1205 & 0.3 & 38 & 25 & 36
\end{tabular}
\end{ruledtabular}
\end{table}

As explained in Sect.~\ref{sec:Xb}, including the tensor interaction causes the scattering amplitude to acquire an additional, spurious, pole, which arises as an artifact of the renormalised singular tensor force. In this appendix, we discuss how this pole can be eliminated from the theory. First, we note that the position of this pole exhibits strong scheme dependence; see Table~\ref{tab:deep} for the position of this spurious, deeply bound pole obtained for several values of the cutoff $\Lambda$. 

The spurious pole can be removed from the spectrum using the projection method of Refs.~\cite{Kukulin:1978he,Epelbaum:2002ji,Nogga:2005hy,Yang:2020pgi}. The procedure consists of two steps.
In the first step, the LSE in Eq.~\eqref{LSEfull2} is solved with the potential in Eq.~\eqref{Vcoupled}. The position of the spurious pole is located, and the two-meson component $\Psi^{\rm (spur)}_\alpha(p)$ ($\alpha=1,2,3$) of its wave function at the pole [normalised to unity using the measure in Eq.~\eqref{eq:P-2body}] is extracted. 
In the second step, the interaction potential is modified according to
\be
V_{\alpha'\alpha}(p',p)\to V_{\alpha'\alpha}(p',p)+\varepsilon\,\Psi^{\rm (spur)}_{\alpha'}(p')\,\Psi^{\rm (spur)}_\alpha(p),
\label{Vspur}
\ee
where the parameter $\varepsilon>0$ is chosen to be large compared to the typical binding scale of the problem, after which the LSE is solved again. The additional energy-independent separable term in the modified potential in Eq.~\eqref{Vspur} shifts the spurious pole far above threshold, leaving other states unaffected provided they do not overlap with $\Psi^{\rm (spur)}$. 
We have verified that within our adopted precision, the physical results remain unchanged when varying $\varepsilon$ from $5$ to $50$~GeV. 

We note that, in the baseline scenario of Table~\ref{tab:results}, the spurious pole lies well below the $\chi_{b1}(1P)$ state (see panel (a) of Fig.~\ref{fig:levels}), and thus has a negligible influence on physical observables. The effect of projecting it out explicitly is also small,
\be
\begin{split}
&\delta M[\chi_{b1}(1P)]=0.9~\mbox{MeV},\\
&\delta M[\chi_{b1}(2P)]\simeq 
\delta M[\chi_{b1}(3P)]<0.33~\mbox{MeV},\\
&\delta {\cal P}_{\bar{b}b}[\chi_{b1}(nP)]<0.2\%,\quad n=1,2,3,\\
&\delta a\simeq \delta r < 4\times 10^{-3}~\mbox{fm},\\
&\delta[\delta_S]\simeq\delta[\delta_D]<0.03^\circ,
\end{split}
\ee
which can also be seen by comparing panels (a) and (b) of Fig.~\ref{fig:levels}.

In the alternative scenario, where the contact term $C$ is tuned to bind the $\Xb$ with a fixed $E_B=1$~MeV, the spurious pole is located at $10.13$~GeV, that is, between $\chi_{b1}(1P)$ and $\chi_{b1}(2P)$ (see panel (a) of Fig.~\ref{fig:levels}). Consequently, it is no longer decoupled and must be projected out. The corresponding results are presented in Table~\ref{tab:comp}.

\bibliography{refs}

@article{Chacko:2024cax,
    author = "Chacko, J. T. and Baru, V. and Hanhart, C. and Krug, S. L.",
    title = "{Two-pion exchange for coupled-channel scattering of two heavy mesons}",
    eprint = "2411.13303",
    archivePrefix = "arXiv",
    primaryClass = "hep-ph",
    doi = "10.1103/PhysRevD.111.034042",
    journal = "Phys. Rev. D",
    volume = "111",
    number = "3",
    pages = "034042",
    year = "2025"
}

@article{Blatt:1952zza,
    author = "Blatt, John M. and Biedenharn, L. C.",
    title = "{Neutron-Proton Scattering with Spin-Orbit Coupling. 1. General Expressions}",
    doi = "10.1103/PhysRev.86.399",
    journal = "Phys. Rev.",
    volume = "86",
    pages = "399--404",
    year = "1952"
}

@article{Eichten:1978tg,
    author = "Eichten, E. and Gottfried, K. and Kinoshita, T. and Lane, K. D. and Yan, Tung-Mow",
    title = "{Charmonium: The Model}",
    reportNumber = "CLNS-375",
    doi = "10.1103/PhysRevD.17.3090",
    journal = "Phys. Rev. D",
    volume = "17",
    pages = "3090",
    year = "1978",
    note = "[Erratum: Phys.Rev.D 21, 313 (1980)]"
}

@article{Epelbaum:2008ga,
    author = "Epelbaum, Evgeny and Hammer, Hans-Werner and Meissner, Ulf-G.",
    title = "{Modern Theory of Nuclear Forces}",
    eprint = "0811.1338",
    archivePrefix = "arXiv",
    primaryClass = "nucl-th",
    reportNumber = "HISKP-TH-08-18, FZJ-IKP-TH-2008-20",
    doi = "10.1103/RevModPhys.81.1773",
    journal = "Rev. Mod. Phys.",
    volume = "81",
    pages = "1773--1825",
    year = "2009"
}

@article{Du:2019pij,
    author = "Du, Meng-Lin and Baru, Vadim and Guo, Feng-Kun and Hanhart, Christoph and Mei{\ss}ner, Ulf-G and Oller, Jos{\'e} A. and Wang, Qian",
    title = "{Interpretation of the LHCb $P_c$ States as Hadronic Molecules and Hints of a Narrow $P_c(4380)$}",
    eprint = "1910.11846",
    archivePrefix = "arXiv",
    primaryClass = "hep-ph",
    doi = "10.1103/PhysRevLett.124.072001",
    journal = "Phys. Rev. Lett.",
    volume = "124",
    number = "7",
    pages = "072001",
    year = "2020"
}

@article{Du:2021zzh,
    author = "Du, Meng-Lin and Baru, Vadim and Dong, Xiang-Kun and Filin, Arseniy and Guo, Feng-Kun and Hanhart, Christoph and Nefediev, Alexey and Nieves, Juan and Wang, Qian",
    title = "{Coupled-channel approach to $T_{cc}^+$ including three-body effects}",
    eprint = "2110.13765",
    archivePrefix = "arXiv",
    primaryClass = "hep-ph",
    doi = "10.1103/PhysRevD.105.014024",
    journal = "Phys. Rev. D",
    volume = "105",
    number = "1",
    pages = "014024",
    year = "2022"
}

@article{Baru:2011rs,
    author = "Baru, V. and Filin, A. A. and Hanhart, C. and Kalashnikova, Yu. S. and Kudryavtsev, A. E. and Nefediev, A. V.",
    title = "{Three-body $D\bar{D}\pi$ dynamics for the X(3872)}",
    eprint = "1108.5644",
    archivePrefix = "arXiv",
    primaryClass = "hep-ph",
    doi = "10.1103/PhysRevD.84.074029",
    journal = "Phys. Rev. D",
    volume = "84",
    pages = "074029",
    year = "2011"
}

@book{Ericson:1988gk,
    author = "Ericson, Torleif Erik Oskar and Weise, W.",
    title = "{Pions and Nuclei}",
    isbn = "978-0-19-852008-5",
    publisher = "Clarendon Press",
    address = "Oxford, UK",
    year = "1988"
}

@article{Braaten:2014qka,
    author = "Braaten, Eric and Langmack, Christian and Smith, D. Hudson",
    title = "{Born-Oppenheimer Approximation for the XYZ Mesons}",
    eprint = "1402.0438",
    archivePrefix = "arXiv",
    primaryClass = "hep-ph",
    doi = "10.1103/PhysRevD.90.014044",
    journal = "Phys. Rev. D",
    volume = "90",
    number = "1",
    pages = "014044",
    year = "2014"
}

@article{Berwein:2024ztx,
    author = "Berwein, Matthias and Brambilla, Nora and Mohapatra, Abhishek and Vairo, Antonio",
    title = "{Hybrids, tetraquarks, pentaquarks, doubly heavy baryons, and quarkonia in Born-Oppenheimer effective theory}",
    eprint = "2408.04719",
    archivePrefix = "arXiv",
    primaryClass = "hep-ph",
    reportNumber = "TUM-EFT 185/23",
    doi = "10.1103/PhysRevD.110.094040",
    journal = "Phys. Rev. D",
    volume = "110",
    number = "9",
    pages = "094040",
    year = "2024"
}

@article{Berwein:2015vca,
    author = "Berwein, Matthias and Brambilla, Nora and Tarr{\'u}s Castell{\`a}, Jaume and Vairo, Antonio",
    title = "{Quarkonium Hybrids with Nonrelativistic Effective Field Theories}",
    eprint = "1510.04299",
    archivePrefix = "arXiv",
    primaryClass = "hep-ph",
    reportNumber = "TUM-EFT-45-14",
    doi = "10.1103/PhysRevD.92.114019",
    journal = "Phys. Rev. D",
    volume = "92",
    number = "11",
    pages = "114019",
    year = "2015"
}

@article{Brambilla:2017uyf,
    author = "Brambilla, Nora and Krein, Gast{\~a}o and Tarr{\'u}s Castell{\`a}, Jaume and Vairo, Antonio",
    title = "{Born-Oppenheimer approximation in an effective field theory language}",
    eprint = "1707.09647",
    archivePrefix = "arXiv",
    primaryClass = "hep-ph",
    reportNumber = "TUM-EFT-69-15",
    doi = "10.1103/PhysRevD.97.016016",
    journal = "Phys. Rev. D",
    volume = "97",
    number = "1",
    pages = "016016",
    year = "2018"
}

@article{Oncala:2017hop,
    author = "Oncala, Rub{\'e}n and Soto, Joan",
    title = "{Heavy Quarkonium Hybrids: Spectrum, Decay and Mixing}",
    eprint = "1702.03900",
    archivePrefix = "arXiv",
    primaryClass = "hep-ph",
    reportNumber = "ICCUB-17-004, NIKHF-2017-005",
    doi = "10.1103/PhysRevD.96.014004",
    journal = "Phys. Rev. D",
    volume = "96",
    number = "1",
    pages = "014004",
    year = "2017"
}

@article{Soto:2020xpm,
    author = "Soto, Joan and Tarr{\'u}s Castell{\`a}, Jaume",
    title = "{Nonrelativistic effective field theory for heavy exotic hadrons}",
    eprint = "2005.00552",
    archivePrefix = "arXiv",
    primaryClass = "hep-ph",
    doi = "10.1103/PhysRevD.102.014012",
    journal = "Phys. Rev. D",
    volume = "102",
    number = "1",
    pages = "014012",
    year = "2020",
    note = "[Erratum: Phys.Rev.D 110, 099901 (2024)]"
}

@misc{Brambilla:2026jqz,
    author = "Brambilla, Nora and Bruschini, Roberto and Mohapatra, Abhishek and Peng, Fang-Zheng and Scirpa, Tommaso",
    title = "{Quarkoniumlike states above open-flavor thresholds in Born-Oppenheimer EFT}",
    eprint = "2608.04105",
    archivePrefix = "arXiv",
    primaryClass = "hep-ph",
    month = "8",
    year = "2026"
}

@article{Braaten:2024tbm,
    author = "Braaten, Eric and Bruschini, Roberto",
    title = "{Exotic hidden-heavy hadrons and where to find them}",
    eprint = "2409.08002",
    archivePrefix = "arXiv",
    primaryClass = "hep-ph",
    doi = "10.1016/j.physletb.2025.139386",
    journal = "Phys. Lett. B",
    volume = "863",
    pages = "139386",
    year = "2025"
}

@article{Lensky:2005hb,
    author = "Lensky, V. and Baru, V. and Haidenbauer, J. and Hanhart, C. and Kudryavtsev, Alexander Evgenyevich and Meissner, U. -G.",
    title = "{Precision calculation of $\gamma d \to \pi^+ nn$ within chiral perturbation theory}",
    eprint = "nucl-th/0505039",
    archivePrefix = "arXiv",
    reportNumber = "FZJ-IKP-TH-2005-08, HISKP-TH-05-06",
    doi = "10.1140/epja/i2005-10154-7",
    journal = "Eur. Phys. J. A",
    volume = "26",
    pages = "107--123",
    year = "2005"
}

@article{Brambilla:1999xf,
    author = "Brambilla, Nora and Pineda, Antonio and Soto, Joan and Vairo, Antonio",
    title = "{Potential NRQCD: An Effective theory for heavy quarkonium}",
    eprint = "hep-ph/9907240",
    archivePrefix = "arXiv",
    reportNumber = "CERN-TH-99-199, HEPHY-PUB-716-99, UB-ECM-PF-99-06, UWTHPH-1999-34, UB-ECM-PF-99-13",
    doi = "10.1016/S0550-3213(99)00693-8",
    journal = "Nucl. Phys. B",
    volume = "566",
    pages = "275",
    year = "2000"
}

@article{Brambilla:2000gk,
    author = "Brambilla, Nora and Pineda, Antonio and Soto, Joan and Vairo, Antonio",
    title = "{The QCD potential at $O(1/m)$}",
    eprint = "hep-ph/0002250",
    archivePrefix = "arXiv",
    reportNumber = "CERN-TH-2000-053, UB-ECM-PF-00-03, UWTHPH-1999-47",
    doi = "10.1103/PhysRevD.63.014023",
    journal = "Phys. Rev. D",
    volume = "63",
    pages = "014023",
    year = "2001"
}

@article{Pineda:2000sz,
    author = "Pineda, Antonio and Vairo, Antonio",
    title = "{The QCD potential at $O(1/m^2)$: Complete spin dependent and spin independent result}",
    eprint = "hep-ph/0009145",
    archivePrefix = "arXiv",
    reportNumber = "CERN-TH-2000-197, HD-THEP-00-31",
    doi = "10.1103/PhysRevD.64.039902",
    journal = "Phys. Rev. D",
    volume = "63",
    pages = "054007",
    year = "2001",
    note = "[Erratum: Phys.Rev.D 64, 039902 (2001)]"
}

@article{Brambilla:2004jw,
    author = "Brambilla, Nora and Pineda, Antonio and Soto, Joan and Vairo, Antonio",
    title = "{Effective Field Theories for Heavy Quarkonium}",
    eprint = "hep-ph/0410047",
    archivePrefix = "arXiv",
    reportNumber = "IFUM-805-FT, UB-ECM-PF-04-24",
    doi = "10.1103/RevModPhys.77.1423",
    journal = "Rev. Mod. Phys.",
    volume = "77",
    pages = "1423",
    year = "2005"
}

@article{Epelbaum:2014sza,
    author = "Epelbaum, E. and Krebs, H. and Mei{\ss}ner, U. G.",
    title = "{Precision nucleon-nucleon potential at fifth order in the chiral expansion}",
    eprint = "1412.4623",
    archivePrefix = "arXiv",
    primaryClass = "nucl-th",
    doi = "10.1103/PhysRevLett.115.122301",
    journal = "Phys. Rev. Lett.",
    volume = "115",
    number = "12",
    pages = "122301",
    year = "2015"
}

@article{Epelbaum:2014efa,
    author = "Epelbaum, E. and Krebs, H. and Mei{\ss}ner, U. G.",
    title = "{Improved chiral nucleon-nucleon potential up to next-to-next-to-next-to-leading order}",
    eprint = "1412.0142",
    archivePrefix = "arXiv",
    primaryClass = "nucl-th",
    doi = "10.1140/epja/i2015-15053-8",
    journal = "Eur. Phys. J. A",
    volume = "51",
    number = "5",
    pages = "53",
    year = "2015"
}

@article{Furnstahl:2014xsa,
    author = "Furnstahl, R. J. and Phillips, D. R. and Wesolowski, S.",
    title = "{A recipe for EFT uncertainty quantification in nuclear physics}",
    eprint = "1407.0657",
    archivePrefix = "arXiv",
    primaryClass = "nucl-th",
    doi = "10.1088/0954-3899/42/3/034028",
    journal = "J. Phys. G",
    volume = "42",
    number = "3",
    pages = "034028",
    year = "2015"
}

@article{Baru:2015nea,
    author = "Baru, V. and Epelbaum, E. and Filin, A. A. and Guo, F. -K. and Hammer, H. -W. and Hanhart, C. and Mei{\ss}ner, U. -G. and Nefediev, A. V.",
    title = "{Remarks on study of X(3872) from effective field theory with pion-exchange interaction}",
    eprint = "1501.02924",
    archivePrefix = "arXiv",
    primaryClass = "hep-ph",
    doi = "10.1103/PhysRevD.91.034002",
    journal = "Phys. Rev. D",
    volume = "91",
    number = "3",
    pages = "034002",
    year = "2015"
}

@article{Belle:2003nnu,
    author = "Choi, S. K. and others",
    collaboration = "Belle",
    title = "{Observation of a narrow charmonium-like state in exclusive $B^\pm \to K^\pm \pi^+ \pi^- J/\psi$ decays}",
    eprint = "hep-ex/0309032",
    archivePrefix = "arXiv",
    doi = "10.1103/PhysRevLett.91.262001",
    journal = "Phys. Rev. Lett.",
    volume = "91",
    pages = "262001",
    year = "2003"
}

@article{Brambilla:2019esw,
    author = "Brambilla, Nora and Eidelman, Simon and Hanhart, Christoph and Nefediev, Alexey and Shen, Cheng-Ping and Thomas, Christopher E. and Vairo, Antonio and Yuan, Chang-Zheng",
    title = "{The $XYZ$ states: experimental and theoretical status and perspectives}",
    eprint = "1907.07583",
    archivePrefix = "arXiv",
    primaryClass = "hep-ex",
    reportNumber = "TUM-EFT 125/19",
    doi = "10.1016/j.physrep.2020.05.001",
    journal = "Phys. Rept.",
    volume = "873",
    pages = "1--154",
    year = "2020"
}

@article{Dai:2026fkg,
    author = "Dai, Xinchen and Jia, Sen and Nefediev, Alexey and Nieves, Juan and Shen, Chengping and Zhang, Liming",
    title = "{Exotic hadrons associated with $b$-quark}",
    eprint = "2603.09315",
    archivePrefix = "arXiv",
    primaryClass = "hep-ph",
    doi = "10.1016/j.physrep.2026.06.002",
    journal = "Phys. Rept.",
    volume = "1191",
    pages = "1--62",
    year = "2026"
}

@article{Ali:2017jda,
      author         = "Ali, Ahmed and Lange, Jens S{\"o}ren and Stone, Sheldon",
      title          = "{Exotics: Heavy pentaquarks and tetraquarks}",
      journal        = "Prog. Part. Nucl. Phys.",
      volume         = "97",
      year           = "2017",
      pages          = "123-198",
      doi            = "10.1016/j.ppnp.2017.08.003",
      eprint         = "1706.00610",
      archivePrefix  = "arXiv",
      primaryClass   = "hep-ph",
      reportNumber   = "DESY-17-071",
      SLACcitation   = "%%CITATION = ARXIV:1706.00610;%%"
}

@article{Olsen:2017bmm,
      author         = "Olsen, Stephen Lars and Skwarnicki, Tomasz and Zieminska,
                        Daria",
      title          = "{Nonstandard heavy mesons and baryons: Experimental
                        evidence}",
      journal        = "Rev. Mod. Phys.",
      volume         = "90",
      year           = "2018",
      number         = "1",
      pages          = "015003",
      doi            = "10.1103/RevModPhys.90.015003",
      eprint         = "1708.04012",
      archivePrefix  = "arXiv",
      primaryClass   = "hep-ph",
      SLACcitation   = "%%CITATION = ARXIV:1708.04012;%%"
}

@article{Albuquerque:2018jkn,
    author = "Albuquerque, Raphael M. and Dias, Jorgivan M. and Khemchandani, K. P. and Mart\'\i{}nez Torres, A. and Navarra, Fernando S. and Nielsen, Marina and Zanetti, Carina M.",
    title = "{QCD sum rules approach to the $X,~Y$ and $Z$ states}",
    eprint = "1812.08207",
    archivePrefix = "arXiv",
    primaryClass = "hep-ph",
    doi = "10.1088/1361-6471/ab2678",
    journal = "J. Phys. G",
    volume = "46",
    number = "9",
    pages = "093002",
    year = "2019"
}

@article{Liu:2019zoy,
    author = "Liu, Yan-Rui and Chen, Hua-Xing and Chen, Wei and Liu, Xiang and Zhu, Shi-Lin",
    title = "{Pentaquark and tetraquark states}",
    eprint = "1903.11976",
    archivePrefix = "arXiv",
    primaryClass = "hep-ph",
    doi = "10.1016/j.ppnp.2019.04.003",
    journal = "Prog. Part. Nucl. Phys.",
    volume = "107",
    pages = "237--320",
    year = "2019"
}

@article{Guo:2019twa,
    author = "Guo, Feng-Kun and Liu, Xiao-Hai and Sakai, Shuntaro",
    title = "{Threshold cusps and triangle singularities in hadronic reactions}",
    eprint = "1912.07030",
    archivePrefix = "arXiv",
    primaryClass = "hep-ph",
    doi = "10.1016/j.ppnp.2020.103757",
    journal = "Prog. Part. Nucl. Phys.",
    volume = "112",
    pages = "103757",
    year = "2020"
}

@article{Chen:2022asf,
    author = "Chen, Hua-Xing and Chen, Wei and Liu, Xiang and Liu, Yan-Rui and Zhu, Shi-Lin",
    title = "{An updated review of the new hadron states}",
    eprint = "2204.02649",
    archivePrefix = "arXiv",
    primaryClass = "hep-ph",
    doi = "10.1088/1361-6633/aca3b6",
    journal = "Rept. Prog. Phys.",
    volume = "86",
    number = "2",
    pages = "026201",
    year = "2023"
}

@article{Liu:2024uxn,
    author = "Liu, Ming-Zhu and Pan, Ya-Wen and Liu, Zhi-Wei and Wu, Tian-Wei and Lu, Jun-Xu and Geng, Li-Sheng",
    title = "{Three ways to decipher the nature of exotic hadrons: Multiplets, three-body hadronic molecules, and correlation functions}",
    eprint = "2404.06399",
    archivePrefix = "arXiv",
    primaryClass = "hep-ph",
    doi = "10.1016/j.physrep.2024.12.001",
    journal = "Phys. Rept.",
    volume = "1108",
    pages = "1--108",
    year = "2025"
}

@article{Chen:2024eaq,
  title = {Production of exotic hadrons in {\emph{pp}} and nuclear collisions},
  author = {Chen, Jinhui and Guo, Feng-Kun and Ma, Yu-Gang and Shen, Cheng-Ping and Shou, Qiye and Wang, Qian and Wu, Jia-Jun and Zou, Bing-Song},
  year = {2025},
  month = feb,
  journal = {Nucl. Sci. Tech.},
  volume = {36},
  number = {4},
  eprint = {2411.18257},
  primaryclass = {hep-ph},
  pages = {55},
  doi = {10.1007/s41365-025-01664-w},
  urldate = {2024-11-28},
  archiveprefix = {arXiv}
}

@article{Wang:2025sic,
    author = "Wang, Zhi-Gang",
    title = "{Review of the QCD sum rules for exotic states}",
    eprint = "2502.11351",
    archivePrefix = "arXiv",
    primaryClass = "hep-ph",
    doi = "10.15302/frontphys.2026.016300",
    journal = "Front. Phys. (Beijing)",
    volume = "21",
    number = "1",
    pages = "016300",
    year = "2026"
}

@article{Lu:2016mbb,
    author = "Lu, Yu and Anwar, Muhammad Naeem and Zou, Bing-Song",
    title = "{Coupled-Channel Effects for the Bottomonium with Realistic Wave Functions}",
    eprint = "1606.06927",
    archivePrefix = "arXiv",
    primaryClass = "hep-ph",
    doi = "10.1103/PhysRevD.94.034021",
    journal = "Phys. Rev. D",
    volume = "94",
    number = "3",
    pages = "034021",
    year = "2016"
}

@article{Baru:2018qkb,
    author = "Baru, V. and Epelbaum, E. and Gegelia, J. and Hanhart, C. and Mei{\ss}ner, U. G. and Nefediev, A. V.",
    title = "{Remarks on the Heavy-Quark Flavour Symmetry for doubly heavy hadronic molecules}",
    eprint = "1810.06921",
    archivePrefix = "arXiv",
    primaryClass = "hep-ph",
    doi = "10.1140/epjc/s10052-019-6560-7",
    journal = "Eur. Phys. J. C",
    volume = "79",
    number = "1",
    pages = "46",
    year = "2019"
}

@article{Baru:2016iwj,
    author = "Baru, V. and Epelbaum, E. and Filin, A. A. and Hanhart, C. and Mei{\ss}ner, Ulf-G. and Nefediev, A. V.",
    title = "{Heavy-quark spin symmetry partners of the X (3872) revisited}",
    eprint = "1605.09649",
    archivePrefix = "arXiv",
    primaryClass = "hep-ph",
    doi = "10.1016/j.physletb.2016.10.008",
    journal = "Phys. Lett. B",
    volume = "763",
    pages = "20--28",
    year = "2016"
}

@article{Cincioglu:2016fkm,
    author = "Cincioglu, E. and Nieves, J. and Ozpineci, A. and Yilmazer, A. U.",
    title = "{Quarkonium Contribution to Meson Molecules}",
    eprint = "1606.03239",
    archivePrefix = "arXiv",
    primaryClass = "hep-ph",
    doi = "10.1140/epjc/s10052-016-4413-1",
    journal = "Eur. Phys. J. C",
    volume = "76",
    number = "10",
    pages = "576",
    year = "2016"
}

@article{Ortega:2021zgk,
    author = "Ortega, Pablo G. and Entem, David R. and Fernandez, Francisco",
    title = "{Symmetries, Partners and Thresholds: The Case of the Xb}",
    eprint = "2107.08451",
    archivePrefix = "arXiv",
    primaryClass = "hep-ph",
    doi = "10.3390/sym13091600",
    journal = "Symmetry",
    volume = "13",
    number = "9",
    pages = "1600",
    year = "2021"
}

@article{Ebert:2005nc,
    author = "Ebert, D. and Faustov, R. N. and Galkin, V. O.",
    title = "{Masses of heavy tetraquarks in the relativistic quark model}",
    eprint = "hep-ph/0512230",
    archivePrefix = "arXiv",
    reportNumber = "HU-EP-05-82",
    doi = "10.1016/j.physletb.2006.01.026",
    journal = "Phys. Lett. B",
    volume = "634",
    pages = "214--219",
    year = "2006"
}

@article{Ebert:2008se,
    author = "Ebert, D. and Faustov, R. N. and Galkin, V. O.",
    title = "{Relativistic model of hidden bottom tetraquarks}",
    eprint = "0812.3477",
    archivePrefix = "arXiv",
    primaryClass = "hep-ph",
    reportNumber = "HU-EP-08-68",
    doi = "10.1142/S0217732309030357",
    journal = "Mod. Phys. Lett. A",
    volume = "24",
    pages = "567--573",
    year = "2009"
}

@article{Ali:2009pi,
    author = "Ali, Ahmed and Hambrock, Christian and Ahmed, Ishtiaq and Aslam, M. Jamil",
    title = "{A case for hidden $b\bar{b}$ tetraquarks based on $e^+e^- \to b\bar{b}$ cross section between $\sqrt{s}=10.54$ and 11.20 GeV}",
    eprint = "0911.2787",
    archivePrefix = "arXiv",
    primaryClass = "hep-ph",
    reportNumber = "DESY-09-192",
    doi = "10.1016/j.physletb.2009.12.053",
    journal = "Phys. Lett. B",
    volume = "684",
    pages = "28--39",
    year = "2010"
}

@article{Matheus:2006xi,
    author = "Matheus, Ricardo D'Elia and Narison, S. and Nielsen, M. and Richard, J. M.",
    title = "{Can the X(3872) be a 1++ four-quark state?}",
    eprint = "hep-ph/0608297",
    archivePrefix = "arXiv",
    doi = "10.1103/PhysRevD.75.014005",
    journal = "Phys. Rev. D",
    volume = "75",
    pages = "014005",
    year = "2007"
}

@article{Zhou:2018hlv,
    author = "Zhou, Zhi-Yong and Chen, Dian-Yong and Xiao, Zhiguang",
    title = "{Does the bottomonium counterpart of $X(3872)$ exist?}",
    eprint = "1810.03452",
    archivePrefix = "arXiv",
    primaryClass = "hep-ph",
    reportNumber = "USTC-ICTS-18-17",
    doi = "10.1103/PhysRevD.99.034005",
    journal = "Phys. Rev. D",
    volume = "99",
    number = "3",
    pages = "034005",
    year = "2019"
}

@article{Tornqvist:1993ng,
    author = "Tornqvist, Nils A.",
    title = "{From the deuteron to deusons, an analysis of deuteron - like meson meson bound states}",
    eprint = "hep-ph/9310247",
    archivePrefix = "arXiv",
    reportNumber = "HU-SEFT-R-1993-12",
    doi = "10.1007/BF01413192",
    journal = "Z. Phys. C",
    volume = "61",
    pages = "525--537",
    year = "1994"
}

@article{Nieves:2011zz,
    author = "Nieves, J. and Valderrama, M. Pavon",
    title = "{Deriving the existence of $B\bar{B}^*$ bound states from the X(3872) and Heavy Quark Symmetry}",
    eprint = "1106.0600",
    archivePrefix = "arXiv",
    primaryClass = "hep-ph",
    doi = "10.1103/PhysRevD.84.056015",
    journal = "Phys. Rev. D",
    volume = "84",
    pages = "056015",
    year = "2011"
}

@article{Guo:2013sya,
    author = "Guo, Feng-Kun and Hidalgo-Duque, Carlos and Nieves, Juan and Valderrama, Manuel Pavon",
    title = "{Consequences of Heavy Quark Symmetries for Hadronic Molecules}",
    eprint = "1303.6608",
    archivePrefix = "arXiv",
    primaryClass = "hep-ph",
    doi = "10.1103/PhysRevD.88.054007",
    journal = "Phys. Rev. D",
    volume = "88",
    pages = "054007",
    year = "2013"
}

@article{Karliner:2013dqa,
    author = "Karliner, Marek and Nussinov, Shmuel",
    title = "{The doubly heavies: $\bar{Q}Q\bar{q}q$ and $QQ\bar{q}\bar{q}$ tetraquarks and $QQq$ baryons}",
    eprint = "1304.0345",
    archivePrefix = "arXiv",
    primaryClass = "hep-ph",
    doi = "10.1007/JHEP07(2013)153",
    journal = "JHEP",
    volume = "07",
    pages = "153",
    year = "2013"
}

@article{Ozpineci:2013zas,
    author = "Ozpineci, A. and Xiao, C. W. and Oset, E.",
    title = "{Hidden beauty molecules within the local hidden gauge approach and heavy quark spin symmetry}",
    eprint = "1306.3154",
    archivePrefix = "arXiv",
    primaryClass = "hep-ph",
    doi = "10.1103/PhysRevD.88.034018",
    journal = "Phys. Rev. D",
    volume = "88",
    pages = "034018",
    year = "2013"
}

@article{Karliner:2015ina,
    author = "Karliner, Marek and Rosner, Jonathan L.",
    title = "{New Exotic Meson and Baryon Resonances from Doubly-Heavy Hadronic Molecules}",
    eprint = "1506.06386",
    archivePrefix = "arXiv",
    primaryClass = "hep-ph",
    reportNumber = "EFI-15-20, TAUP-2997-15",
    doi = "10.1103/PhysRevLett.115.122001",
    journal = "Phys. Rev. Lett.",
    volume = "115",
    number = "12",
    pages = "122001",
    year = "2015"
}

@article{Karliner:2014lta,
    author = "Karliner, Marek and Rosner, Jonathan L.",
    title = "{$X(3872)$, $X_b$, and the $\chi_{b1}(3P)$ state}",
    eprint = "1410.7729",
    archivePrefix = "arXiv",
    primaryClass = "hep-ph",
    reportNumber = "EFI-14-39, TAUP-2989-14",
    doi = "10.1103/PhysRevD.91.014014",
    journal = "Phys. Rev. D",
    volume = "91",
    number = "1",
    pages = "014014",
    year = "2015"
}

@article{Yamaguchi:2019vea,
    author = "Yamaguchi, Yasuhiro and Hosaka, Atsushi and Takeuchi, Sachiko and Takizawa, Makoto",
    title = "{Heavy hadronic molecules with pion exchange and quark core couplings: a guide for practitioners}",
    eprint = "1908.08790",
    archivePrefix = "arXiv",
    primaryClass = "hep-ph",
    reportNumber = "RIKEN-QHP-425",
    doi = "10.1088/1361-6471/ab72b0",
    journal = "J. Phys. G",
    volume = "47",
    number = "5",
    pages = "053001",
    year = "2020"
}

@article{AlFiky:2005jd,
    author = "AlFiky, Mohammad T. and Gabbiani, Fabrizio and Petrov, Alexey A.",
    title = "{X(3872): Hadronic molecules in effective field theory}",
    eprint = "hep-ph/0506141",
    archivePrefix = "arXiv",
    reportNumber = "WSU-HEP-0502",
    doi = "10.1016/j.physletb.2006.07.069",
    journal = "Phys. Lett. B",
    volume = "640",
    pages = "238--245",
    year = "2006"
}

@article{Brambilla:2024imu,
    author = "Brambilla, Nora and Mohapatra, Abhishek and Scirpa, Tommaso and Vairo, Antonio",
    title = "{Nature of ${\chi}_{c1}(3872)$ and $T_{cc}^+(3875)$}",
    eprint = "2411.14306",
    archivePrefix = "arXiv",
    primaryClass = "hep-ph",
    reportNumber = "TUM-EFT 193/24",
    doi = "10.1103/pdy7-hvg7",
    journal = "Phys. Rev. Lett.",
    volume = "135",
    number = "13",
    pages = "131902",
    year = "2025"
}

@misc{Brambilla:2026ujo,
  author        = {Brambilla, Nora and Butenschoen, Mathias and Hibler, Simon
                   and Mohapatra, Abhishek and Vairo, Antonio and Wang, Xiangpeng},
  title         = {Inclusive hadroproduction of {$\chi_{c1}(3872)$}, {$X_b$} and pentaquarks},
  eprint        = {2602.14916},
  archivePrefix = {arXiv},
  primaryClass  = {hep-ph},
  reportNumber  = {TUM-EFT 203/26},
  year          = {2026}
}

@article{Baru:2010ww,
    author = "Baru, V. and Hanhart, C. and Kalashnikova, Yu. S. and Kudryavtsev, A. E. and Nefediev, A. V.",
    title = "{Interplay of quark and meson degrees of freedom in a near-threshold resonance}",
    eprint = "1001.0369",
    archivePrefix = "arXiv",
    primaryClass = "hep-ph",
    reportNumber = "FZJ-IKP-TH-2009-39",
    doi = "10.1140/epja/i2010-10929-7",
    journal = "Eur. Phys. J. A",
    volume = "44",
    pages = "93--103",
    year = "2010"
}

@article{Hanhart:2011jz,
    author = "Hanhart, C. and Kalashnikova, Yu. S. and Nefediev, A. V.",
    title = "{Interplay of quark and meson degrees of freedom in a near-threshold resonance: multi-channel case}",
    eprint = "1106.1185",
    archivePrefix = "arXiv",
    primaryClass = "hep-ph",
    doi = "10.1140/epja/i2011-11101-9",
    journal = "Eur. Phys. J. A",
    volume = "47",
    pages = "101--110",
    year = "2011"
}

@article{Guo:2016bjq,
    author = "Guo, F. -K. and Hanhart, C. and Kalashnikova, Yu. S. and Matuschek, P. and Mizuk, R. V. and Nefediev, A. V. and Wang, Q. and Wynen, J. -L.",
    title = "{Interplay of quark and meson degrees of freedom in near-threshold states: A practical parametrization for line shapes}",
    eprint = "1602.00940",
    archivePrefix = "arXiv",
    primaryClass = "hep-ph",
    doi = "10.1103/PhysRevD.93.074031",
    journal = "Phys. Rev. D",
    volume = "93",
    number = "7",
    pages = "074031",
    year = "2016"
}

@article{Hammer:2016prh,
    author = "Hammer, I. K. and Hanhart, C. and Nefediev, A. V.",
    title = "{Remarks on meson loop effects on quark models}",
    eprint = "1607.06971",
    archivePrefix = "arXiv",
    primaryClass = "hep-ph",
    doi = "10.1140/epja/i2016-16330-8",
    journal = "Eur. Phys. J. A",
    volume = "52",
    number = "11",
    pages = "330",
    year = "2016"
}

@article{Hanhart:2022qxq,
    author = "Hanhart, Christoph and Nefediev, Alexey",
    title = "{Do near-threshold molecular states mix with neighboring $Q\bar Q$ states?}",
    eprint = "2209.10165",
    archivePrefix = "arXiv",
    primaryClass = "hep-ph",
    doi = "10.1103/PhysRevD.106.114003",
    journal = "Phys. Rev. D",
    volume = "106",
    number = "11",
    pages = "114003",
    year = "2022"
}

@article{ParticleDataGroup:2026mpi,
    author = "Takahashi, F. and others",
    collaboration = "Particle Data Group",
    title = "{Review of Particle Physics$^\ast$}",
    doi = "10.1142/s0217751x26300115",
    journal = "Int. J. Mod. Phys. A",
    volume = "41",
    number = "22",
    pages = "2630011",
    year = "2026"
}

@article{Hyodo:2013nka,
    author = "Hyodo, Tetsuo",
    title = "{Structure and compositeness of hadron resonances}",
    eprint = "1310.1176",
    archivePrefix = "arXiv",
    primaryClass = "hep-ph",
    reportNumber = "YITP-13-109",
    doi = "10.1142/S0217751X13300457",
    journal = "Int. J. Mod. Phys. A",
    volume = "28",
    pages = "1330045",
    year = "2013"
}

@article{Bulava:2024jpj,
    author = "Bulava, John and Knechtli, Francesco and Koch, Vanessa and Morningstar, Colin and Peardon, Michael",
    title = "{The quark-mass dependence of the potential energy between static colour sources in the QCD vacuum with light and strange quarks}",
    eprint = "2403.00754",
    archivePrefix = "arXiv",
    primaryClass = "hep-lat",
    reportNumber = "WUB/24-00",
    doi = "10.1016/j.physletb.2024.138754",
    journal = "Phys. Lett. B",
    volume = "854",
    pages = "138754",
    year = "2024"
}

@article{Guo:2017jvc,
    author = "Guo, Feng-Kun and Hanhart, Christoph and Mei{\ss}ner, Ulf-G. and Wang, Qian and Zhao, Qiang and Zou, Bing-Song",
    title = "{Hadronic molecules}",
    eprint = "1705.00141",
    archivePrefix = "arXiv",
    primaryClass = "hep-ph",
    doi = "10.1103/RevModPhys.90.015004",
    journal = "Rev. Mod. Phys.",
    volume = "90",
    number = "1",
    pages = "015004",
    year = "2018",
    note = "[Erratum: Rev.Mod.Phys. 94, 029901 (2022)]"
}

@article{Bruschini:2023zkb,
    author = "Bruschini, R.",
    title = "{Heavy-quark spin symmetry breaking in the Born-Oppenheimer approximation}",
    eprint = "2303.17533",
    archivePrefix = "arXiv",
    primaryClass = "hep-ph",
    doi = "10.1007/JHEP08(2023)219",
    journal = "JHEP",
    volume = "08",
    pages = "219",
    year = "2023"
}

@article{Bulava:2019iut,
    author = {Bulava, John and H{\"o}rz, Ben and Knechtli, Francesco and Koch, Vanessa and Moir, Graham and Morningstar, Colin and Peardon, Mike},
    title = "{String breaking by light and strange quarks in QCD}",
    eprint = "1902.04006",
    archivePrefix = "arXiv",
    primaryClass = "hep-lat",
    reportNumber = "WUB/19-00, CP3-Origins-2019-002 DNRF90, MITP/19-010",
    doi = "10.1016/j.physletb.2019.05.018",
    journal = "Phys. Lett. B",
    volume = "793",
    pages = "493--498",
    year = "2019"
}

@article{Eichten:1980mw,
    author = "Eichten, E. and Feinberg, F.",
    title = "{Spin Dependent Forces in QCD}",
    reportNumber = "HUTP-80-A053",
    doi = "10.1103/PhysRevD.23.2724",
    journal = "Phys. Rev. D",
    volume = "23",
    pages = "2724",
    year = "1981"
}

@article{Gromes:1984ma,
    author = "Gromes, Dieter",
    title = "{Spin Dependent Potentials in QCD and the Correct Long Range Spin Orbit Term}",
    reportNumber = "HD-THEP-84-5",
    doi = "10.1007/BF01452566",
    journal = "Z. Phys. C",
    volume = "26",
    pages = "401",
    year = "1984"
}

@article{Schnitzer:1983tt,
    author = "Schnitzer, Howard J.",
    title = "{Systematics of the Spin Dependent Quarkonium Forces}",
    reportNumber = "BRX-TH-142, HUTP-83-A064",
    doi = "10.1016/0370-2693(84)90682-8",
    journal = "Phys. Lett. B",
    volume = "134",
    pages = "253",
    year = "1984"
}

@article{Bondar:2011ev,
    author = "Bondar, A. E. and Garmash, A. and Milstein, A. I. and Mizuk, R. and Voloshin, M. B.",
    title = "{Heavy quark spin structure in $Z_b$ resonances}",
    eprint = "1105.4473",
    archivePrefix = "arXiv",
    primaryClass = "hep-ph",
    reportNumber = "FTPI-MINN-11-06, UMN-TH-2939-11",
    doi = "10.1103/PhysRevD.84.054010",
    journal = "Phys. Rev. D",
    volume = "84",
    pages = "054010",
    year = "2011"
}

@article{Isgur:1999rf,
    author = "Isgur, Nathan",
    title = "{Meson - like baryons and the spin orbit puzzle}",
    eprint = "hep-ph/9910272",
    archivePrefix = "arXiv",
    reportNumber = "JLAB-THY-99-30",
    doi = "10.1103/PhysRevD.62.014025",
    journal = "Phys. Rev. D",
    volume = "62",
    pages = "014025",
    year = "2000"
}

@article{Kalashnikova:2005ui,
    author = "Kalashnikova, Yu. S.",
    title = "{Coupled-channel model for charmonium levels and an option for X(3872)}",
    eprint = "hep-ph/0506270",
    archivePrefix = "arXiv",
    doi = "10.1103/PhysRevD.72.034010",
    journal = "Phys. Rev. D",
    volume = "72",
    pages = "034010",
    year = "2005"
}

@article{Franck:1926,
  author  = {Franck, J. and Dymond, E. G.},
  title   = {Elementary processes of photochemical reactions},
  journal = {Transactions of the Faraday Society},
  volume  = {21},
  pages   = {536--542},
  year    = {1926},
  doi     = {10.1039/TF9262100536},
}

@article{Condon:1926,
  author  = {Condon, E. U.},
  title   = {A Theory of Intensity Distribution in Band Systems},
  journal = {Physical Review},
  volume  = {28},
  pages   = {1182--1201},
  year    = {1926},
  doi     = {10.1103/PhysRev.28.1182},
}

@article{LlanesEstrada:2008fc,
  author        = {Llanes-Estrada, Felipe J. and Cotanch, Stephen R. and General, Ignacio G. and Wang, Ping},
  title         = {Franck--Condon principle for heavy-quark hadron decays},
  eprint        = {0803.0806},
  archivePrefix = {arXiv},
  primaryClass  = {hep-ph},
  doi           = {10.1063/1.2973494},
  journal       = {AIP Conf. Proc.},
  volume        = {1030},
  pages         = {171--176},
  year          = {2008},
}

@article{TorresRincon:2010fc,
  author        = {Torres-Rincon, Juan M. and Llanes-Estrada, Felipe J.},
  title         = {Heavy Quark Fluorescence},
  eprint        = {1003.5989},
  archivePrefix = {arXiv},
  primaryClass  = {hep-ph},
  doi           = {10.1103/PhysRevLett.105.022003},
  journal       = {Phys. Rev. Lett.},
  volume        = {105},
  pages         = {022003},
  year          = {2010},
}

@article{Wise:1992hn,
  author  = {Wise, Mark B.},
  title   = {Chiral perturbation theory for hadrons containing a heavy quark},
  doi     = {10.1103/PhysRevD.45.R2188},
  journal = {Phys. Rev. D},
  volume  = {45},
  pages   = {R2188},
  year    = {1992},
}

@article{Burdman:1992gh,
  author  = {Burdman, Gustavo and Donoghue, John F.},
  title   = {Union of chiral and heavy quark symmetries},
  doi     = {10.1016/0370-2693(92)90068-F},
  journal = {Phys. Lett. B},
  volume  = {280},
  pages   = {287--291},
  year    = {1992},
}

@article{Yan:1992gz,
  author  = {Yan, Tung-Mow and Cheng, Hai-Yang and Cheung, Chi-Yee and Lin, Guey-Lin and Lin, Y. C. and Yu, Hoi-Lai},
  title   = {Heavy quark symmetry and chiral dynamics},
  doi     = {10.1103/PhysRevD.46.1148},
  journal = {Phys. Rev. D},
  volume  = {46},
  pages   = {1148--1164},
  year    = {1992},
  note    = {[Erratum: Phys. Rev. D 55, 5851 (1997)]},
}

@article{Flynn:2015xna,
  author        = {Flynn, J. M. and Fritzsch, P. and Kawanai, T. and Lehner, C. and Samways, B. and Sachrajda, C. T. and Van de Water, R. S. and Witzel, O.},
  title         = {{The $B^*B\pi$ Coupling Using Relativistic Heavy Quarks}},
  eprint        = {1506.06413},
  archivePrefix = {arXiv},
  primaryClass  = {hep-lat},
  doi           = {10.1103/PhysRevD.93.014510},
  journal       = {Phys. Rev. D},
  volume        = {93},
  pages         = {014510},
  year          = {2016},
}

@article{Gell-Mann:1953dcn,
  author  = {Gell-Mann, M. and Goldberger, M. L.},
  title   = {The Formal Theory of Scattering},
  doi     = {10.1103/PhysRev.91.398},
  journal = {Phys. Rev.},
  volume  = {91},
  pages   = {398},
  year    = {1953},
}

@article{Du:2024gzw,
    author = "Du, Meng-Lin and Guo, Feng-Kun and Wu, Bing",
    title = "{Effective-Range Expansion with a Long-Range Force}",
    eprint = "2408.09375",
    archivePrefix = "arXiv",
    primaryClass = "hep-ph",
    doi = "10.1103/cwdt-dj6z",
    journal = "Phys. Rev. Lett.",
    volume = "135",
    number = "1",
    pages = "011903",
    year = "2025"
}

@article{Wang:2026ups,
    author = "Wang, Wen-Jia and Wu, Bing and Du, Meng-Lin and Guo, Feng-Kun",
    title = "{Effective range expansion with the left-hand cut: Higher order improvements}",
    eprint = "2601.04989",
    archivePrefix = "arXiv",
    primaryClass = "hep-ph",
    doi = "10.1103/tdyn-422c",
    journal = "Phys. Rev. D",
    volume = "114",
    number = "1",
    pages = "014003",
    year = "2026"
}

@article{Bethe:1949yr,
    author = "Bethe, H. A.",
    title = "{Theory of the Effective Range in Nuclear Scattering}",
    doi = "10.1103/PhysRev.76.38",
    journal = "Phys. Rev.",
    volume = "76",
    pages = "38--50",
    year = "1949"
}

@article{Alasiri:2024nue,
    author = "Alasiri, Fareed and Braaten, Eric and Mohapatra, Abhishek",
    title = "{Born-Oppenheimer potentials for SU(3) gauge theory}",
    eprint = "2406.05123",
    archivePrefix = "arXiv",
    primaryClass = "hep-ph",
    reportNumber = "TUM-EFT 188/24",
    doi = "10.1103/PhysRevD.110.054029",
    journal = "Phys. Rev. D",
    volume = "110",
    number = "5",
    pages = "054029",
    year = "2024"
}

@article{D0:2012pig,
    author = "Abazov, Victor Mukhamedovich and others",
    collaboration = "D0",
    title = "{Observation of a narrow mass state decaying into $\Upsilon(1S) + \gamma$ in $p\bar{p}$ collisions at $\sqrt{s} = 1.96$ TeV}",
    eprint = "1203.6034",
    archivePrefix = "arXiv",
    primaryClass = "hep-ex",
    reportNumber = "FERMILAB-PUB-12-084-E",
    doi = "10.1103/PhysRevD.86.031103",
    journal = "Phys. Rev. D",
    volume = "86",
    pages = "031103",
    year = "2012"
}

@article{CMS:2018vgf,
    author = "Sirunyan, A. M. and others",
    collaboration = "CMS",
    title = "{Observation of the $\chi_\mathrm{b1}$(3P) and $\chi_\mathrm{b2}$(3P) and measurement of their masses}",
    eprint = "1805.11192",
    archivePrefix = "arXiv",
    primaryClass = "hep-ex",
    reportNumber = "CMS-BPH-17-008, CERN-EP-2018-134",
    doi = "10.1103/PhysRevLett.121.092002",
    journal = "Phys. Rev. Lett.",
    volume = "121",
    pages = "092002",
    year = "2018"
}

@article{Doring:2009bi,
    author = "Doring, M. and Hanhart, C. and Huang, F. and Krewald, S. and Meissner, U. -G.",
    title = "{The Role of the background in the extraction of resonance contributions from meson-baryon scattering}",
    eprint = "0903.1781",
    archivePrefix = "arXiv",
    primaryClass = "nucl-th",
    reportNumber = "FZJ-IKP-TH-2009-8, HISKP-TH-09-08",
    doi = "10.1016/j.physletb.2009.09.052",
    journal = "Phys. Lett. B",
    volume = "681",
    pages = "26--31",
    year = "2009"
}

@article{Hanhart:2012wi,
    author = "Hanhart, C.",
    title = "{A New Parameterization for the Pion Vector Form Factor}",
    eprint = "1203.6839",
    archivePrefix = "arXiv",
    primaryClass = "hep-ph",
    doi = "10.1016/j.physletb.2012.07.038",
    journal = "Phys. Lett. B",
    volume = "715",
    pages = "170--177",
    year = "2012"
}

@article{Machleidt:1987hj,
    author = "Machleidt, R. and Holinde, K. and Elster, C.",
    title = "{The Bonn Meson Exchange Model for the Nucleon Nucleon Interaction}",
    doi = "10.1016/S0370-1573(87)80002-9",
    journal = "Phys. Rept.",
    volume = "149",
    pages = "1--89",
    year = "1987"
}

@article{Epelbaum:2019kcf,
    author = "Epelbaum, Evgeny and Krebs, Hermann and Reinert, Patrick",
    title = "{High-precision nuclear forces from chiral EFT: State-of-the-art, challenges and outlook}",
    eprint = "1911.11875",
    archivePrefix = "arXiv",
    primaryClass = "nucl-th",
    doi = "10.3389/fphy.2020.00098",
    journal = "Front. in Phys.",
    volume = "8",
    pages = "98",
    year = "2020"
}

@article{Nakano:1982bc,
    author = "Nakano, K.",
    title = "{Two potential formalisms and the Coulomb--nuclear interference}",
    doi = "10.1103/PhysRevC.26.1123",
    journal = "Phys. Rev. C",
    volume = "26",
    pages = "1123--1131",
    year = "1982"
}

@article{Eichten:2004uh,
    author = "Eichten, Estia J. and Lane, Kenneth and Quigg, Chris",
    title = "{Charmonium levels near threshold and the narrow state $X(3872) \to \pi^{+}\pi^{-}J/\psi$}",
    eprint = "hep-ph/0401210",
    archivePrefix = "arXiv",
    reportNumber = "FERMILAB-PUB-04-001-T, BUHEP-04-01",
    doi = "10.1103/PhysRevD.69.094019",
    journal = "Phys. Rev. D",
    volume = "69",
    pages = "094019",
    year = "2004"
}

@article{Liu:2011yp,
    author = "Liu, Jia-Feng and Ding, Gui-Jun",
    title = "{Bottomonium Spectrum with Coupled-Channel Effects}",
    eprint = "1105.0855",
    archivePrefix = "arXiv",
    primaryClass = "hep-ph",
    doi = "10.1140/epjc/s10052-012-1981-6",
    journal = "Eur. Phys. J. C",
    volume = "72",
    pages = "1981",
    year = "2012"
}

@article{Shi:2024llv,
    author = "Shi, Pan-Pan and Albaladejo, Miguel and Du, Meng-Lin and Guo, Feng-Kun and Nieves, Juan",
    title = "{P-wave charmonium contribution to hidden-charm states from a reanalysis of lattice QCD data}",
    eprint = "2410.19563",
    archivePrefix = "arXiv",
    primaryClass = "hep-ph",
    doi = "10.1103/PhysRevD.111.074043",
    journal = "Phys. Rev. D",
    volume = "111",
    number = "7",
    pages = "074043",
    year = "2025"
}

@article{Ortega:2009hj,
    author = "Ortega, P. G. and Segovia, J. and Entem, D. R. and Fern{\\'a}ndez, F.",
    title = "{Coupled channel approach to the structure of the $X(3872)$}",
    eprint = "0907.3997",
    archivePrefix = "arXiv",
    primaryClass = "hep-ph",
    doi = "10.1103/PhysRevD.81.054023",
    journal = "Phys. Rev. D",
    volume = "81",
    pages = "054023",
    year = "2010"
}

@article{Ortega:2017qmg,
    author = "Ortega, Pablo G. and Segovia, Jorge and Entem, David R. and Fern{\\'a}ndez, Francisco",
    title = "{Charmonium resonances in the 3.9 GeV/$c^2$ energy region and the $X(3915)/X(3930)$ puzzle}",
    eprint = "1706.02639",
    archivePrefix = "arXiv",
    primaryClass = "hep-ph",
    doi = "10.1016/j.physletb.2018.01.005",
    journal = "Phys. Lett. B",
    volume = "778",
    pages = "1--5",
    year = "2018"
}

@article{Zhou:2017dwj,
    author = "Zhou, Zhi-Yong and Xiao, Zhiguang",
    title = "{Understanding $X(3862)$, $X(3872)$, and $X(3930)$ in a Friedrichs-model-like scheme}",
    eprint = "1704.04438",
    archivePrefix = "arXiv",
    primaryClass = "hep-ph",
    doi = "10.1103/PhysRevD.96.054031",
    journal = "Phys. Rev. D",
    volume = "96",
    number = "5",
    pages = "054031",
    year = "2017",
    note = "[Erratum: Phys. Rev. D 96, 099905 (2017)]"
}

@article{Wang:2023ovj,
    author = "Wang, Guang-Juan and Yang, Zhi and Wu, Jia-Jun and Oka, Makoto and Zhu, Shi-Lin",
    title = "{New insight into the exotic states strongly coupled with the $D\bar{D}^*$ from the $T_{cc}^+$}",
    eprint = "2306.12406",
    archivePrefix = "arXiv",
    primaryClass = "hep-ph",
    doi = "10.1016/j.scib.2024.07.012",
    journal = "Sci. Bull.",
    volume = "69",
    pages = "3036--3041",
    year = "2024"
}

@article{Xiao:2023lpv,
    author = "Xiao, Zhiguang and Zhou, Zhi-Yong",
    title = "{On the generalized Friedrichs-Lee model with multiple discrete and continuous states}",
    eprint = "2310.14937",
    archivePrefix = "arXiv",
    primaryClass = "hep-ph",
    doi = "10.1088/1674-1137/adcd4b",
    journal = "Chin. Phys. C",
    volume = "49",
    number = "8",
    pages = "083102",
    year = "2025"
}

@article{Brisudova:1998wq,
    author = "Brisudova, Martina M. and Burakovsky, L. and Goldman, J. Terrance",
    title = "{Effect of color screening on heavy quarkonia Regge trajectories}",
    eprint = "hep-ph/9810296",
    archivePrefix = "arXiv",
    reportNumber = "LA-UR-98-4357",
    doi = "10.1016/S0370-2693(99)00732-7",
    journal = "Phys. Lett. B",
    volume = "460",
    pages = "1--7",
    year = "1999"
}

@article{Tornqvist:1984xz,
    author = "Tornqvist, Nils A.",
    title = "{Quarkonium and Quark Loops}",
    reportNumber = "HU-TFT-84-47",
    journal = "Acta Phys. Polon. B",
    volume = "16",
    pages = "503",
    year = "1985",
    note = "[Erratum: Acta Phys.Polon.B 16, 683 (1985)]"
}

@article{Brisudova:1999ut,
    author = "Brisudova, Martina M. and Burakovsky, L. and Goldman, J. Terrance",
    title = "{Effective functional form of Regge trajectories}",
    eprint = "hep-ph/9906293",
    archivePrefix = "arXiv",
    reportNumber = "LA-UR-99-2736",
    doi = "10.1103/PhysRevD.61.054013",
    journal = "Phys. Rev. D",
    volume = "61",
    pages = "054013",
    year = "2000"
}

@article{Hashimoto:2004fv,
    author = "Hashimoto, Shoji and Onogi, Tetsuya",
    title = "{Heavy quarks on the lattice}",
    eprint = "hep-ph/0407221",
    archivePrefix = "arXiv",
    reportNumber = "KEK-CP-155, YITP-04-40",
    doi = "10.1146/annurev.nucl.53.041002.110550",
    journal = "Ann. Rev. Nucl. Part. Sci.",
    volume = "54",
    pages = "451--486",
    year = "2004"
}

@article{Bicudo:2022cqi,
    author = "Bicudo, Pedro",
    title = "{Tetraquarks and pentaquarks in lattice QCD with light and heavy quarks}",
    eprint = "2212.07793",
    archivePrefix = "arXiv",
    primaryClass = "hep-lat",
    doi = "10.1016/j.physrep.2023.10.001",
    journal = "Phys. Rept.",
    volume = "1039",
    pages = "1--49",
    year = "2023"
}

@article{Swanson:2003tb,
    author = "Swanson, Eric S.",
    title = "{Short range structure in the X(3872)}",
    eprint = "hep-ph/0311229",
    archivePrefix = "arXiv",
    reportNumber = "JLAB-THY-03-227",
    doi = "10.1016/j.physletb.2004.03.033",
    journal = "Phys. Lett. B",
    volume = "588",
    pages = "189--195",
    year = "2004"
}

@article{Suzuki:2005ha,
    author = "Suzuki, Mahiko",
    title = "{The X(3872) boson: Molecule or charmonium}",
    eprint = "hep-ph/0508258",
    archivePrefix = "arXiv",
    reportNumber = "LBL-58724",
    doi = "10.1103/PhysRevD.72.114013",
    journal = "Phys. Rev. D",
    volume = "72",
    pages = "114013",
    year = "2005"
}

@article{Wang:2018jlv,
    author = "Wang, Q. and Baru, V. and Filin, A. A. and Hanhart, C. and Nefediev, A. V. and Wynen, J. -L",
    title = "{Line shapes of the $Z_b(10610)$ and $Z_b(10650)$ in the elastic and inelastic channels revisited}",
    eprint = "1805.07453",
    archivePrefix = "arXiv",
    primaryClass = "hep-ph",
    doi = "10.1103/PhysRevD.98.074023",
    journal = "Phys. Rev. D",
    volume = "98",
    number = "7",
    pages = "074023",
    year = "2018"
}

@article{Thomas:2008ja,
    author = "Thomas, C. E. and Close, F. E.",
    title = "{Is X(3872) a molecule?}",
    eprint = "0805.3653",
    archivePrefix = "arXiv",
    primaryClass = "hep-ph",
    reportNumber = "OUTP-08-08P",
    doi = "10.1103/PhysRevD.78.034007",
    journal = "Phys. Rev. D",
    volume = "78",
    pages = "034007",
    year = "2008"
}

@article{Meng:2022ozq,
    author = "Meng, Lu and Wang, Bo and Wang, Guang-Juan and Zhu, Shi-Lin",
    title = "{Chiral perturbation theory for heavy hadrons and chiral effective field theory for heavy hadronic molecules}",
    eprint = "2204.08716",
    archivePrefix = "arXiv",
    primaryClass = "hep-ph",
    doi = "10.1016/j.physrep.2023.04.003",
    journal = "Phys. Rept.",
    volume = "1019",
    pages = "1--149",
    year = "2023"
}

@article{Guo:2010ak,
    author = "Guo, Feng-Kun and Hanhart, Christoph and Li, Gang and Mei{\ss}ner, Ulf-G. and Zhao, Qiang",
    title = "{Effect of charmed meson loops on charmonium transitions}",
    eprint = "1008.3632",
    archivePrefix = "arXiv",
    primaryClass = "hep-ph",
    reportNumber = "FZJ-IKP-TH-2010-08, HISKP-TH-10-09",
    doi = "10.1103/PhysRevD.83.034013",
    journal = "Phys. Rev. D",
    volume = "83",
    pages = "034013",
    year = "2011"
}

@article{Baru:2017gwo,
    author = "Baru, V. and Epelbaum, E. and Filin, A. A. and Hanhart, C. and Nefediev, A. V.",
    title = "{Spin partners of the Z$_{b}$ (10610) and Z$_{b}$ (10650) revisited}",
    eprint = "1704.07332",
    archivePrefix = "arXiv",
    primaryClass = "hep-ph",
    doi = "10.1007/JHEP06(2017)158",
    journal = "JHEP",
    volume = "06",
    pages = "158",
    year = "2017"
}

@article{Baru:2019xnh,
    author = "Baru, V. and Epelbaum, E. and Filin, A. A. and Hanhart, C. and Nefediev, A. V. and Wang, Q.",
    title = "{Spin partners $W_{bJ}$ from the line shapes of the $Z_b(10610)$ and $Z_b(10650)$}",
    eprint = "1901.10319",
    archivePrefix = "arXiv",
    primaryClass = "hep-ph",
    doi = "10.1103/PhysRevD.99.094013",
    journal = "Phys. Rev. D",
    volume = "99",
    number = "9",
    pages = "094013",
    year = "2019"
}

@article{Bedaque:2003wa,
    author = "Bedaque, P. F. and Hammer, H. W. and van Kolck, U.",
    title = "{Narrow resonances in effective field theory}",
    eprint = "nucl-th/0304007",
    archivePrefix = "arXiv",
    doi = "10.1016/j.physletb.2003.07.049",
    journal = "Phys. Lett. B",
    volume = "569",
    pages = "159--167",
    year = "2003"
}

@article{Alhakami:2017ntb,
    author = "Alhakami, Mohammad H.",
    title = "{Short-range interactions and narrow resonances in effective field theory}",
    eprint = "1706.05622",
    archivePrefix = "arXiv",
    primaryClass = "nucl-th",
    doi = "10.1103/PhysRevD.96.056019",
    journal = "Phys. Rev. D",
    volume = "96",
    number = "5",
    pages = "056019",
    year = "2017"
}

@article{Habashi:2020qgw,
    author = "Habashi, J. Balal and Sen, S. and Fleming, S. and van Kolck, U.",
    title = "{Effective Field Theory for Two-Body Systems with Shallow S-Wave Resonances}",
    eprint = "2007.07360",
    archivePrefix = "arXiv",
    primaryClass = "nucl-th",
    doi = "10.1016/j.aop.2020.168283",
    journal = "Annals Phys.",
    volume = "422",
    pages = "168283",
    year = "2020"
}

@article{Habashi:2020ofb,
    author = "Habashi, J. B. and Fleming, S. and van Kolck, U.",
    title = "{Nonrelativistic Effective Field Theory with a Resonance Field}",
    eprint = "2012.14995",
    archivePrefix = "arXiv",
    primaryClass = "hep-ph",
    doi = "10.1140/epja/s10050-021-00452-5",
    journal = "Eur. Phys. J. A",
    volume = "57",
    number = "5",
    pages = "169",
    year = "2021"
}

@book{Manohar:2000dt,
    author = "Manohar, Aneesh V. and Wise, Mark B.",
    title = "{Heavy quark physics}",
    doi = "10.1017/9781009402125",
    isbn = "978-0-521-03757-0, 978-1-009-40212-5",
    volume = "10",
    year = "2000",
    publisher = "Cambridge University Press"
}

@article{Hu:2005gf,
    author = "Hu, Jie and Mehen, Thomas",
    title = "{Chiral Lagrangian with heavy quark-diquark symmetry}",
    eprint = "hep-ph/0511321",
    archivePrefix = "arXiv",
    reportNumber = "JLAB-THY-05-452",
    doi = "10.1103/PhysRevD.73.054003",
    journal = "Phys. Rev. D",
    volume = "73",
    pages = "054003",
    year = "2006"
}

@article{Nogga:2005hy,
    author = "Nogga, A. and Timmermans, R. G. E. and van Kolck, U.",
    title = "{Renormalization of one-pion exchange and power counting}",
    eprint = "nucl-th/0506005",
    archivePrefix = "arXiv",
    reportNumber = "FZJ-IKP-TH-2005-19",
    doi = "10.1103/PhysRevC.72.054006",
    journal = "Phys. Rev. C",
    volume = "72",
    pages = "054006",
    year = "2005"
}

@article{Ji:2025hjw,
    author = "Ji, Teng and Dong, Xiang-Kun and Guo, Feng-Kun and Hanhart, Christoph and Mei{\ss}ner, Ulf-G.",
    title = "{Precise determination of the properties of $X(3872)$ and of its isovector partner $W_{c1}$}",
    eprint = "2502.04458",
    archivePrefix = "arXiv",
    primaryClass = "hep-ph",
    month = "2",
    year = "2026",
    journal = "Sci. Bull.",
    pages = "in print"
}

@article{Du:2023hlu,
    author = "Du, Meng-Lin and Filin, Arseniy and Baru, Vadim and Dong, Xiang-Kun and Epelbaum, Evgeny and Guo, Feng-Kun and Hanhart, Christoph and Nefediev, Alexey and Nieves, Juan and Wang, Qian",
    title = "{Role of Left-Hand Cut Contributions on Pole Extractions from Lattice Data: Case Study for $T_{cc}(3875)^+$}",
    eprint = "2303.09441",
    archivePrefix = "arXiv",
    primaryClass = "hep-ph",
    doi = "10.1103/PhysRevLett.131.131903",
    journal = "Phys. Rev. Lett.",
    volume = "131",
    number = "13",
    pages = "131903",
    year = "2023"
}

@article{Ni:2025gvx,
    author = "Ni, Ru-Hui and Deng, Qian and Wu, Jia-Jun and Zhong, Xian-Hui",
    title = "{Bottomonia in an unquenched quark model}",
    eprint = "2501.15110",
    archivePrefix = "arXiv",
    primaryClass = "hep-ph",
    doi = "10.1103/6x3t-x15s",
    journal = "Phys. Rev. D",
    volume = "111",
    number = "11",
    pages = "114027",
    year = "2025"
}

@article{Bai:2026atm,
    author = "Bai, Zi-Yue and Chen, Dian-Yong and Huang, Qi and Liu, Xiang and Luo, Si-Qiang and Wang, Jun-Zhang",
    title = "{Unquenched charmonium and beyond}",
    eprint = "2602.19887",
    archivePrefix = "arXiv",
    primaryClass = "hep-ph",
    doi = "10.1016/j.physrep.2026.08.002",
    journal = "Phys. Rept.",
    volume = "1204",
    pages = "1--162",
    year = "2027"
}

@article{LHCb:2024vfz,
    author = "Aaij, Roel and others",
    collaboration = "LHCb",
    title = "{Observation of New Charmonium or Charmoniumlike States in $B^+\to D^{*\pm}D^\mp K^+$ Decays}",
    eprint = "2406.03156",
    archivePrefix = "arXiv",
    primaryClass = "hep-ex",
    reportNumber = "LHCb-PAPER-2023-047, CERN-EP-2024-096",
    doi = "10.1103/PhysRevLett.133.131902",
    journal = "Phys. Rev. Lett.",
    volume = "133",
    number = "13",
    pages = "131902",
    year = "2024"
}

@article{Fleming:2008yn,
    author = {Fleming, Sean and Mehen, Thomas},
    title = {{Hadronic Decays of the X(3872) to chi(cJ) in Effective Field Theory}},
    eprint = {0807.2674},
    archivePrefix = {arXiv},
    primaryClass = {hep-ph},
    doi = {10.1103/PhysRevD.78.094019},
    journal = {Phys. Rev. D},
    volume = {78},
    pages = {094019},
    year = {2008}
}

@other{Li:2024pfg,
    author = {Li, Haozheng and Shi, Chunjiang and Chen, Ying and Gong, Ming and Liang, Juzheng and Liu, Zhaofeng and Sun, Wei},
    title = {{$X(3872)$ Relevant $D\bar{D}^*$ Scattering in $N_f=2$ Lattice QCD}},
    eprint = {2402.14541},
    archivePrefix = {arXiv},
    primaryClass = {hep-lat},
    month = {2},
    year = {2024}
}

@article{Gamermann:2009uq,
    author = {Gamermann, D. and Nieves, J. and Oset, E. and Ruiz Arriola, E.},
    title = {{Couplings in coupled channels versus wave functions: application to the X(3872) resonance}},
    eprint = {0911.4407},
    archivePrefix = {arXiv},
    primaryClass = {hep-ph},
    doi = {10.1103/PhysRevD.81.014029},
    journal = {Phys. Rev. D},
    volume = {81},
    pages = {014029},
    year = {2010}
}

@article{Garcia-Recio:2015jsa,
    author = {Garcia-Recio, C. and Hidalgo-Duque, C. and Nieves, J. and Salcedo, L. L. and Tolos, L.},
    title = {{Compositeness of the strange, charm, and beauty odd parity $\Lambda$ states}},
    eprint = {1506.04235},
    archivePrefix = {arXiv},
    primaryClass = {hep-ph},
    doi = {10.1103/PhysRevD.92.034011},
    journal = {Phys. Rev. D},
    volume = {92},
    number = {3},
    pages = {034011},
    year = {2015}
}

@article{Born:1927,
    author  = {M. Born and R. Oppenheimer},
    title   = {Zur Quantentheorie der Molekeln},
    journal = {Ann. Phys.},
    volume  = {389},
    number  = {20},
    pages   = {457--484},
    year    = {1927},
    doi     = {10.1002/andp.19273892002}
}

@article{Chao:1979tg,
    author = "Chao, Kuang-Ta",
    title = "{The S Wave $Q \bar{Q} q \bar{q}$ States in the Adiabatic Approximation}",
    reportNumber = "OXFORD-TP 53/79",
    doi = "10.1016/0550-3213(81)90143-7",
    journal = "Nucl. Phys. B",
    volume = "183",
    pages = "435--444",
    year = "1981"
}

@article{Juge:1999ie,
    author = "Juge, K. J. and Kuti, J. and Morningstar, C. J.",
    title = "{Ab initio study of hybrid anti-b g b mesons}",
    eprint = "hep-ph/9902336",
    archivePrefix = "arXiv",
    reportNumber = "FERMILAB-PUB-99-022-T, UCSD-PTH-99-01",
    doi = "10.1103/PhysRevLett.82.4400",
    journal = "Phys. Rev. Lett.",
    volume = "82",
    pages = "4400--4403",
    year = "1999"
}

@article{Guo:2008yz,
    author = "Guo, Peng and Szczepaniak, Adam P. and Galata, Giuseppe and Vassallo, Andrea and Santopinto, Elena",
    title = "{Heavy quarkonium hybrids from Coulomb gauge QCD}",
    eprint = "0807.2721",
    archivePrefix = "arXiv",
    primaryClass = "hep-ph",
    doi = "10.1103/PhysRevD.78.056003",
    journal = "Phys. Rev. D",
    volume = "78",
    pages = "056003",
    year = "2008"
}

@article{Soto:2017one,
    author = "Soto, Joan",
    editor = "Narison, Stephan",
    title = "{Heavy Quarkonium Hybrids}",
    eprint = "1709.08038",
    archivePrefix = "arXiv",
    primaryClass = "hep-ph",
    doi = "10.1016/j.nuclphysbps.2018.03.020",
    journal = "Nucl. Part. Phys. Proc.",
    volume = "294-296",
    pages = "87--94",
    year = "2018"
}

@article{Brambilla:2018pyn,
    author = "Brambilla, Nora and Lai, Wai Kin and Segovia, Jorge and Tarr{\'u}s Castell{\`a}, Jaume and Vairo, Antonio",
    title = "{Spin structure of heavy-quark hybrids}",
    eprint = "1805.07713",
    archivePrefix = "arXiv",
    primaryClass = "hep-ph",
    reportNumber = "TUM-EFT-95-17, TUM-EFT 95/17",
    doi = "10.1103/PhysRevD.99.014017",
    journal = "Phys. Rev. D",
    volume = "99",
    number = "1",
    pages = "014017",
    year = "2019",
    note = "[Erratum: Phys.Rev.D 101, 099902 (2020)]"
}

@article{Brambilla:2019jfi,
    author = "Brambilla, Nora and Lai, Wai Kin and Segovia, Jorge and Tarr{\'u}s Castell{\`a}, Jaume",
    title = "{QCD spin effects in the heavy hybrid potentials and spectra}",
    eprint = "1908.11699",
    archivePrefix = "arXiv",
    primaryClass = "hep-ph",
    reportNumber = "TUM-EFT 129/19",
    doi = "10.1103/PhysRevD.101.054040",
    journal = "Phys. Rev. D",
    volume = "101",
    number = "5",
    pages = "054040",
    year = "2020"
}

@article{Bicudo:2015kna,
    author = "Bicudo, Pedro and Cichy, Krzysztof and Peters, Antje and Wagner, Marc",
    title = "{BB interactions with static bottom quarks from Lattice QCD}",
    eprint = "1510.03441",
    archivePrefix = "arXiv",
    primaryClass = "hep-lat",
    doi = "10.1103/PhysRevD.93.034501",
    journal = "Phys. Rev. D",
    volume = "93",
    number = "3",
    pages = "034501",
    year = "2016"
}

@article{Griffiths:1983ah,
    author = "Griffiths, L. A. and Michael, Christopher and Rakow, Paul E. L.",
    title = "{Mesons With Excited Glue}",
    reportNumber = "LTH-106",
    doi = "10.1016/0370-2693(83)90680-9",
    journal = "Phys. Lett. B",
    volume = "129",
    pages = "351--356",
    year = "1983"
}

@article{Juge:1997nc,
    author = "Juge, K. J. and Kuti, J. and Morningstar, C. J.",
    editor = "Davies, C. T. H. and Barbour, I. M. and Bowler, K. C. and Kenway, R. D. and Pendleton, B. J. and Richards, D. G.",
    title = "{Gluon excitations of the static quark potential and the hybrid quarkonium spectrum}",
    eprint = "hep-lat/9709131",
    archivePrefix = "arXiv",
    reportNumber = "UCSD-PTH-97-31",
    doi = "10.1016/S0920-5632(97)00759-7",
    journal = "Nucl. Phys. B Proc. Suppl.",
    volume = "63",
    pages = "326--331",
    year = "1998"
}

@article{Juge:1997nd,
    author = "Juge, K. J. and Kuti, J. and Morningstar, C. J.",
    editor = "Davies, C. T. H. and Barbour, I. M. and Bowler, K. C. and Kenway, R. D. and Pendleton, B. J. and Richards, D. G.",
    title = "{Bag picture of the excited QCD vacuum with static Q anti-Q source}",
    eprint = "hep-lat/9709132",
    archivePrefix = "arXiv",
    reportNumber = "UCSD-PTH-97-32",
    doi = "10.1016/S0920-5632(97)00828-1",
    journal = "Nucl. Phys. B Proc. Suppl.",
    volume = "63",
    pages = "543--545",
    year = "1998"
}

@article{Bicudo:2015vta,
    author = {Bicudo, Pedro and Cichy, Krzysztof and Peters, Antje and Wagenbach, Bj{\"o}rn and Wagner, Marc},
    title = "{Evidence for the existence of $u d \bar{b} \bar{b}$ and the non-existence of $s s \bar{b} \bar{b}$ and $c c \bar{b} \bar{b}$ tetraquarks from lattice QCD}",
    eprint = "1505.00613",
    archivePrefix = "arXiv",
    primaryClass = "hep-lat",
    doi = "10.1103/PhysRevD.92.014507",
    journal = "Phys. Rev. D",
    volume = "92",
    number = "1",
    pages = "014507",
    year = "2015"
}

@article{Bicudo:2016ooe,
    author = "Bicudo, Pedro and Scheunert, Jonas and Wagner, Marc",
    title = "{Including heavy spin effects in the prediction of a $\bar{b} \bar{b} u d$ tetraquark with lattice QCD potentials}",
    eprint = "1612.02758",
    archivePrefix = "arXiv",
    primaryClass = "hep-lat",
    doi = "10.1103/PhysRevD.95.034502",
    journal = "Phys. Rev. D",
    volume = "95",
    number = "3",
    pages = "034502",
    year = "2017"
}

@article{Bicudo:2017szl,
    author = "Bicudo, Pedro and Cardoso, Marco and Peters, Antje and Pflaumer, Martin and Wagner, Marc",
    title = "{$u d \bar{b} \bar{b}$ tetraquark resonances with lattice QCD potentials and the Born-Oppenheimer approximation}",
    eprint = "1704.02383",
    archivePrefix = "arXiv",
    primaryClass = "hep-lat",
    doi = "10.1103/PhysRevD.96.054510",
    journal = "Phys. Rev. D",
    volume = "96",
    number = "5",
    pages = "054510",
    year = "2017"
}

@article{Bicudo:2019ymo,
    author = "Bicudo, Pedro and Cardoso, Marco and Cardoso, Nuno and Wagner, Marc",
    title = "{Bottomonium resonances with $I = 0$ from lattice QCD correlation functions with static and light quarks}",
    eprint = "1910.04827",
    archivePrefix = "arXiv",
    primaryClass = "hep-lat",
    doi = "10.1103/PhysRevD.101.034503",
    journal = "Phys. Rev. D",
    volume = "101",
    number = "3",
    pages = "034503",
    year = "2020"
}

@article{Bicudo:2020qhp,
    author = "Bicudo, Pedro and Cardoso, Nuno and Mueller, Lasse and Wagner, Marc",
    title = "{Computation of the quarkonium and meson-meson composition of the $\Upsilon(nS)$  states and of the new $\Upsilon(10753)$ Belle resonance from lattice QCD static potentials}",
    eprint = "2008.05605",
    archivePrefix = "arXiv",
    primaryClass = "hep-lat",
    doi = "10.1103/PhysRevD.103.074507",
    journal = "Phys. Rev. D",
    volume = "103",
    number = "7",
    pages = "074507",
    year = "2021"
}

@article{Bicudo:2022ihz,
    author = "Bicudo, Pedro and Cardoso, Nuno and Mueller, Lasse and Wagner, Marc",
    title = "{Study of I=0 bottomonium bound states and resonances in S, P, D, and F waves with lattice QCD static-static-light-light potentials}",
    eprint = "2205.11475",
    archivePrefix = "arXiv",
    primaryClass = "hep-lat",
    doi = "10.1103/PhysRevD.107.094515",
    journal = "Phys. Rev. D",
    volume = "107",
    number = "9",
    pages = "094515",
    year = "2023"
}

@other{Brambilla:2026zfw,
    author = "Brambilla, Nora and Mohapatra, Abhishek and Scirpa, Tommaso and Vairo, Antonio",
    title = "{Open-flavor threshold effects on quarkonium spectrum in the BOEFT}",
    eprint = "2604.12603",
    archivePrefix = "arXiv",
    primaryClass = "hep-ph",
    reportNumber = "TUM-EFT 199/25",
    month = "4",
    year = "2026"
}

@other{Alasiri:2026vfz,
    author = "Alasiri, Fareed and Braaten, Eric and Bruschini, Roberto",
    title = "{$\boldsymbol{\chi_{c1}}(3872)$ and its Partners in the Diabatic Born-Oppenheimer Approximation for QCD}",
    eprint = "2606.12309",
    archivePrefix = "arXiv",
    primaryClass = "hep-ph",
    month = "6",
    year = "2026"
}

@article{Bali:2005fu,
    author = {Bali, Gunnar S. and Neff, Hartmut and Duessel, Thomas and Lippert, Thomas and Schilling, Klaus},
    collaboration = {SESAM},
    title = {{Observation of string breaking in QCD}},
    eprint = {hep-lat/0505012},
    archivePrefix = {arXiv},
    doi = {10.1103/PhysRevD.71.114513},
    journal = {Phys. Rev. D},
    volume = {71},
    pages = {114513},
    year = {2005}
}

@article{Beane:2000wh,
    author = "Beane, S. R. and Bedaque, Paulo F. and Childress, L. and Kryjevski, A. and McGuire, J. and van Kolck, U.",
    title = "{Singular potentials and limit cycles}",
    eprint = "quant-ph/0010073",
    archivePrefix = "arXiv",
    reportNumber = "NT-UW-00-023, DOE-ER-41132-102-INT-00, RBRC-140, KRL-MAP-271, NT@UW-00-023",
    doi = "10.1103/PhysRevA.64.042103",
    journal = "Phys. Rev. A",
    volume = "64",
    pages = "042103",
    year = "2001"
}

@article{Beane:2001bc,
    author = "Beane, S. R. and Bedaque, Paulo F. and Savage, M. J. and van Kolck, U.",
    title = "{Towards a perturbative theory of nuclear forces}",
    eprint = "nucl-th/0104030",
    archivePrefix = "arXiv",
    reportNumber = "NT-UW-01-06, LBNL-47692, RBRC-186, JLAB-THY-02-84",
    doi = "10.1016/S0375-9474(01)01324-0",
    journal = "Nucl. Phys. A",
    volume = "700",
    pages = "377--402",
    year = "2002"
}

@article{Epelbaum:2002ji,
    author = "Epelbaum, E. and Nogga, A. and Gl{\"o}ckle, Walter and Kamada, H. and Meissner, U. G. and Witala, H.",
    title = "{Few nucleon systems with two nucleon forces from chiral effective field theory}",
    eprint = "nucl-th/0201064",
    archivePrefix = "arXiv",
    reportNumber = "FZJ-IKP-TH-2002-03",
    doi = "10.1140/epja/i2002-10048-2",
    journal = "Eur. Phys. J. A",
    volume = "15",
    pages = "543--563",
    year = "2002"
}

@article{Yang:2020pgi,
    author = {Yang, C.-J. and Ekstr{\"o}m, A. and Forss{\'e}n, C. and Hagen, G.},
    title = "{Power counting in chiral effective field theory and nuclear binding}",
    eprint = "2011.11584",
    archivePrefix = "arXiv",
    primaryClass = "nucl-th",
    doi = "10.1103/PhysRevC.103.054304",
    journal = "Phys. Rev. C",
    volume = "103",
    number = "5",
    pages = "054304",
    year = "2021"
}

@article{Birse:2005um,
    author = "Birse, Michael C.",
    title = "{Power counting with one-pion exchange}",
    eprint = "nucl-th/0507077",
    archivePrefix = "arXiv",
    doi = "10.1103/PhysRevC.74.014003",
    journal = "Phys. Rev. C",
    volume = "74",
    pages = "014003",
    year = "2006"
}

@article{Kukulin:1978he,
    author = "Kukulin, V. I. and Pomerantsev, V. N.",
    title = "{The Orthogonal Projection Method in Scattering Theory}",
    doi = "10.1016/0003-4916(78)90069-6",
    journal = "Annals Phys.",
    volume = "111",
    pages = "330--363",
    year = "1978"
}

@article{Wilson:1970ag,
    author = "Wilson, Kenneth G.",
    title = "{The Renormalization Group and Strong Interactions}",
    reportNumber = "SLAC-PUB-0807",
    doi = "10.1103/PhysRevD.3.1818",
    journal = "Phys. Rev. D",
    volume = "3",
    pages = "1818",
    year = "1971"
}

@article{Bedaque:1998kg,
    author = "Bedaque, Paulo F. and Hammer, H.-W. and van Kolck, U.",
    title = "{Renormalization of the three-body system with short range interactions}",
    eprint = "nucl-th/9809025",
    archivePrefix = "arXiv",
    reportNumber = "DOE-ER-40561-28, TRI-PP-98-17, KRL-MAP-235, NT-UW-98-24",
    doi = "10.1103/PhysRevLett.82.463",
    journal = "Phys. Rev. Lett.",
    volume = "82",
    pages = "463--467",
    year = "1999"
}

@article{Jacob:1959at,
    author = {Jacob, M. and Wick, G. C.},
    title = {{On the General Theory of Collisions for Particles with Spin}},
    doi = {10.1016/0003-4916(59)90051-X},
    journal = {Annals Phys.},
    volume = {7},
    pages = {404--428},
    year = {1959}
}

@article{TarrusCastella:2022rxb,
    author = {Tarr{\'u}s Castell{\`a}, Jaume},
    title = {{Heavy meson thresholds in Born-Oppenheimer effective field theory}},
    eprint = {2207.09365},
    archivePrefix = {arXiv},
    primaryClass = {hep-ph},
    doi = {10.1103/PhysRevD.106.094020},
    journal = {Phys. Rev. D},
    volume = {106},
    number = {9},
    pages = {094020},
    year = {2022}
}

@article{Wang:2019mxn,
    author = "Wang, Zhi-Gang",
    title = "{Analysis of the hidden-bottom tetraquark mass spectrum with the QCD sum rules}",
    eprint = "1903.10895",
    archivePrefix = "arXiv",
    primaryClass = "hep-ph",
    doi = "10.1140/epjc/s10052-019-7019-6",
    journal = "Eur. Phys. J. C",
    volume = "79",
    number = "6",
    pages = "489",
    year = "2019"
}

@misc{nullius,
author = "Guo, Feng-Kun",
title = "Nullius: an evidence-first harness for agent-assisted research",
url = "https://github.com/fkguo/nullius",
note = "https://github.com/fkguo/nullius"
}

@article{Zhang:2026nmz,
    author = "Zhang, Zi-Zhao and Liu, Bo-Chao",
    title = "{Mass spectrum and decay widths of charmonium-like mesons: A diabatic approach with complex scaling}",
    eprint = "2609.29909",
    archivePrefix = "arXiv",
    primaryClass = "hep-ph",
    month = "9",
    year = "2026",
    journal = ""
}

@article{Bali:2003jq,
    author = "Bali, Gunnar S. and Pineda, Antonio",
    title = "{QCD phenomenology of static sources and gluonic excitations at short distances}",
    eprint = "hep-ph/0310130",
    archivePrefix = "arXiv",
    reportNumber = "UB-ECM-PF-03-22, GUTPA-03-10-01",
    doi = "10.1103/PhysRevD.69.094001",
    journal = "Phys. Rev. D",
    volume = "69",
    pages = "094001",
    year = "2004"
}

\end{document}